\documentclass[10pt,preprint2]{aastex}

\usepackage{amsmath}

\def\p/{\mbox{$^1$}}
\def\pp/{\mbox{$^2$}}
\def\ppp/{\mbox{$^3$}}
\def\pppp/{\mbox{$^4$}}
\def\m/{\mbox{$^{-1}$}}
\def\mm/{\mbox{$^{-2}$}}
\def\mmm/{\mbox{$^{-3}$}}
\def\mmmm/{\mbox{$^{-4}$}}
\def\Ms/{\mbox{M$_\odot$}}

\def\rv{\mbox{$R_{5495}$}}

\def\Bap{\mbox{$B_{\rm ap}$}}
\def\Jap{\mbox{$J_{\rm ap}$}}
\newcommand{\HeI}[1]{\mbox{He\,{\sc i}~$\lambda${#1}}}
\newcommand{\HeII}[1]{\mbox{He\,{\sc ii}~$\lambda${#1}}}

\newcommand{\CIII}[1]{\mbox{C\,{\sc iii}~$\lambda${#1}}}

\newcommand{\NIII}[1]{\mbox{N\,{\sc iii}~$\lambda${#1}}}

\newcommand{\SiIV}[1]{\mbox{Si\,{\sc iv}~$\lambda${#1}}}
\newcommand{\MgII}[1]{\mbox{Mg\,{\sc ii}~$\lambda${#1}}}

\shorttitle{The Galactic O-Star Spectroscopic Survey. III}
\shortauthors{Ma\'{\i}z Apell\'aniz et al.}

\slugcomment{Submitted for publication to the Astrophysical Journal Supplement Series}

\begin{document}

\title{\vspace{-3cm}The Galactic O-Star Spectroscopic Survey (GOSSS). \linebreak III. 142 additional O-type systems\altaffilmark{1}}

\author{\vspace{-5mm}J. Ma\'{\i}z Apell\'aniz\altaffilmark{2,3,4,5}}
\affil{Centro de Astrobiolog\'{\i}a, CSIC-INTA, campus ESAC, camino bajo del castillo s/n, E-28\,692 Madrid, Spain}
\author{\vspace{-5mm}A. Sota\altaffilmark{2,3,4}}
\affil{Instituto de Astrof\'{\i}sica de Andaluc\'{\i}a-CSIC, Glorieta de la Astronom\'{\i}a s/n, E-18\,008 Granada, Spain}
\author{\vspace{-5mm}J. I. Arias, R. H. Barb\'a\altaffilmark{2}}
\affil{Departamento de F\'{\i}sica, Universidad de La Serena, Av. Cisternas 1200 Norte, La Serena, Chile}
\author{\vspace{-5mm}N. R. Walborn}
\affil{Space Telescope Science Institute, 3700 San Martin Drive, Baltimore, MD 21\,218, USA}
\author{\vspace{-5mm}S. Sim\'on-D{\'\i}az}
\affil{Instituto de Astrof{\'\i}sica de Canarias, E-38\,200 La Laguna, Tenerife, Spain}
\affil{Departamento de Astrof{\'\i}sica, Universidad de La Laguna, E-38\,205 La Laguna, Tenerife, Spain}
\author{\vspace{-5mm}I. Negueruela}
\affil{DFISTS, EPS, Universidad de Alicante, carretera San Vicente del Raspeig s/n, E-03\,690 Alicante, Spain}
\author{\vspace{-5mm}A. Marco}
\affil{DFISTS, EPS, Universidad de Alicante, carretera San Vicente del Raspeig s/n, E-03\,690 Alicante, Spain}
\affil{Dept. of Astronomy, University of Florida, 211 Bryant Space Science Center, Gainesville, FL 32\,611, USA}
\author{\vspace{-5mm}J. R. S. Le{\~a}o}
\affil{Univ. Federal do Rio Grande do Norte - UFRN, Caixa Postal 1524, CEP 59\,078-970, Natal - RN, Brazil}
\author{\vspace{-5mm}A. Herrero}
\affil{Instituto de Astrof{\'\i}sica de Canarias, E-38\,200 La Laguna, Tenerife, Spain}
\affil{Departamento de Astrof{\'\i}sica, Universidad de La Laguna, E-38\,205 La Laguna, Tenerife, Spain}
\author{\vspace{-5mm}R. C. Gamen}
\affil{Instituto de Astrof\'{\i}sica de La Plata (CONICET, UNLP), Paseo del Bosque s/n, 1900 La Plata, Argentina}
\author{\vspace{-5mm}E. J. Alfaro}
\affil{Instituto de Astrof\'{\i}sica de Andaluc\'{\i}a-CSIC, Glorieta de la Astronom\'{\i}a s/n, E-18\,008 Granada, Spain}
\affil{\vspace{-1.2cm}$\,\!$}


\altaffiltext{1}{The GOSSS spectroscopic data in this article were gathered with five facilities: 
the 1.5~m Telescope at the \facility{Observatorio de Sierra Nevada} (OSN), 
the 2.5~m du Pont Telescope at \facility{Las Campanas Observatory} (LCO), 
the 3.5~m Telescope at \facility{Calar Alto Observatory} (CAHA), 
and the 4.2~m William Herschel Telescope (WHT) and 10.4~m Gran Telescopio Canarias (GTC) at \facility{Observatorio del Roque de los Muchachos} (ORM). 
}

\altaffiltext{2}{Visiting Astronomer, LCO, Chile.}
\altaffiltext{3}{Visiting Astronomer, CAHA, Spain.}
\altaffiltext{4}{Visiting Astronomer, WHT, Spain.}
\altaffiltext{5}{e-mail contact: {\tt jmaiz@cab.inta-csic.es}.}

\begin{abstract}
This is the third installment of GOSSS, a massive spectroscopic survey of Galactic O stars, based on new homogeneous, high signal-to-noise ratio, 
$R\sim 2500$ digital observations selected from the Galactic O-Star Catalog (GOSC). In this paper we present 142 additional stellar systems with O stars from both 
hemispheres, bringing the total of O-type systems published within the project to 590. Among the new objects there are 20 new O stars. We also identify 11 new double-lined 
spectroscopic binaries (SB2s), of which 6 are of O+O type and 5 of O+B type, and an additional new tripled-lined spectroscopic binary (SB3) of O+O+B type. We also revise 
some of the previous GOSSS classifications, present some egregious examples of stars erroneously classified as O-type in the past, introduce the use of luminosity class 
IV at spectral types O4-O5.5, and adapt the classification scheme to the work of \citet{Ariaetal16}.  
\end{abstract}

\keywords{binaries:spectroscopic --- binaries:visual --- stars:early type --- stars:emission line,Be --- surveys}

\section{Introduction}
\label{sec:Intr}

The Galactic O-Star Spectroscopic Survey (GOSSS) is a long-term project that is obtaining homogeneous, high SNR, $R\sim2500$, blue-violet 
spectra of a large number (1000+) of O stars in the Milky Way and deriving accurate and self-consistent spectral types for all of them
\citep{Maizetal11}. In \citet{Sotaetal11a}, from now on paper I, we presented the first installment of the survey, which was comprised of 
the results for 178 northern ($\delta > -20^{\circ}$) O stars. In \citet{Sotaetal14}, from now on paper II, we extended the sample to the 
southern hemisphere for a total of 448 O stars. Papers I and II concentrated on the brightest O stars with the aim of achieving completeness 
down to $B = 8$ but they also included many dimmer stars. This third paper continues the previous work by adding 142 new stars and raising
the sample size to 590. Most of the new stars are of luminosity class V, which are relevant to the OVz phenomenon \citep{Ariaetal16}, but objects of 
other luminosity classes are also included. 

This paper is organized as follows. We first present the changes in the observational setup and the classification scheme in section~2. Then, the spectral 
classifications are shown in section~3, divided in updates to O stars present in papers I and II, new O stars, and late-type stars previously misclassified as 
of O type. Finally, in section~4 we analyze the status of the project based on the new spectral classifications.

\section{Data and methods}
\label{sec:Data}

\subsection{Blue-violet spectroscopy with $R$~$\sim$~2500}
\label{sec:GOSSS}

The GOSSS data were described in papers I and II and the reader is referred there for further information. Here we detail the changes from those previous works.

Most of the spectra presented in paper I were obtained with the Albireo (Observatorio de Sierra Nevada - OSN - 1.5~m telescope) and TWIN (Calar Alto - CAHA - 3.5~m telescope) 
spectrographs. On the other hand, most of the spectra in paper II were obtained with the Boller \& Chivens spectrograph at the Las Campanas  (LCO) 2.5 m du Pont telescope). 
Starting with paper II, some spectra were also obtained with the ISIS spectrograph at the 4.2~m William Herschel Telescope (WHT) at the Observatorio del Roque de los Muchachos (ORM) 
in La Palma, Spain. 
In this paper we use data from all of the above instruments and we also add a new one, OSIRIS, at the 10.4~m Gran Telescopio Canarias (GTC) at the
ORM (Table~\ref{settings}). Only a few GTC spectra are used here but the number will increase in future GOSSS papers, as we have already 
acquired data for over 200 stars, most of them too dim to be accessible with the other telescopes mentioned above. One difference between the GTC setup and the 
rest is that we use two volume-phased holographic gratings, R2500U and R2500V, allowing us to cover a larger wavelength range than with the other instruments. The exposures for the 
two gratings are taken consecutively, with the time difference between the first one and the last one being always less than one hour in order to avoid changes in the phase of rapidly 
moving spectroscopic binaries. As we do with the rest of the spectrographs, we use checks to compare that the quality of the data from all the spectrographs is uniform and, in those 
cases where the spectral resolution is higher than 2500, we degrade it to that value.

We have also started taking GOSSS data with [a] the GMOS spectrograph at the 8.1~m Gemini South telescope, [b] the Goodman High Throughput Spectrograph at the 4.1~m SOAR Telescope (both 
at Cerro Pach{\'o}n, Chile), and [c] FRODOspec at the 2.0~m Liverpool Telescope (at the ORM) but we do not use them in this paper; their first GOSSS spectra will likely appear in a future paper IV.

The GOSSS data in this paper were obtained between 2007 and 2015. The OSN and LCO observations were obtained in visitor mode, the GTC observations in service mode, and the CAHA
observations in a combination of both. For some SB2 and SB3 spectroscopic binaries, multiple epochs were obtained to observe the different orbit phases. In cases with 
known orbits, observations near quadrature were attempted. 

The spectral classifications which are the main content of this paper are presented in Tables~\ref{spectralclasold},~\ref{spectralclasnew},~and~\ref{spectralclasbad} and they were
obtained with MGB \cite{Maizetal12,Maizetal15b}. 

\begin{table*}
\caption{Telescopes, instruments, and settings used in this paper.}
\centerline{
\begin{tabular}{lccccc}
\\
\hline
\multicolumn{1}{c}{Telescope} & Spectrograph      & Grating & Spectral scale & Spatial scale & Wav. range  \\
                             &                   &         & (\AA/px)       & (\arcsec/px)  & (\AA)       \\
\hline
LCO 2.5 m (du Pont)           & Boller \& Chivens &  1200   & 0.80           & 0.71          & 3900$-$5500 \\
OSN 1.5 m                     & Albireo           &  1800   & 0.62           & 0.83          & 3750$-$5070 \\
CAHA 3.5 m                    & TWIN (blue arm)   &  1200   & 0.55           & 0.58          & 3930$-$5020 \\
ORM 4.2 m (WHT)               & ISIS (blue arm)   &   600   & 0.44           & 0.20          & 3900$-$5600 \\
ORM 10.4 m (GTC)              & OSIRIS            &  2500U  & 0.62           & 0.25          & 3440$-$4610 \\
                              &                   &  2500V  & 0.80           & 0.25          & 4500$-$6000 \\
\hline
\end{tabular}
}
\label{settings}
\end{table*}

\subsection{Cataloguing}
\label{sec:cat}

The spectral types are available through the latest version (currently v3.2) of the Galactic O-Star Catalog (GOSC, \citealt{Maizetal04b}), 
accessible at \url{http://gosc.iaa.es}. Starting in 
version 3, the GOSSS spectral types are the default ones and the basis for the catalog selection, though older classifications and those 
obtained with high-resolution spectra are also kept as possible additional columns. $B$- and $J$-band photometry are also provided in GOSC 
for all stars\footnote{See \citet{Maizetal13} for details, we use \Bap\ and \Jap, respectively, to refer to the photometry in GOSC, where ``ap'' 
refers to approximate and is intended to be significant only to one tenth of a magnitude.}. The rectified GOSSS spectra can also be obtained from 
GOSC as FITS tables. 

The previous version of GOSC (3.1.2) had only entries in the main catalog, the part corresponding to O stars. Versions 2.0-2.4 of GOSC \citep{Sotaetal08}
did include four supplements for WR and WR+O systems, other early-type stars, late-type stars, and extragalactic massive stars, respectively. 
Since here we are presenting some late-type stars (see below), we are reintroducing the third supplement in GOSC 3.2. In future papers, we plan to 
release GOSSS spectral types for the other three supplements and we will populate GOSC accordingly.

The GOSSS spectral types were first made available at GOSC in June 2013 as part of GOSSS Data Release 1.0 (GOSSS-DR1.0, \citealt{Sotaetal13b}). 
Later changes led to Data Releases 1.1 (with the spectral types from \citealt{Sotaetal14}) and 1.1.1 (minor changes in August 2014). 
The addition of new spectral types and changes to previous ones presented here constitutes GOSSS-DR2.0.

\subsection{Spectral classification methodology and the new standard grid}
\label{sec:Clas}

The spectral classification methodology was laid out in paper I and some changes were presented in paper II. Here we introduce two additional changes.

The first one concerns the OVz phenomenon, described in detail in \citet{SabSetal14} and \citet{Ariaetal16}. OVz stars have deep \HeII{4686} absorption lines,
likely caused primarily by their extreme youth, though additional factors may also play a role. In paper II, we defined the $z$ ratio as:

\begin{equation}
z = \frac{{\rm EW}(\HeII{4686})}{{\rm Max}[{\rm EW}(\HeI{4471}),{\rm EW}(\HeII{4542})]}
\end{equation}

\noindent by measuring the equivalent widths of \HeI{4471}, \HeII{4542}, and \HeII{4686} in our main-sequence standard stars and classified a star
as OVz when $z$ was greater than $\sim 1.0$. \citet{Ariaetal16} analyzed the OVz phenomenon and decided to raise the critical value of $z$ to 1.1
to avoid the borderline, often unclear, cases. Here we adopt that criterion and reclassify some of the stars in papers I and II.

The second one concerns the range of spectral types for which luminosity class IV is defined.
When the O-type luminosity classification was introduced and calibrated by \citet{Walb71a,Walb72,Walb73a} only three classes (V, III, I) were defined 
for spectral types earlier than O6. The luminosity range is relatively small at those types, and there was concern about possible differences in the strong 
stellar winds unrelated to luminosity affecting the Of criteria at the earliest types. Nevertheless, bright giants (class II) and supergiant subcategories 
(Ia, Iab, and Ib) were defined as early as types O6-O6.5, while \citet{Sotaetal11a} extended the use of class IV as well to those types in the larger, 
higher quality GOSSS sample. It is also relevant that subsequent UV work demonstrated a high degree of correlation (and, by implication, robustness) between 
the stellar-wind profiles and the optical spectral types throughout the entire O-type range \citep{Walbetal85}. 

In the context of the quantification of the 
$z$ ratio by \citet{Ariaetal16}, we have become aware of several spectra classified as type V with excessively small values of the ratio, i.e. in the range 
0.5-0.7. A number of them are of types O4-O5.5, and inspection from this viewpoint reveals Of morphologies in the GOSSS data well intermediate between most 
class V spectra and the class III standards, in terms of both the \HeII{4686} absorption and \NIII{4634-41-42} emission strengths. Multiple observations of several 
of these spectra demonstrate robustness in these features. Hence we now define luminosity class IV at types O4-O5.5 to describe such spectra. In view of the 
strengths of the Of features in most of them, we adopt the (f)-parameter for them, as opposed to ((f)) for class IV at types O6 and later. We also note that 
HD~93\,250~AB was previously assigned to class III by \citet{Sotaetal14}, whereas it is clearly more similar to the other class IV objects; originally it had 
been classified as V. Thus, this classification enhancement improves the consistency of the luminosity classification at these types.

Calibration work to determine whether this classification development corresponds to a consistent luminosity difference remains for the future, likely depending 
upon Gaia \citep{Perretal01} measurements for reliable results. However, regardless of how that turns out, the development provides a more precise systematic description of the 
spectra themselves.

The two changes above (the critical value for the OVz phenomenon and the definition of luminosity class IV for O4-O5.5 stars) imply updates on the standard grid
used for spectral classification. The first GOSSS-based standard grid (OB2500 v1.0) was presented in \citet{Sotaetal11a} and the second one (OB2500 v2.0), which 
filled some gaps, in \citet{Maizetal15b}. Here we present OB2500 v3.0 in Table~\ref{standards}, with an additional column for OVz standards and some gaps in the grid
filled with respect to the previous version. We plan to keep filling those gaps when we find additional appropriate stars within our survey. Note that most of the 
gaps occur at the earliest types (O5.5 and earlier for the III to Vz columns, O7.5 and earlier for the supergiants), which is expected given the scarcity of those types
in the solar neighborhood. One possibility we will consider in the future is the use of LMC O stars.

\begin{table*}
\caption{The OB2500 v3.0 grid of classification standards.}
\tiny
\begin{tabular}{lllllllll}
\\
\hline
 & \multicolumn{1}{c}{Vz} & \multicolumn{1}{c}{V} & \multicolumn{1}{c}{IV} & \multicolumn{1}{c}{III} & \multicolumn{1}{c}{II} & \multicolumn{1}{c}{Ib} & \multicolumn{1}{c}{Iab/I} & \multicolumn{1}{c}{Ia} \\
\hline
O2   &                    &                       &                        &                          &                         &                   & {\it HD 93\,129 AaAb} &                    \\
\hline
O3   & {\it HD 64\,568}   & \nodata               &                        & \nodata                  &                         &                   &      Cyg OB2-7        &                    \\
\hline
O3.5 & {\it HD 93\,128}   & \nodata               &                        & {\it Pismis 24-17}       &                         &                   & {\it NGC 3603 HST-48} &                    \\
\hline
O4   & {\it HD 96\,715}   & {\bf HD 46\,223}      & {\bf HD 168\,076 AB}   & \nodata                  &                         &                   &      HD 15\,570       &                    \\
     &                    &                       & {\it HD 93\,250 AB}    &                          &                         &                   &      HD 16\,691       &                    \\
     &                    &                       &                        &                          &                         &                   &      HD 190\,429 A    &                    \\
\hline
O4.5 & \nodata            &      HD 15\,629       &      HD 193\,682       & \nodata                  &                         &                   &      HD 14\,947       &                    \\
     &                    & {\it HDE 303\,308 AB} &                        &                          &                         &                   &      Cyg OB2-9        &                    \\
\hline
O5   & {\bf HD 46\,150}   & {\it HDE 319\,699}    & {\bf HD 168\,112 AB}   & {\it HD 93\,843}         &                         &                   & {\it CPD -47 2963 AB} &                    \\
\hline
O5.5 & \nodata            & {\it HD 93\,204}      & \nodata                & \nodata                  &                         &                   &      Cyg OB2-11       &                    \\
     &                    &                       &                        &                          &                         &                   & {\it ALS 18\,747}     &                    \\
\hline
O6   &      HD 42\,088    & {\bf ALS 4880}        & {\it HD 101\,190 AaAb} &      HDE 338\,931        &      HDE 229\,196       & \nodata           & \nodata               & {\bf HD 169\,582}  \\
     & {\it HDE 303\,311} & {\it CPD -59 2600}    &                        &                          &                         &                   &                       &                    \\
\hline
O6.5 & {\it HD 91\,572}   & {\bf HD 167\,633}     & {\it HDE 322\,417}     &      HD 190\,864         & {\bf HD 157\,857}       & \nodata           & \nodata               & {\it HD 163\,758}  \\
     &                    &      HD 12\,993       &                        & {\it HD 96\,946}         &                         &                   &                       &                    \\
     &                    &                       &                        & {\it HD 152\,723 AaAb}   &                         &                   &                       &                    \\
     &                    &                       &                        & {\it HD 156\,738 AB}     &                         &                   &                       &                    \\
\hline
O7   & {\it HD 97\,966}   & {\it HD 93\,146 A}    &      ALS 12\,320       &      Cyg OB2-4 A         & {\it HD 94\,963}        & {\it HD 69\,464}  & \nodata               & \nodata            \\
     & {\it CPD -58 2620} & {\it HD 93\,222 AB}   &                        & {\it HD 93\,160 AB}      & {\it HD 151\,515}       &      HD 193\,514  &                       &                    \\
     &      HDE 242\,926  &                       &                        &                          &                         &                   &                       &                    \\
     & {\it HD 91\,824}   &                       &                        &                          &                         &                   &                       &                    \\
\hline
O7.5 & {\it HD 152\,590}  &      HD 35\,619       & {\it HD 97\,319}       & {\it HD 163\,800}        &      HD 34\,656         &      HD 17\,603   &      HD 192\,639      & \nodata            \\
     &                    &                       &                        &                          & {\bf HD 171\,589}       & {\it HD 156\,154} & {\bf 9 Sge}           &                    \\
\hline
O8   & {\it HDE 305\,539} & {\it HD 101\,223}     & {\it HD 94\,024}       & {\it HDE 319\,702}       & {\it 63 Oph}            & {\bf BD -11 4586} &      HD 225\,160      & {\it HD 151\,804}  \\
     & {\it HDE 305\,438} & {\it HD 97\,848}      & {\it HD 135\,591}      & {\bf $\lambda$ Ori A}    &                         &                   &                       &                    \\
     &                    &      HD 191\,978      &                        &                          &                         &                   &                       &                    \\
\hline
O8.5 &                    & {\it HDE 298\,429}    & {\bf HD 46\,966 AaAb}  & {\it HD 114\,737 AB}     & {\it HD 75\,211}        & {\it HD 125\,241} & \nodata               & {\it HDE 303\,492} \\
     &                    &      HD 14\,633 AaAb  &                        &      HD 218\,195 A       &      HD 207\,198        &                   &                       &                    \\
     &                    & {\bf HD 46\,149}      &                        &                          &                         &                   &                       &                    \\
     &                    & {\it HD 57\,236}      &                        &                          &                         &                   &                       &                    \\
     &                    & {\it Trumpler 14-9}   &                        &                          &                         &                   &                       &                    \\
\hline
O9   &                    &      10 Lac           & {\it HD 93\,028}       & {\it HD 93\,249 A}       & {\it HD 71\,304}        &      19 Cep       &      HD 202\,124      &      $\alpha$ Cam  \\
     &                    &      HD 216\,898      & {\it CPD -41 7733}     &      HD 24\,431          & {\it $\tau$ CMa AaAb}   &                   & {\it HD 152\,249}     &                    \\
     &                    & {\it CPD -59 2551}    &                        &                          &                         &                   &      HD 210\,809      &                    \\
\hline
O9.2 &                    & {\bf HD 46\,202}      & {\it HD 96\,622}       & {\it CPD -35 2105 AaAbB} &      ALS 11\,761        & {\it HD 76\,968}  & {\it HD 154\,368}     & {\it HD 152\,424}  \\
     &                    &      HD 12\,323       &                        &      HD 16\,832          &                         &                   & {\it HD 123\,008}     &                    \\
     &                    &                       &                        &                          &                         &                   &      HD 218\,915      &                    \\
\hline
O9.5 &                    &      AE Aur           &      HD 192\,001       & {\it HD 96\,264}         & {\bf $\delta$ Ori AaAb} & \nodata           &      HD 188\,209      & \nodata            \\
     &                    & {\it $\mu$ Col}       & {\it HD 93\,027}       &                          &                         &                   &                       &                    \\
     &                    &                       & {\it HD 155\,889 AB}   &                          &                         &                   &                       &                    \\
\hline
O9.7 &                    & {\bf $\upsilon$ Ori}  &      HD 207\,538       &      HD 189\,957         & {\it HD 68\,450}        & {\bf HD 47\,432}  &      HD 225\,146      &      HD 195\,592   \\
     &                    &                       &                        & {\it HD 154\,643}        & {\it HD 152\,405}       & {\it HD 154\,811} & {\it $\mu$ Nor}       & {\it GS Mus}       \\
     &                    &                       &                        &                          &      HD 10\,125         & {\it HD 152\,147} & {\it HD 104\,565}     &                    \\
     &                    &                       &                        &                          &                         &                   &      HD 191\,781      &                    \\
\hline
Notes & \multicolumn{8}{l}{Normal, {\it italic}, and {\bf bold} typefaces are used for stars with $\delta > +20\degr$, $\delta < -20\degr$, and the equatorial intermediate region, respectively.}
\end{tabular}

\label{standards}
\end{table*}

\section{Results}
\label{sec:Res}

This section constitutes the main body of the paper, the spectral classifications, and is divided in three parts. First, we revise some of the results 
from papers I and II. Second, we present new spectral classifications for O stars not included in those papers. Third, we show some cases previously 
classified in the literature or in Simbad as being O stars which turn out to be of late type. The information is given in 
Tables~\ref{spectralclasold},~\ref{spectralclasnew},~and~\ref{spectralclasbad}, with details about each star (sorted by GOS ID within each subsection) 
provided in the text.

\subsection{Stellar systems from papers I and II}
\label{sec:old}

\subsubsection{Name changes due to the discovery of companions}

In the last two years, several papers have used high-spatial resolution techniques to detect new bright visual close companions to stars present in papers I and II. The 
proximity of those companions (with separations of tens of miliarcseconds or less) do not allow us 
to obtain separate GOSSS spectra for them but the nomenclature used in GOSSS (a companion must be included in the star name if it contributes a significant fraction of
light to the blue-violet spectrum, established as $\Delta B\le 2.0$) makes us change the name of the star from the one listed in our previous papers. The name changes 
are given in Table~\ref{literature}, where no spectral types are listed since they remain unchanged unless the system is also listed in the next subsubsection. 
Table~\ref{literature} gives the current GOSSS name (with all the currently known bright components included in the Washington Double Star Catalog, WDS, \citealt{Masoetal01}),
the GOSSS ID, the reference, the new component, and possibly a comment. Note that most of the new components are from \citet{Sanaetal14}.

\subsubsection{Spectral type changes}

In Table~\ref{spectralclasold} we list the stellar systems already included in papers I and II for which we have obtained revised spectral types. They are ordered by 
GOSSS ID, which corresponds for practical purposes to an ordering by Galactic longitude $l$\footnote{The only exceptions being systems with very similar values of $l$.}
The corresponding spectrograms are shown in the same order in Fig.~\ref{fig:old}. The majority of the systems in this section are included because of a change in the
z suffix, as described in subsection~\ref{sec:Clas}. A minority are included due to the definition of luminosity class IV for types O4-O5.5 (see subsection~\ref{sec:Clas})
or the assignment of an SB2 status to the system in the GOSSS data that was not possible at the time of papers I and II. Since all of these systems have been already
discussed in the previous GOSSS papers, we only present additional details about some of them in this section. In particular, we discuss several SB2 systems for which we
have fine-tuned the spectral classification using the new standard grid.

\begin{figure*}
\centerline{\includegraphics*[width=\linewidth]{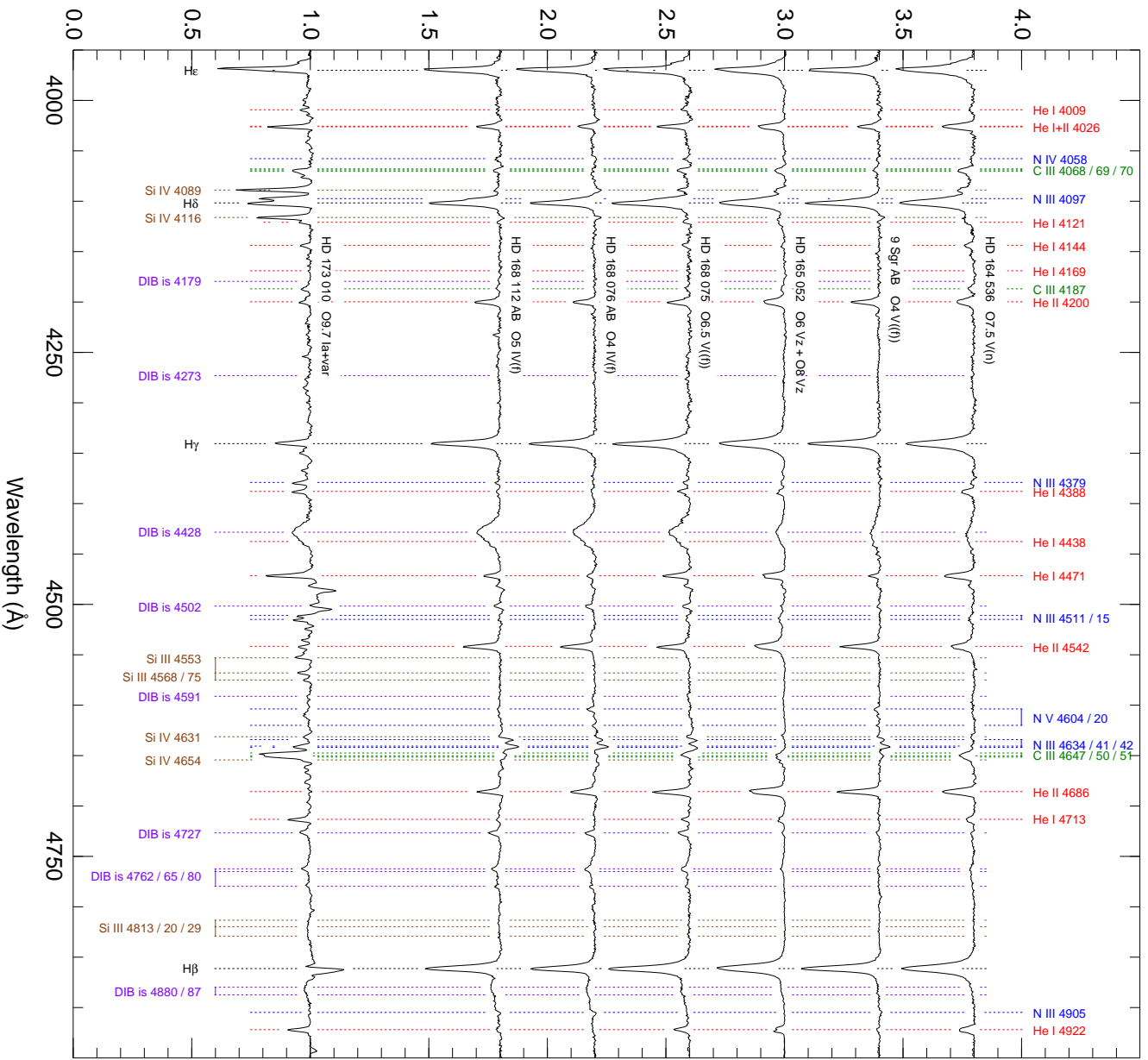}}
\caption{Spectrograms for stars already present in papers I and II. The targets are sorted by GOS ID.}
\label{fig:old}
\end{figure*}	

\addtocounter{figure}{-1}

\begin{figure*}
\centerline{\includegraphics*[width=\linewidth]{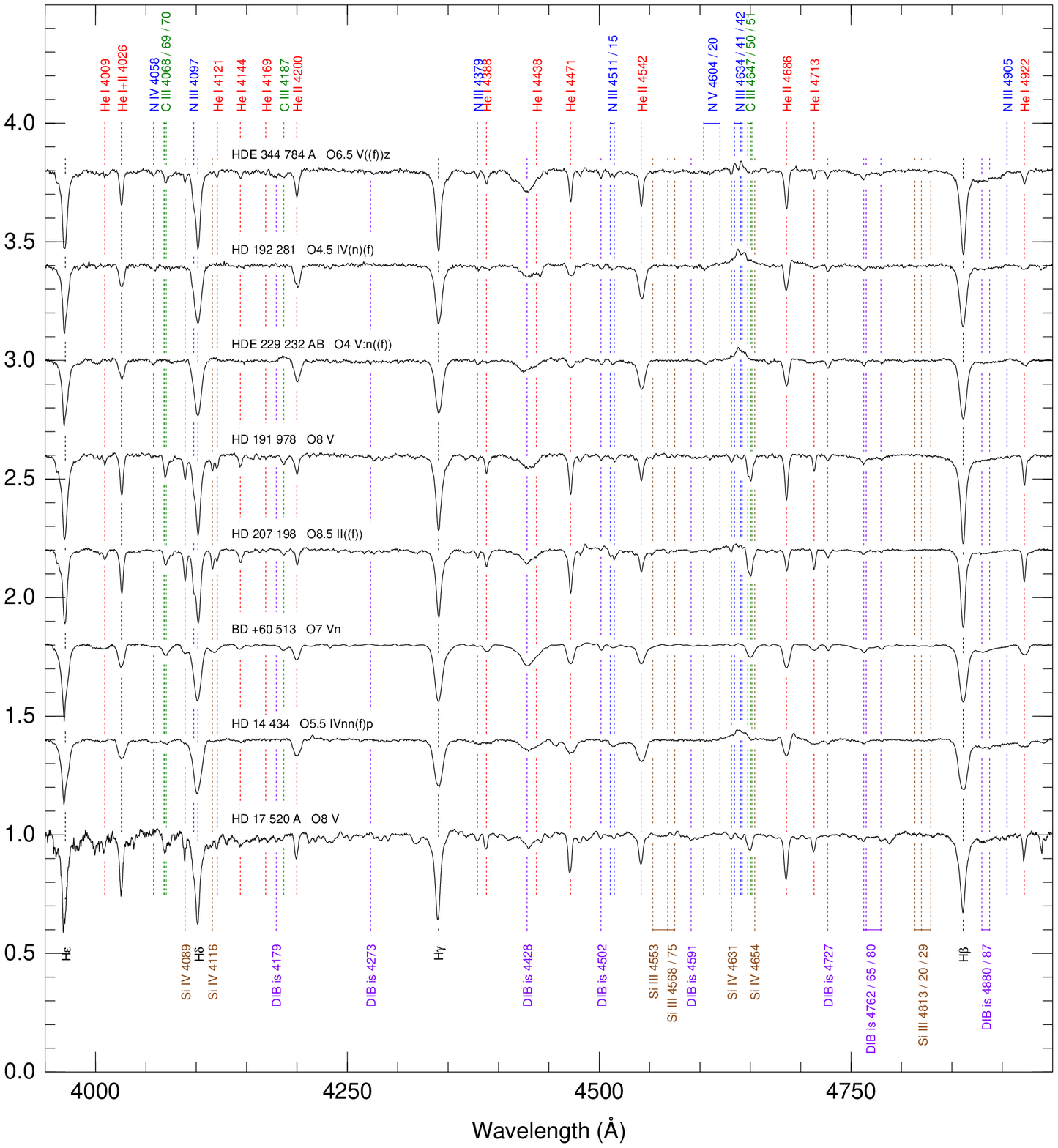}}
\caption{(continued).}
\end{figure*}

\addtocounter{figure}{-1}

\begin{figure*}
\centerline{\includegraphics*[width=\linewidth]{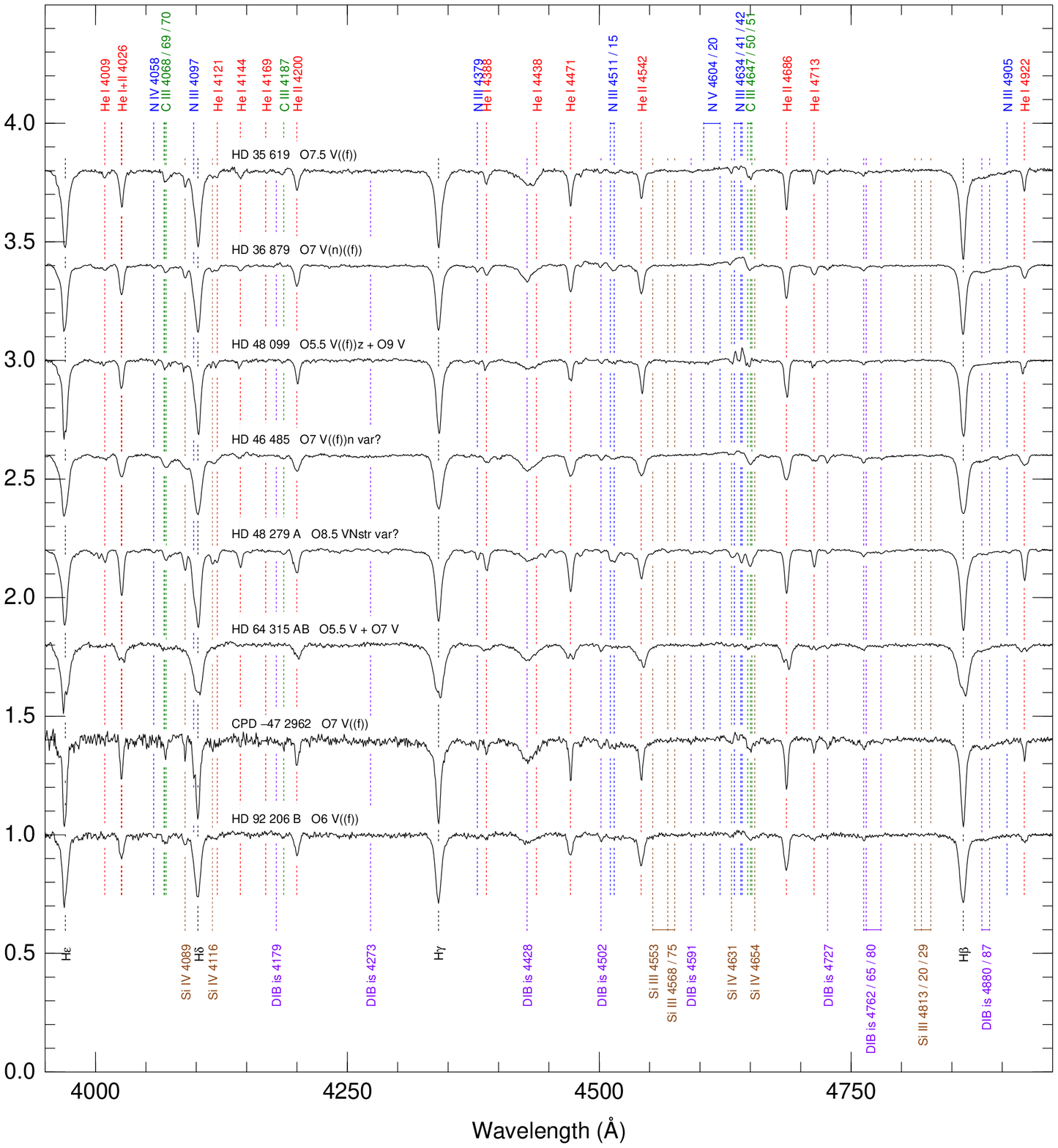}}
\caption{(continued).}
\end{figure*}

\addtocounter{figure}{-1}

\begin{figure*}
\centerline{\includegraphics*[width=\linewidth]{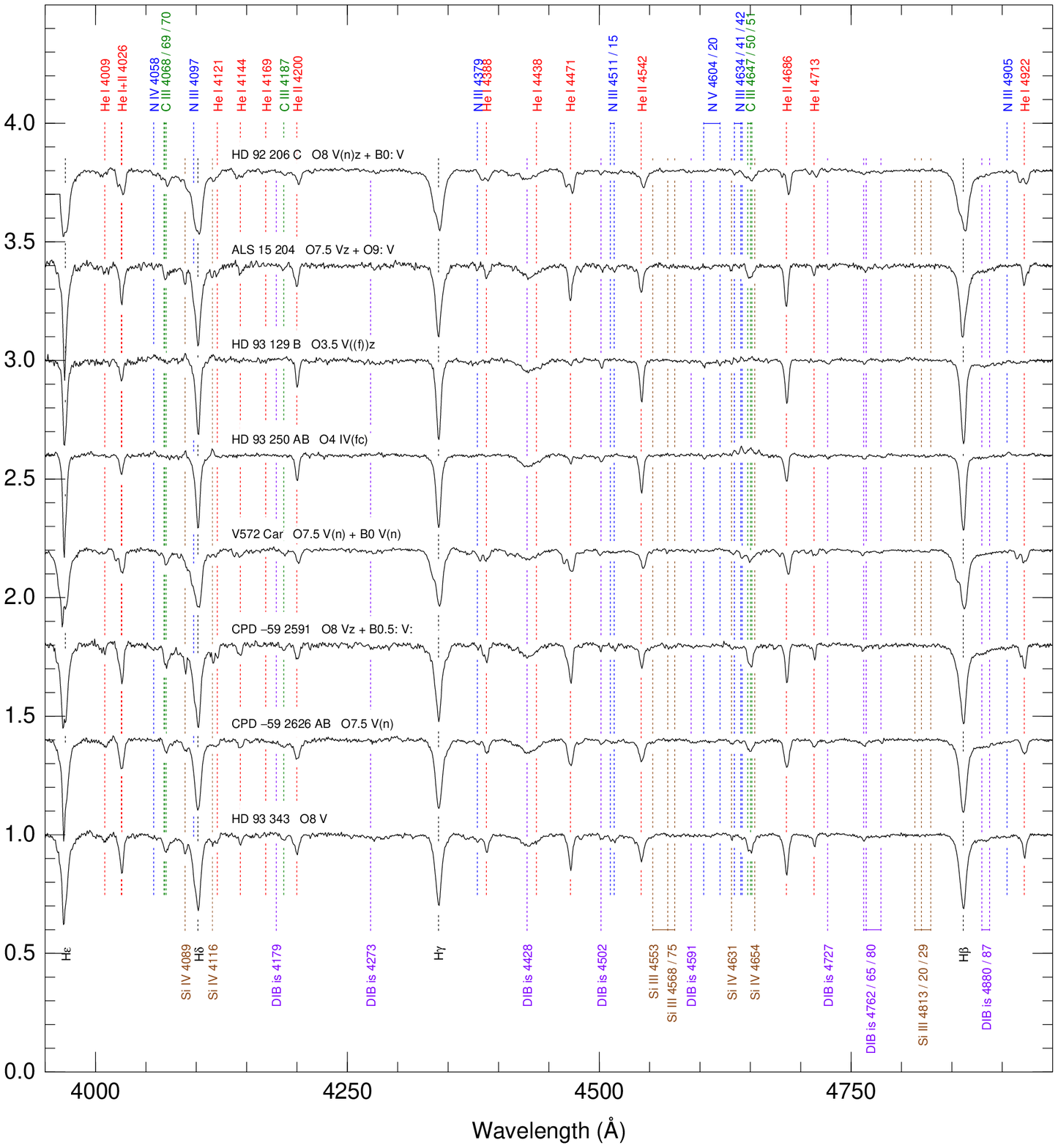}}
\caption{(continued).}
\end{figure*}

\addtocounter{figure}{-1}

\begin{figure*}
\centerline{\includegraphics*[width=\linewidth]{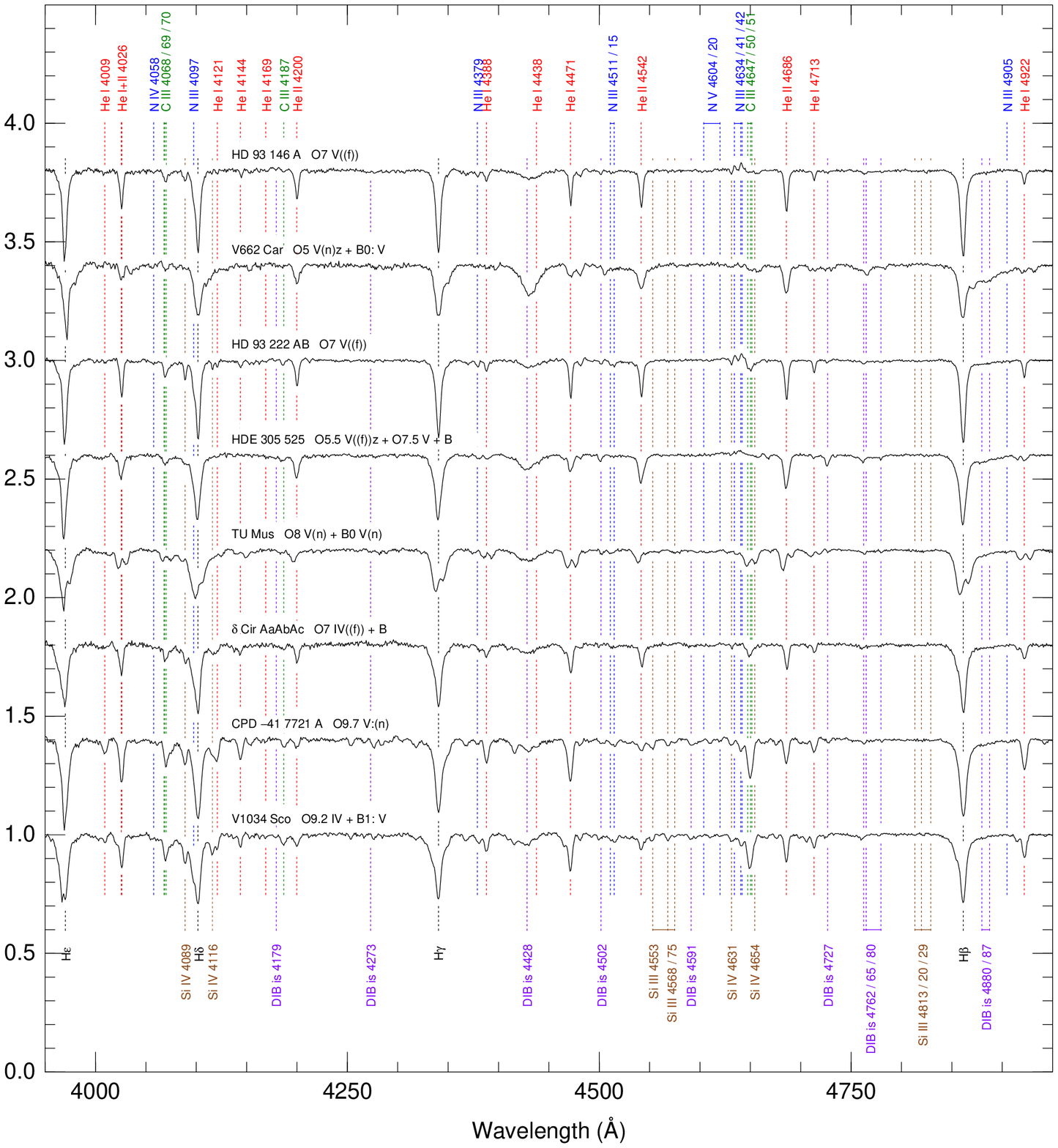}}
\caption{(continued).}
\end{figure*}

\addtocounter{figure}{-1}

\begin{figure*}
\centerline{\includegraphics*[width=\linewidth]{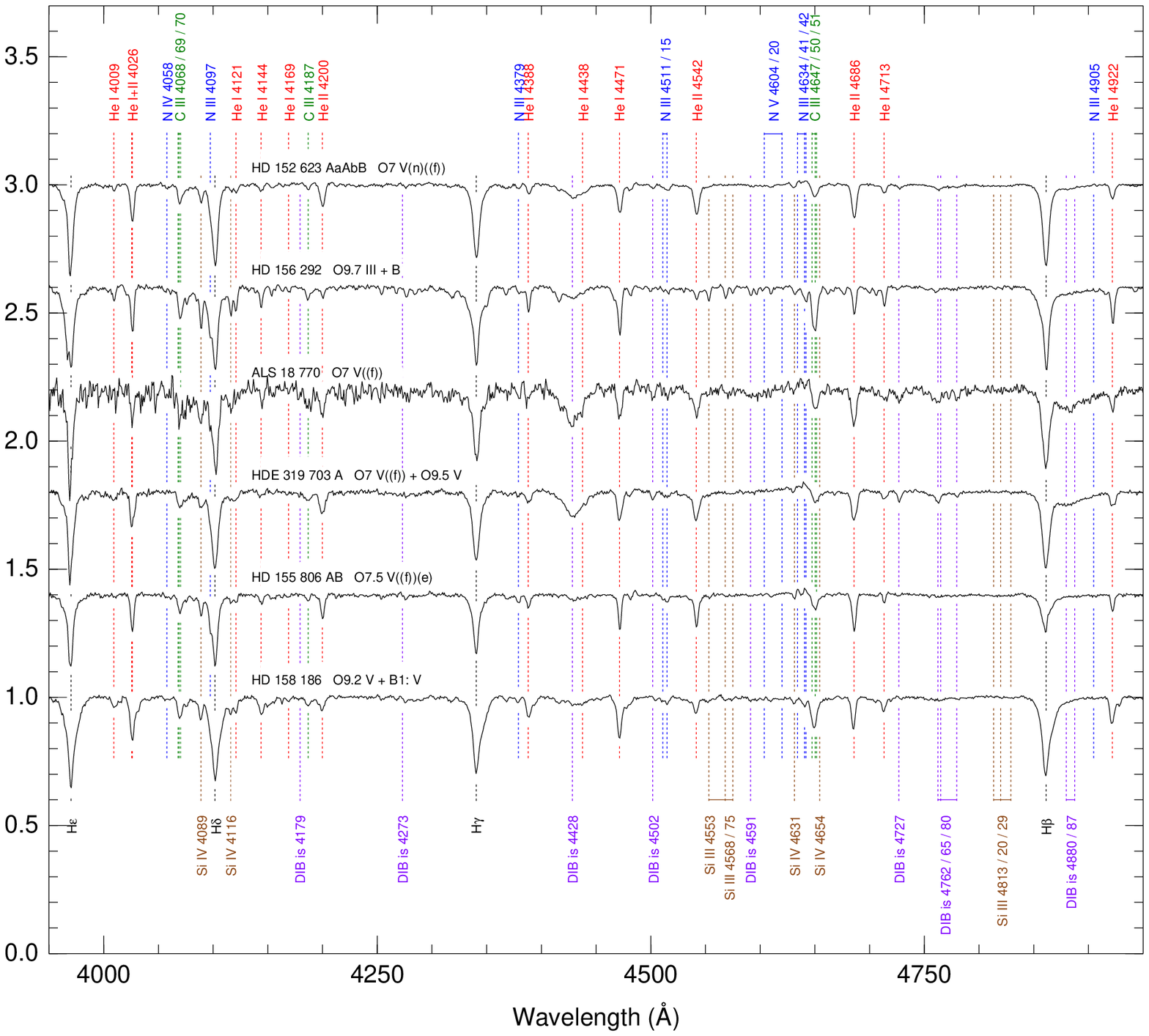}}
\caption{(continued).}
\end{figure*}	

\object[HD 164536]{} 

\object[9 Sgr AB]{} 

\paragraph{HD~165\,052 = ALS~4635.}
\object[HD 165052]{} 
We have obtained observations at a new epoch for this SB2 that has allowed us to improve the classification of O5.5:~Vz~+~O8:~V of paper II to O6~Vz~+~O8~Vz. The system was caught at
a velocity separation $\Delta v$ between the two spectroscopic components of $\sim$200~km/s.

\paragraph{HD~168\,075 = NGC~6611-197.}
\object[HD 168075]{} 
We have obtained observations at a new epoch for this star and derived a spectral type of O6.5~V((f)), slightly earlier than the one in paper I. The change is not
surprising, given that this object is an SB2 classified as O6.5~V((f))~+~B0-1~V by \citet{Sanaetal09}. 

\paragraph{HD~168\,076~AB = ALS~4908~AB.}
\object[HD 168076]{} 
This system is one of the luminosity class IV standards due to the introduction of that class for the spectral subtypes O4-O5.5.

\paragraph{HD~168\,112~AB = ALS~4912~AB.}
\object[HD 168112]{} 
This star is one of the luminosity class IV standards due to the introduction of that class for the spectral subtypes O4-O5.5.

\paragraph{HD~173\,010 = BD~-09~4805 = ALS~9901.}
\object[HD 173010]{} 
This object was classified as O9.7~Ia. \citet{Walbetal16} changed the luminosity class to Ia+ based on GOSSS data.

\object[HDE 344784 A]{} 

\paragraph{HD~192\,281 = V2011~Cyg = ALS~10\,943.}
\object[HD 192281]{} 
This object is one of the new O4-O5.5~IV stars.

\paragraph{HDE~229\,232~AB = BD~+38~4070~AB = ALS~11\,296~AB.}
\object[HDE 229232]{} 
This object has anomalous and broad line profiles, possibly originating in a companion. Note that \citet{Aldoetal15} found a bright
companion which is spatially unresolved in the GOSSS data, hence the AB component designation. \citet{Willetal13} identify this system as an SB1 with a preliminary
period of 6.2 d.

\object[HD 191978]{} 

\object[HD 207198]{} 
The ((f)) suffix for this object was omitted by mistake in paper II. 

\object[BD +60 513]{} 

\paragraph{HD~14\,434 = ALS~7124.}
\object[HD 14434]{} 
This star is one of the new O4-O5.5~IV stars.

\object[HD 17520 A]{} 

\object[HD 35619]{} 

\object[HD 36879]{} 

\paragraph{HD~48\,099 = ALS~9098.}
\object[HD 48099]{} 
We have reanalyzed the data for this SB2 to change the classification of O5~V((f))z~+~O9:~V of paper II to O5.5~V((f))z~+~O9~V. The system was caught at
a $\Delta v$ of $\sim$140~km/s.

\object[HD 46485]{} 

\object[HD 48279 A]{} 

\object[HD 64315]{} 

\object[CPD -47 2962]{} 


\object[HD 92206 B]{} 

\paragraph{HD~92\,206~C = CPD~-57~3580 = ALS~1695.}
\object[CPD -57 3580]{} 
We have reanalyzed the data for this SB2 to change the classification of O8~Vz~+~O9.7~V of paper II to O8~V(n)z~+~B0:~V. The system was caught at
a $\Delta v$ of $\sim$425~km/s.

\paragraph{ALS~15\,204 = CPD~-56~2608~A = Trumpler~14~MJ~92.}
\object[ALS 15204]{} 
\object[ALS 15203]{} 
In paper II we identified this object as an O star for the first time and we hinted it may be an SB2. For this paper we obtained a new epoch and we indeed detect it is an SB2 caught with a $\Delta v$ of $\sim$150 km/s (which is a low
separation for GOSSS but it is possible because of the small magnitude difference between the two components and their low $v\sin i$). The GOSSS spectral classification is O7.5~Vz~+~O9:~V. We also placed ALS~15\,203 (=~CPD~-56~2608~B), 
located 4\farcs8 away, on the slit and found it is an early-B star.

\object[HD 93129 B]{} 

\paragraph{HD~93\,250~AB = ALS~1859~AB.}
\object[HD 93250]{} 
This object is one of the new O4-O5.5~IV stars.

\object[V572 Car]{} 

\paragraph{CPD~-59~2591 = Trumpler~16-21 = ALS~15\,217.}
\object[CPD -59 2591]{} 
We have reanalyzed the data for this SB2 to change the classification of O8.5~V~+~B0.5:~V: of paper II to O8~Vz~+~B0.5:~V:. The system was caught at
a $\Delta v$ of $\sim$200~km/s.

\object[CPD -59 2626]{} 

\object[HD 93343]{} 

\object[HD 93146 A]{} 

\paragraph{V662~Car = FO~15 = ALS~16\,081.}
\object[V662 Car]{} 
We have reanalyzed the data for this SB2 to change the classification of O5~Vz~+~B0:~V of paper II to O5~V(n)z~+~B0:~V:. The system was caught at
a $\Delta v$ of $\sim$625~km/s.

\object[HD 93222]{} 

\paragraph{HDE~305\,525.}
\object[HDE 305525]{} 
In paper II we classified this object as O5.5~V(n)((f))z, that is, with broad lines but without identifying it as a spectroscopic binary. A reanalysis of the data allowed us to discover that [a] the broad lines were caused
by another O star blueshifted by $\sim$175~km/s and [b] the He\,{\sc i} lines (but not the He\,{\sc ii} ones) had a third absorption component redshifted by $\sim$400~km/s, indicating the presence of a B star as an
additional spectroscopic component. Therefore, we classify this is a new SB3 system (the first one identified as such with GOSSS) with spectral types O5.5~V((f))z~+~O7.5~V~+~B. A revision of OWN data confirms the SB3 nature,
as it appears that way in at least one epoch. We looked at the ASAS All-Star Catalogue ({\tt http://www.astrouw.edu.pl/asas/?page=aasc}) and we identified HDE~305\,525 as an eclipsing binary with a 1.9018~d period.

\object[TU Mus]{} 


\paragraph{$\delta$~Cir~AaAbAc = HD~135\,240~AaAbAc = ALS~3331~AaAbAc.}
\object[delta Cir A]{} 
In paper II we did not detect double lines in this object in the GOSSS spectra, even though it is an SB3 system \citep{Pennetal01}, who classified it as 
O7~III-V~+~O9.5~V~+~B0.5~V. A reanalysis of the data with MGB has allowed us to classify it as an SB2 with O7~IV((f))~+~B, which is consistent with the 
\citet{Pennetal01} result considering that we see only two components and they detect three. The system was caught at a $\Delta v$ of $\sim$300~km/s. In the component
nomenclature AaAb corresponds to the inner, short period 3.9 d binary (though the system remains unresolved at the time of the writing) while Ac is the third outer
component resolved by \citet{Sanaetal14} with a period of 1644~d measured by \citet{Mayeetal14b}.

\object[HD 150135 A]{} 

\object[CPD -41 7721 A]{} 

\paragraph{V1034~Sco = CPD~-41~7742 = ALS~15\,757.}
\object[V1034 Sco]{} 
In paper II we did not detect double lines in this object in the GOSSS spectra, even though it is an SB2 system classified by \citet{Sanaetal08b} as O9.5~V~+~B1.5~V.
A reanalysis of the data with MGB has allowed us to classify it as an SB2 with O9.2~IV~+~B1:~V, similar to the \citet{Sanaetal08b} result. The system was caught at a $\Delta v$ of 
$\sim$450~km/s.

\object[HD 152623]{} 

\paragraph{HD~156\,292 = ALS~4055.}
\object[HD 156292]{} 
In paper II we did not detect the SB2 nature of this target in the GOSSS spectra, even though it is a system of that nature as revealed by OWN \citep{Barbetal10}.
A reanalysis of the data with MGB has allowed us to derive an O9.7~III~+~B spectral classification with a $\Delta v$ of $\sim$350~km/s. 

\object[ALS 18770]{} 

\paragraph{HDE~319\,703~A = ALS~4081~A.}
\object[HDE 319703 A]{} 
In paper II we did not detect the SB2 nature of this target in the GOSSS spectra, even though it is a system of that nature as revealed by OWN \citep{Barbetal10}.
A reanalysis of the data with MGB has allowed us to derive an O7~V((f))~+~O9.5~V spectral classification with a $\Delta v$ of $\sim$175~km/s. 

\object[HD 155806]{} 

\paragraph{HD~158\,186 = V1081~Sco = ALS~4182.}
\object[HD 158186]{} 
In paper II we did not detect double lines in this target in the GOSSS spectra, even though it is an SB3 system as revealed by OWN \citep{Barbetal10}.
A reanalysis of the data with MGB has allowed us to derive an SB2 O9.2~V~+~B1:~V spectral classification with a $\Delta v$ of $\sim$375~km/s. 

\subsection{New systems in GOSSS}
\label{sec:new}

In Table~\ref{spectralclasnew} we list the new O-type stellar systems incorporated into GOSSS in this paper, ordered by GOSSS ID. The corresponding spectrograms are shown in 
the same order in Fig.~\ref{fig:new}. We present here additional information about each one of the systems.

\begin{figure*}
\centerline{\includegraphics*[width=\linewidth]{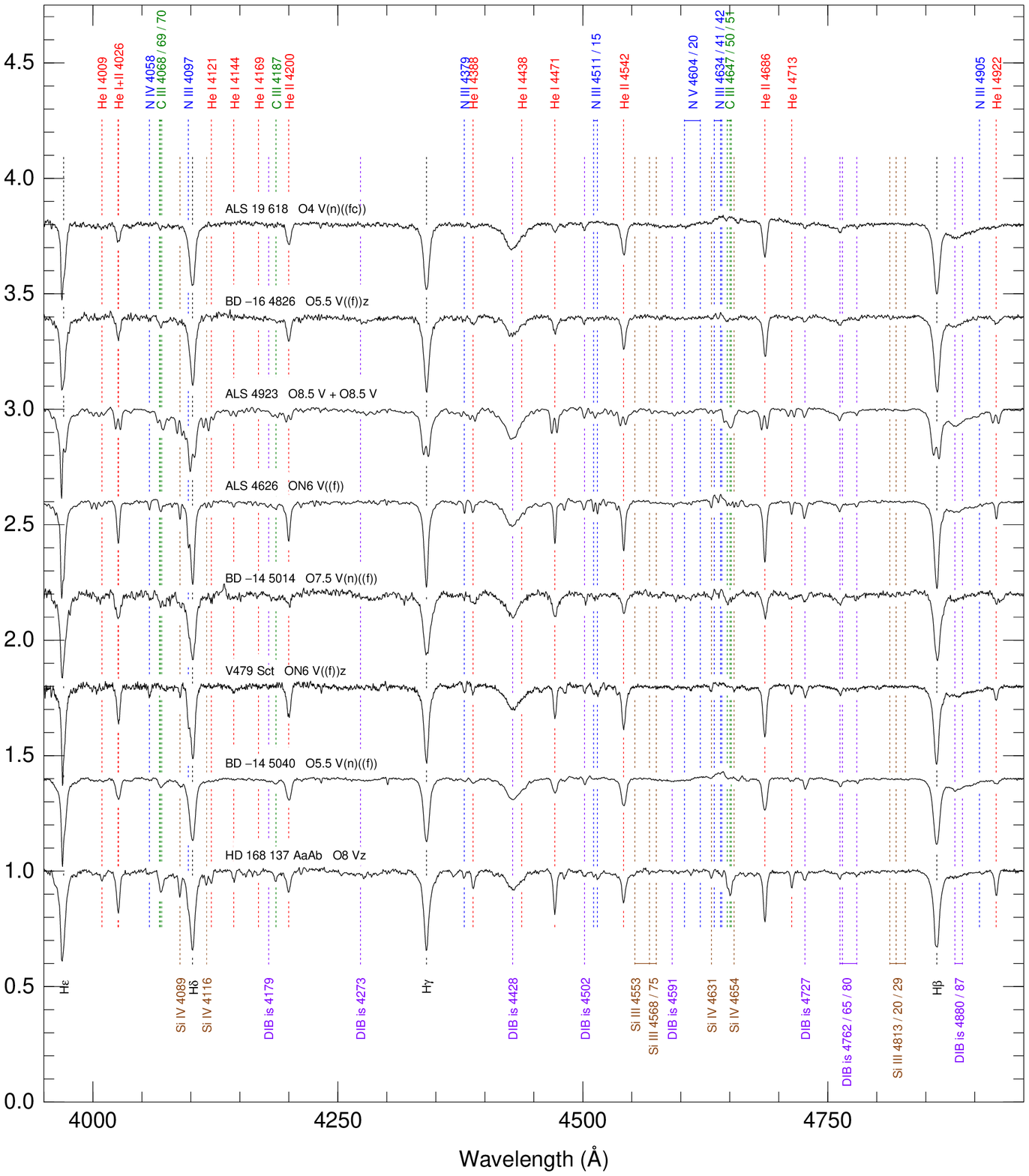}}
\caption{Spectrograms for new GOSSS stars. The targets are sorted by GOS ID.
[See the electronic version of the journal for a color version of this figure.]}
\label{fig:new}
\end{figure*}	

\addtocounter{figure}{-1}

\begin{figure*}
\centerline{\includegraphics*[width=\linewidth]{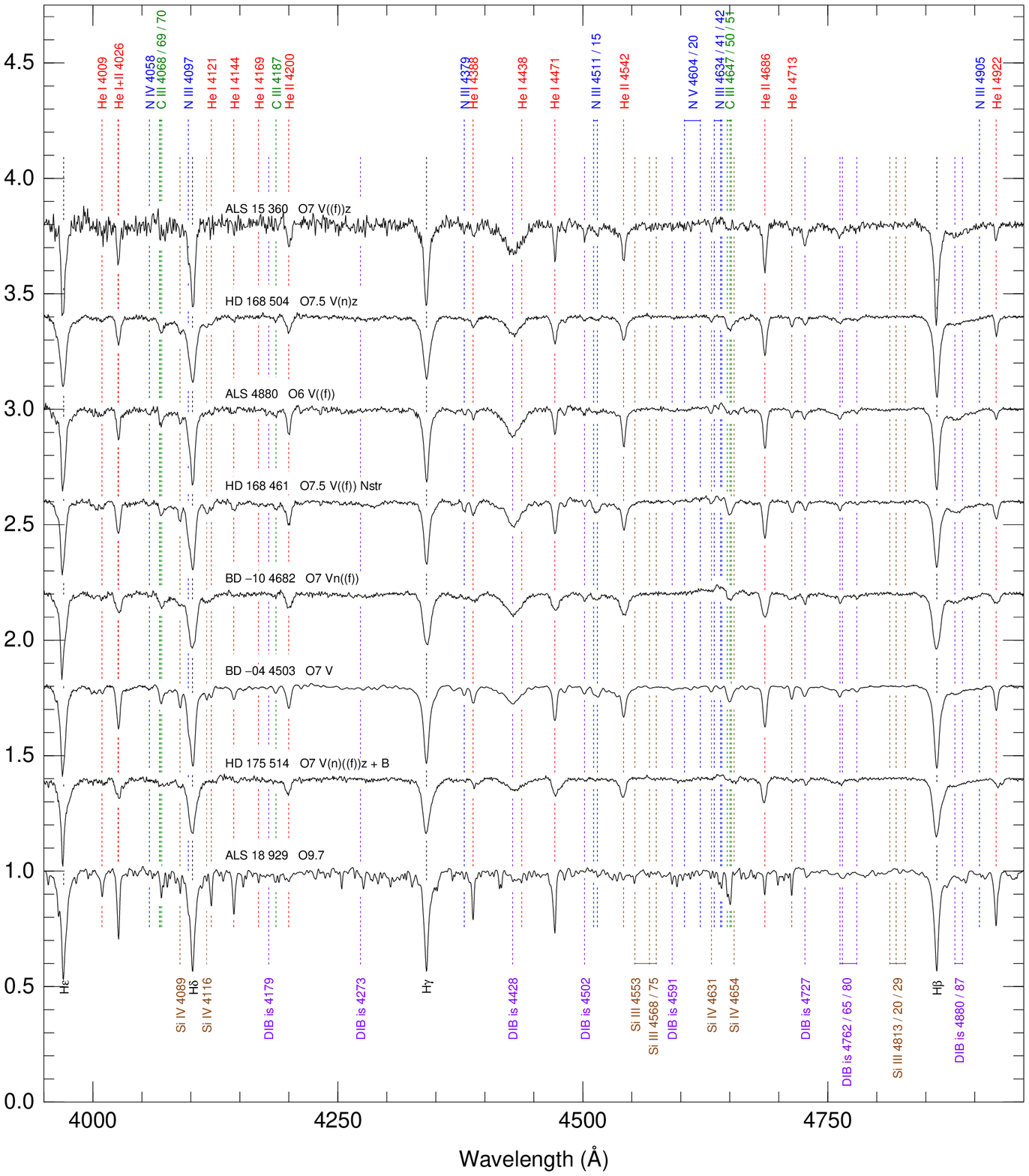}}
\caption{(continued).}
\end{figure*}	

\addtocounter{figure}{-1}

\begin{figure*}
\centerline{\includegraphics*[width=\linewidth]{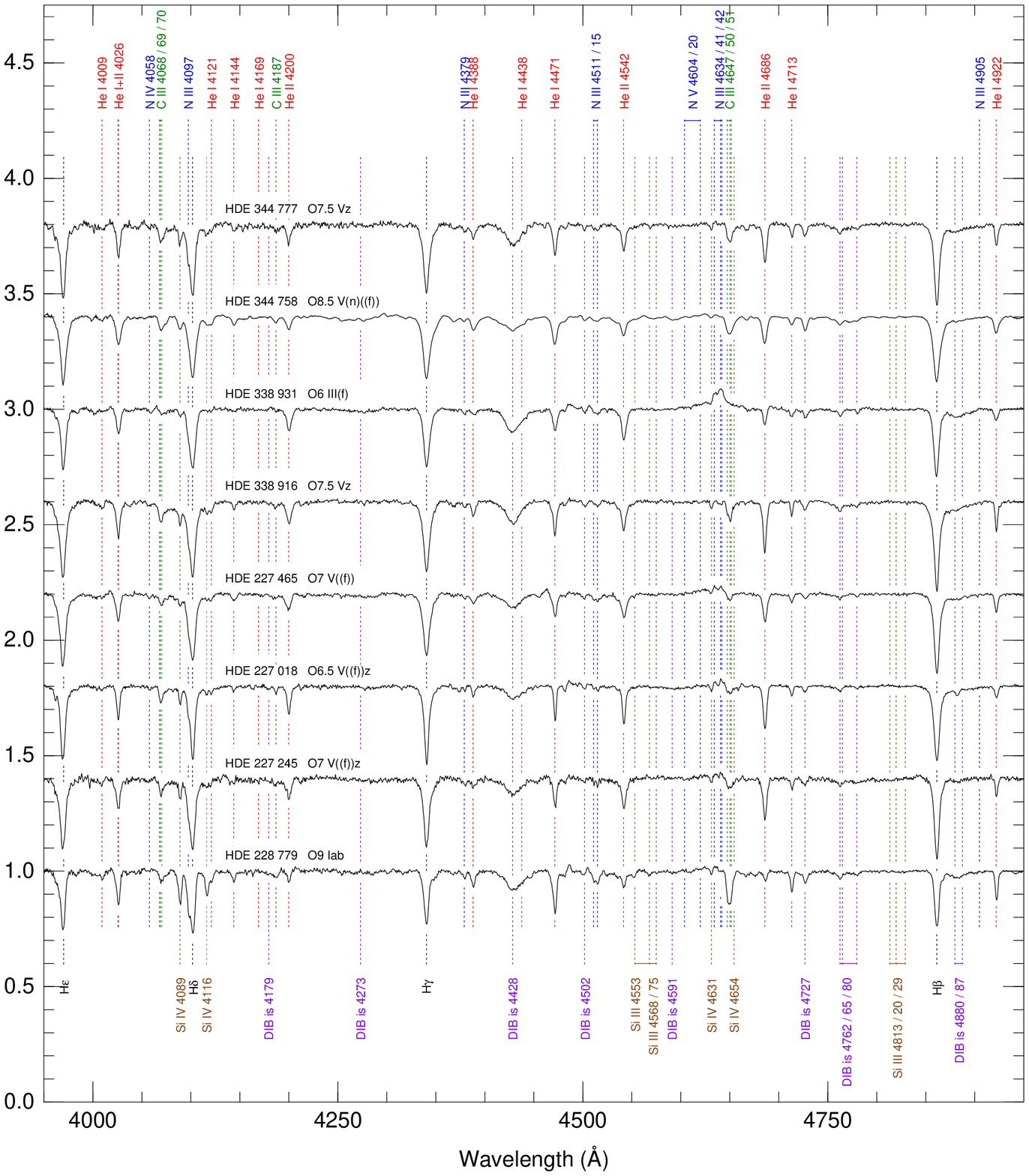}}
\caption{(continued).}
\end{figure*}	

\addtocounter{figure}{-1}

\begin{figure*}
\centerline{\includegraphics*[width=\linewidth]{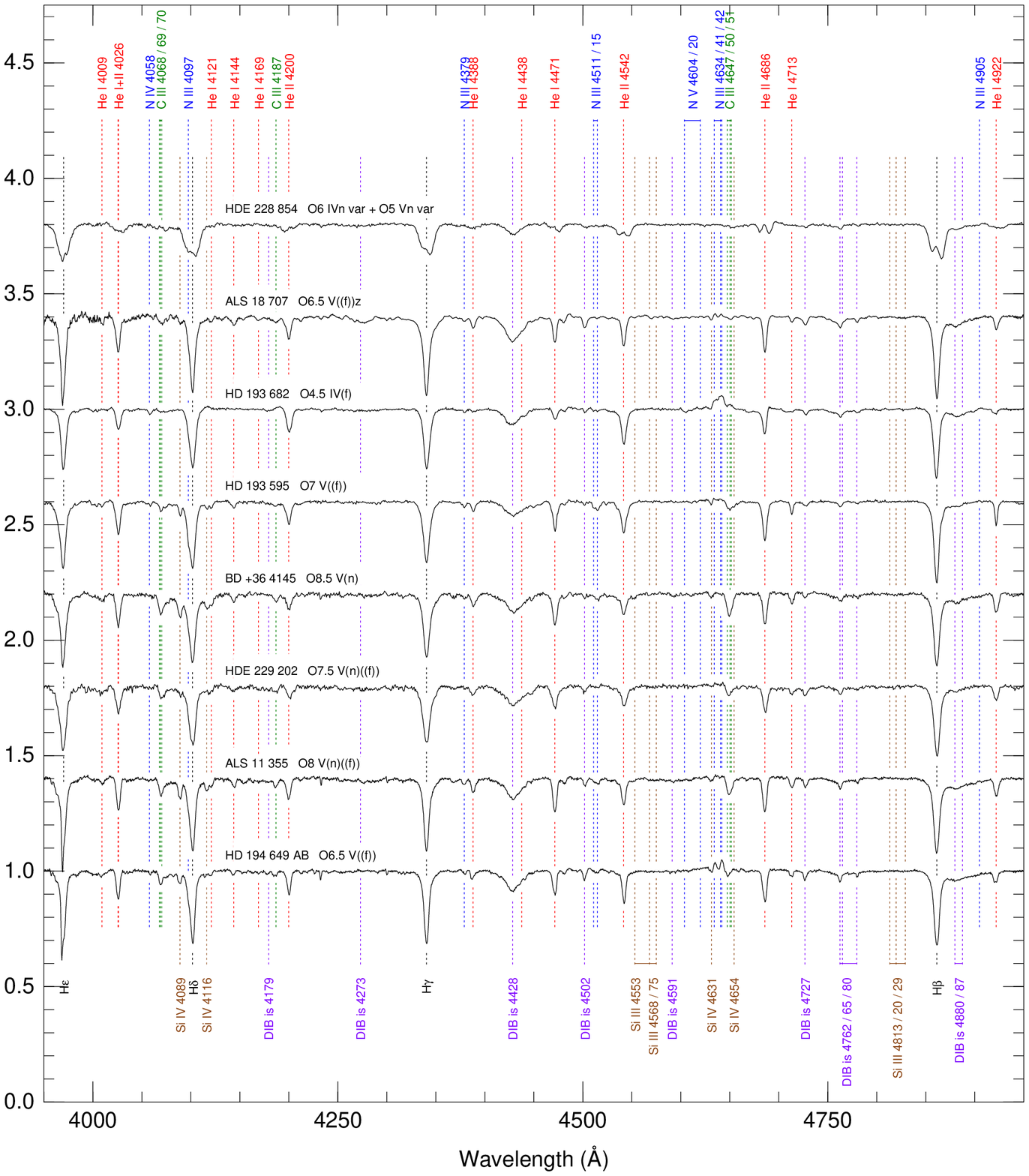}}
\caption{(continued).}
\end{figure*}	

\addtocounter{figure}{-1}

\begin{figure*}
\centerline{\includegraphics*[width=\linewidth]{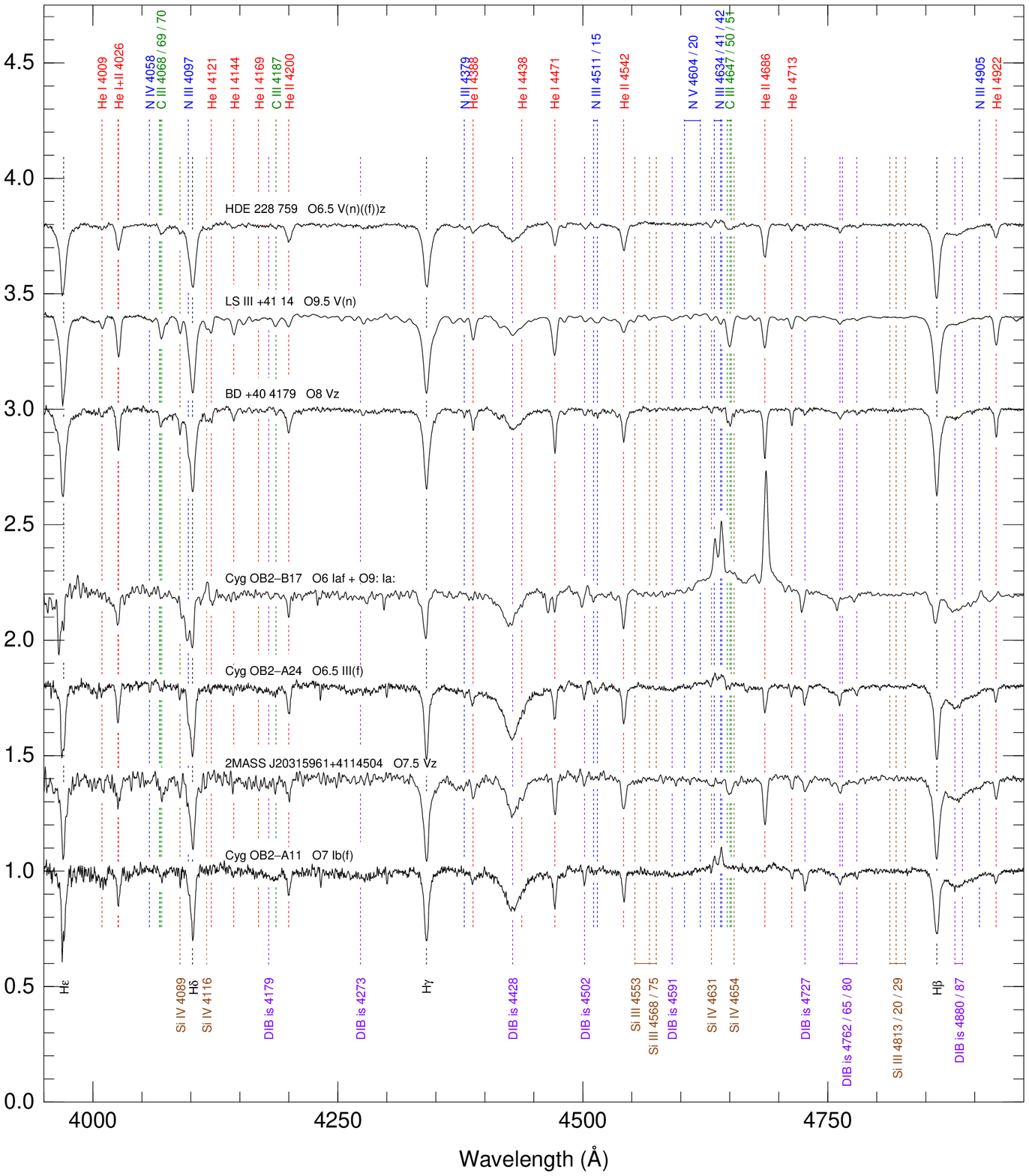}}
\caption{(continued).}
\end{figure*}	

\addtocounter{figure}{-1}

\begin{figure*}
\centerline{\includegraphics*[width=\linewidth]{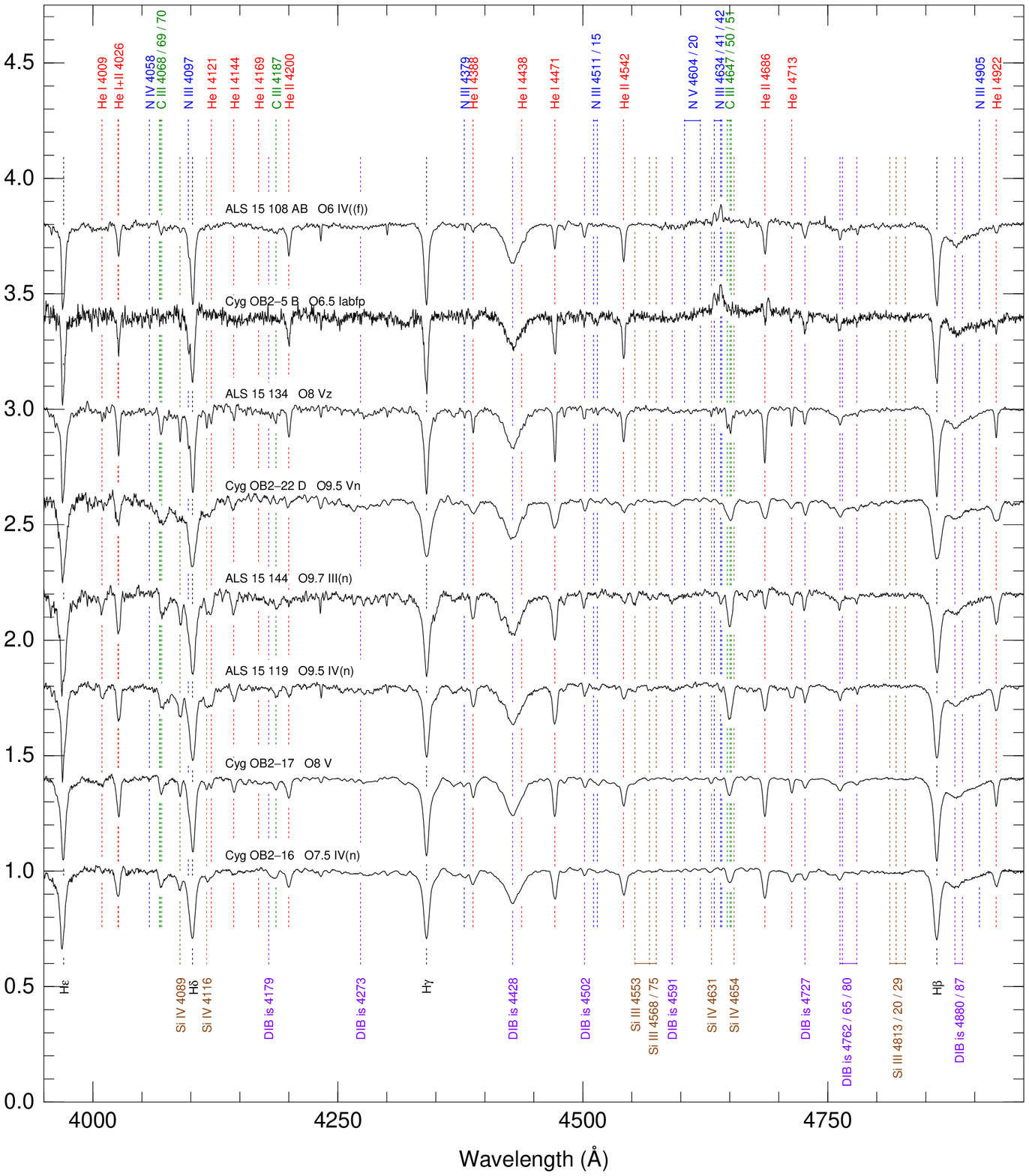}}
\caption{(continued).}
\end{figure*}	

\addtocounter{figure}{-1}

\begin{figure*}
\centerline{\includegraphics*[width=\linewidth]{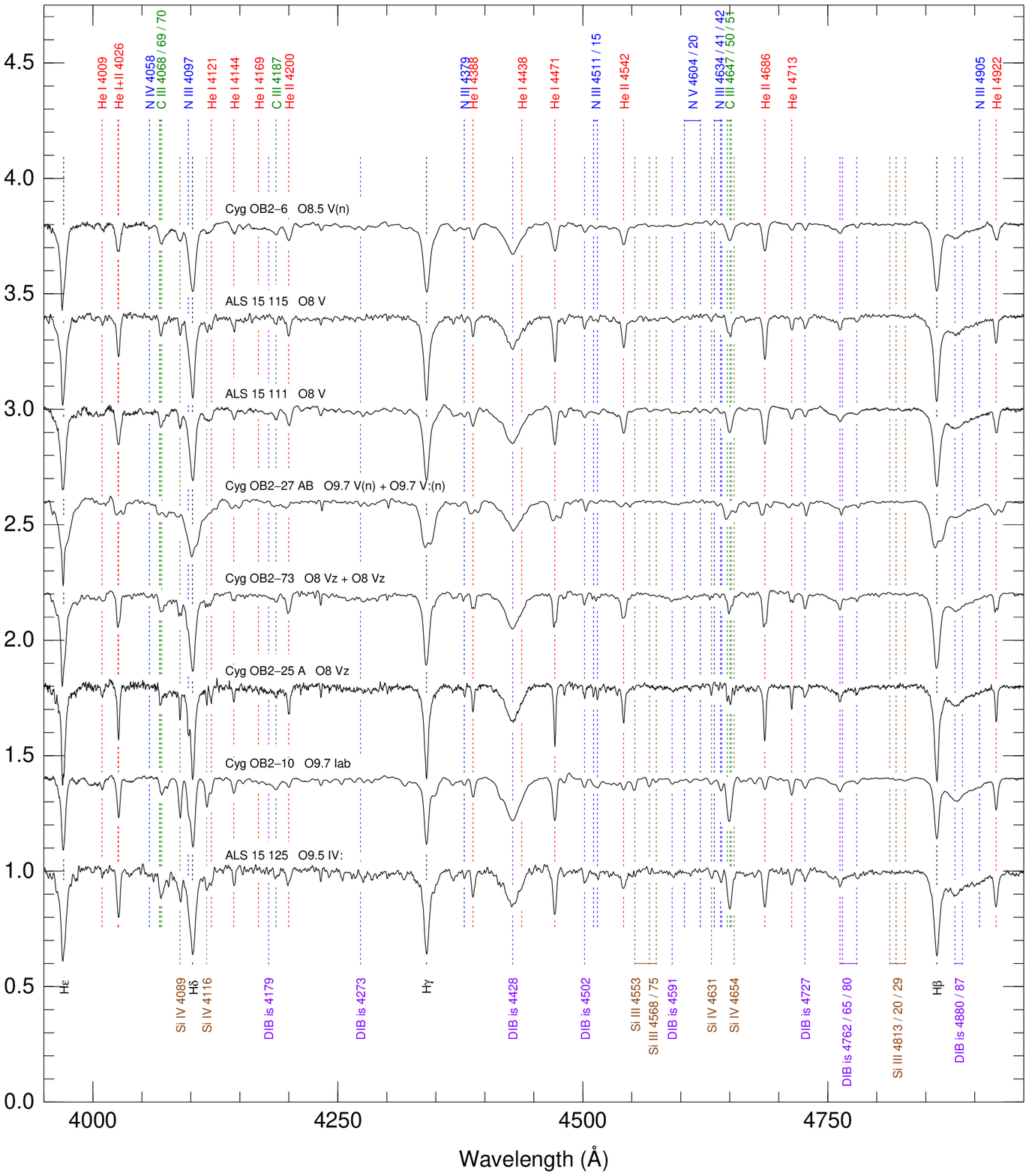}}
\caption{(continued).}
\end{figure*}	

\addtocounter{figure}{-1}

\begin{figure*}
\centerline{\includegraphics*[width=\linewidth]{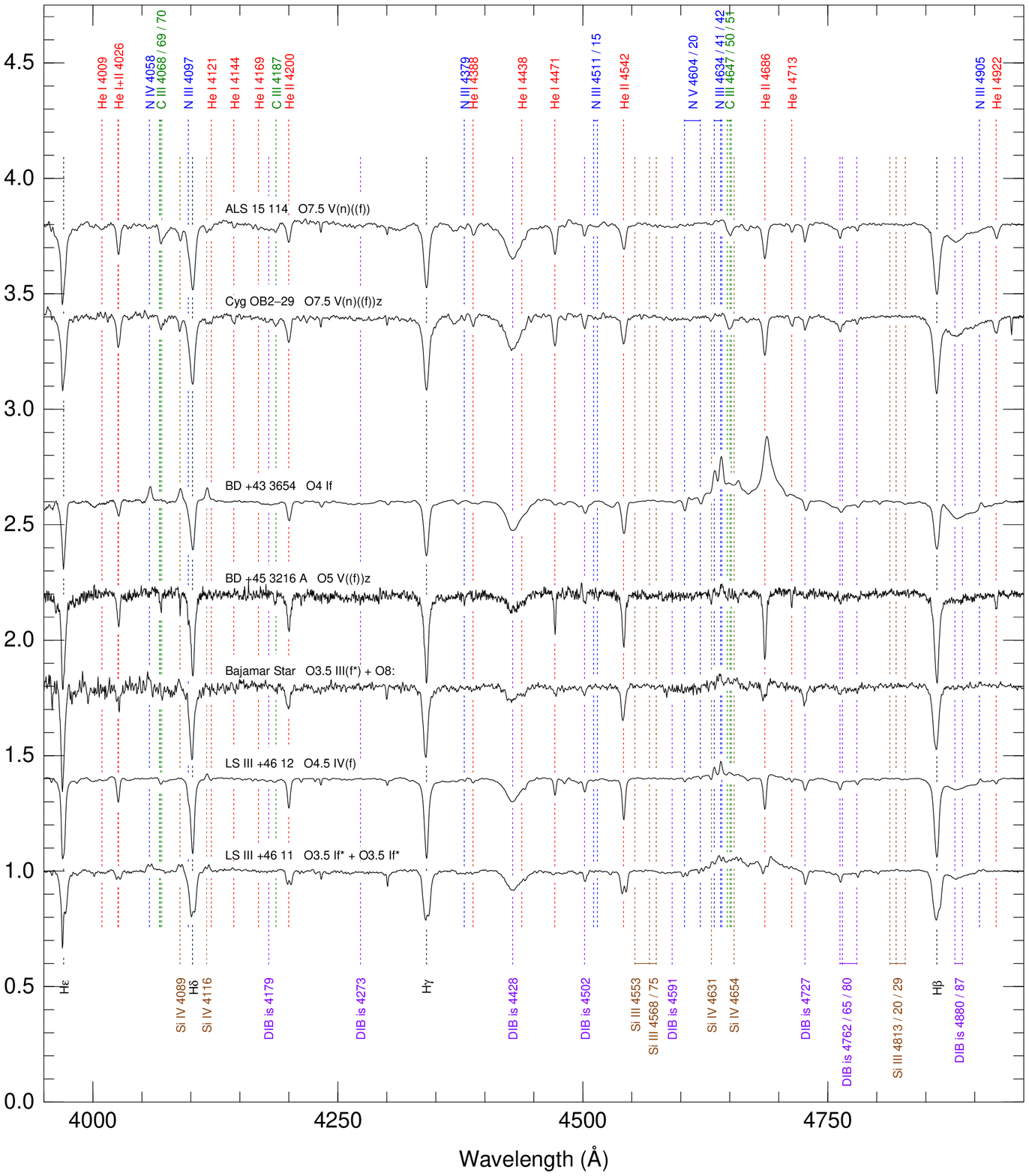}}
\caption{(continued).}
\end{figure*}	

\addtocounter{figure}{-1}

\begin{figure*}
\centerline{\includegraphics*[width=\linewidth]{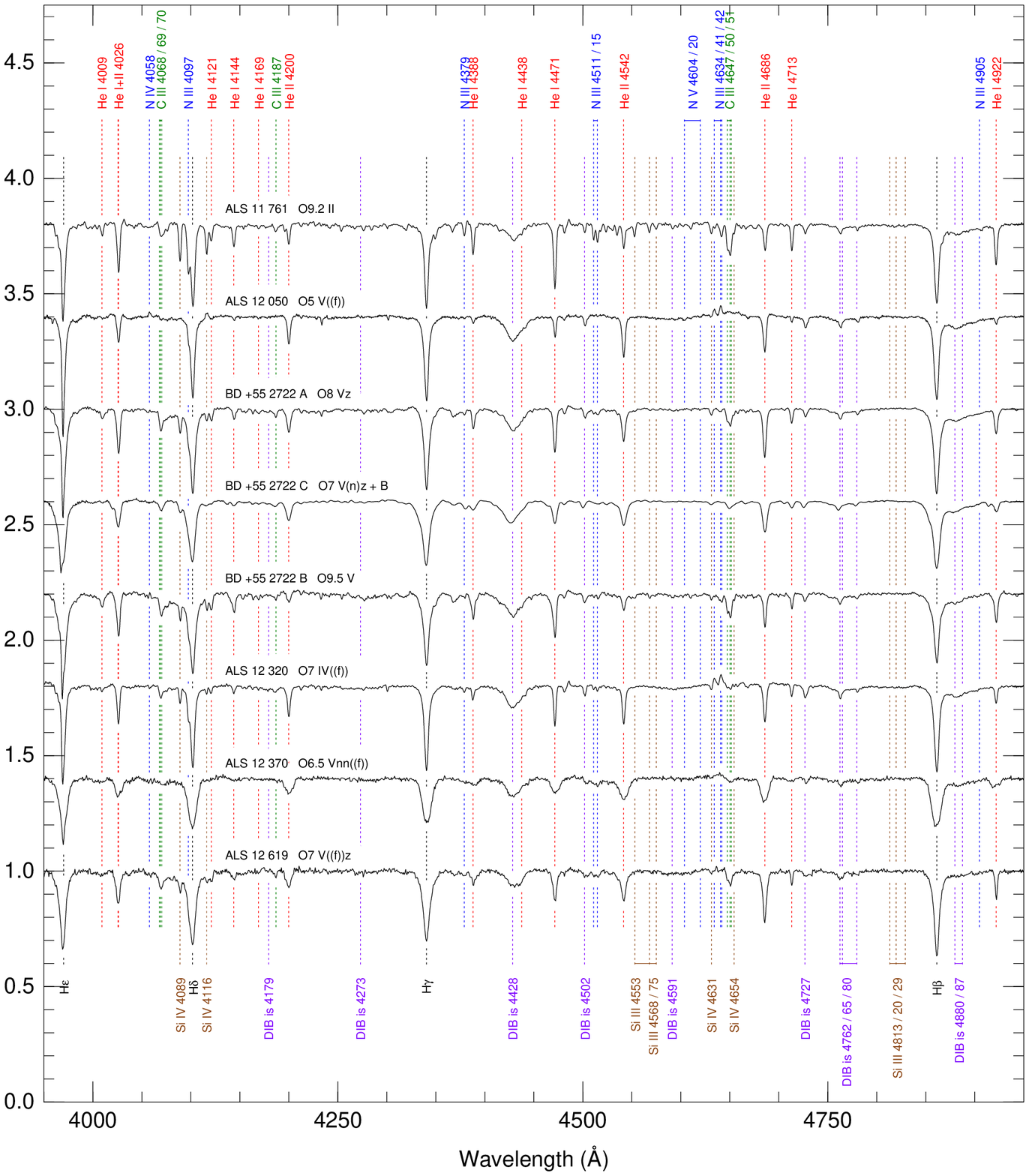}}
\caption{(continued).}
\end{figure*}	

\addtocounter{figure}{-1}

\begin{figure*}
\centerline{\includegraphics*[width=\linewidth]{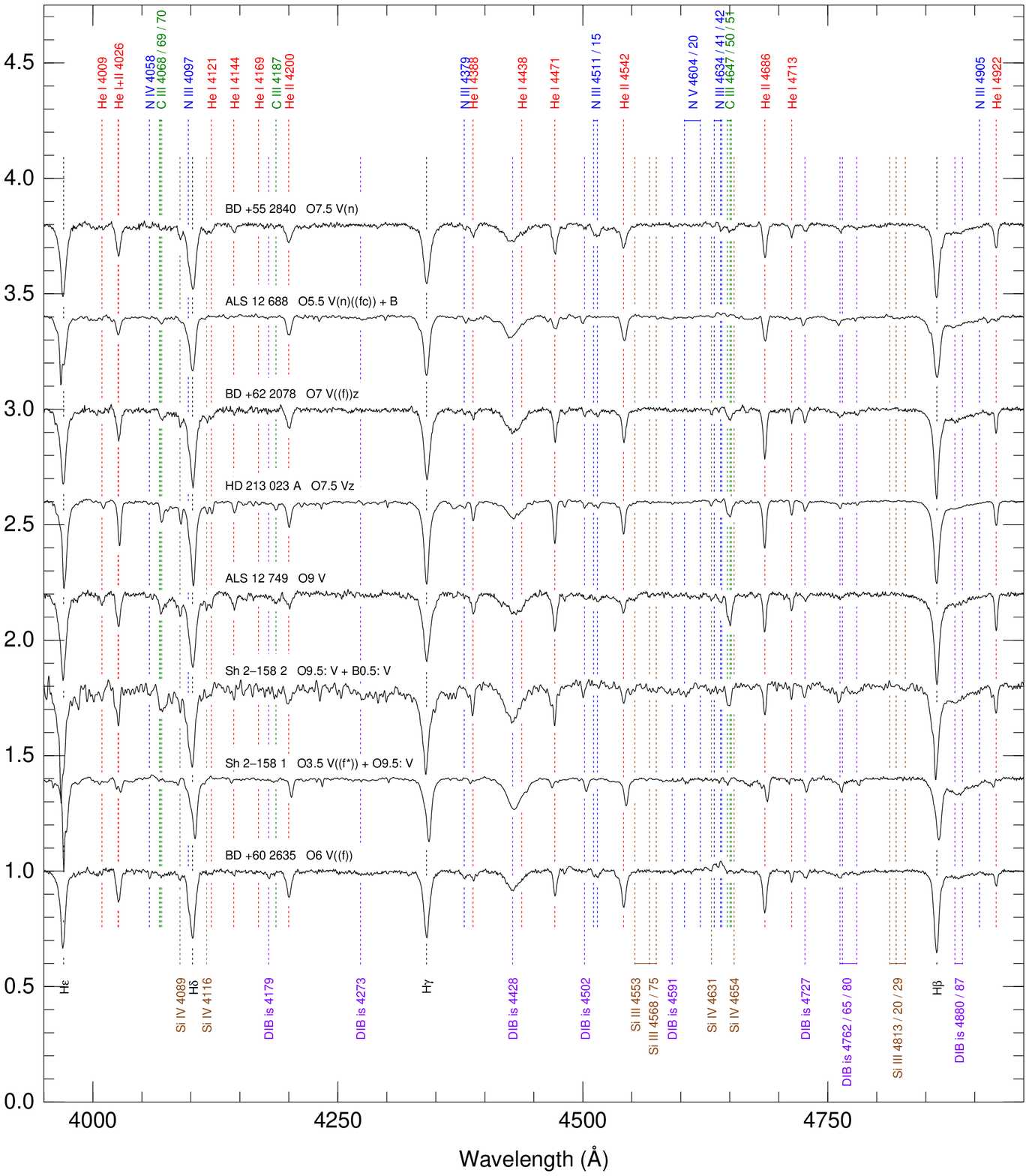}}
\caption{(continued).}
\end{figure*}	

\paragraph{ALS~19\,618 = NGC~6618~Sch-1.}
\object[ALS 19618]{} 
\citet{Crametal78} classified this NGC~6618 star as O5~V. We obtain an earlier spectral type of O4~V(n)((fc)).

\paragraph{BD~-16~4826 = ALS~4944.}
\object[BD -16 4826]{} 
\citet{HiltJohn56} classified this NGC~6618 star as O5 and \citet{GarmVacc91} as O3~V. Our GOSSS data yield O5.5~V((f))z. Some lines are asymmetric, indicating the possiblity
of the system being an SB2. \citet{Willetal13} classify this system as an SB1 and give a period of 15.8 d but note that their observations span less than a full orbit.

\paragraph{ALS~4923.}
\object[ALS 4923]{} 
\citet{VijaDril93} classified this star as a rapid-rotator giant, O9.5~IIInn. GOSSS data reveal that it is really an SB2 composed of two nearly identical main-sequence stars,
with spectral types of O8.5~V~+~O8.5~V. The system was caught with a $\Delta v$ of $\sim$400~km/s, so it is likely to have a short period.

\paragraph{ALS~4626.}
\object[ALS 4626]{} 
\citet{DrilPerr81} classified this star as O5~V. We obtain a classification of ON6~V((f)).

\paragraph{BD~-14~5014 = ALS~4981.}
\object[BD -14 5014]{} 
\citet{Crametal78} classified this star as O8:~V: and \citet{VijaDril93} as O9.5~V. We obtain O7.5~V(n)((f)). The star has a visual companion
listed in the WDS (2MASS~J18222221-1437151) with a last known separation in 2007 of 7\farcs2. We obtained a separate spectrum of the companion and we found that it is a K 
star, so the proximity is likely to be the product of a chance alignment, which agrees with the difference in proper motions between the two objects.

\paragraph{V479~Sct = ALS~5039.}
\object[V479 Sct]{} 
This $\gamma$-ray binary \citep{Aharetal05} was classified as O7~V((f)) by \citet{Motcetal97} and as O6.5~V((f)) by \citet{Claretal01}. We derive a slightly earlier classification of ON6~V((f))z from GOSSS data.

\paragraph{BD~-14~5040 = ALS~5025 = LS~IV~-14~57.}
\object[BD -14 5040]{} 
We classify this star as O5.5~V(n)((f)) which, to our knowledge, is the first spectroscopic identification of this target as an O star. Note that the characterization of the 
object by \citet{Kilk93} was based purely on photometry. 

\paragraph{HD~168\,137~AaAb = ALS~4915~AaAb = NGC~6611-401~AaAb.}
\object[HD 168137 A]{} 
This NGC~6611 system was classified as O8~V by \citet{HiltMorg69}. \citet{Sanaetal09} found it to be a long-period SB2 with spectral types O7~V~+~O8~V and \citet{Sanaetal14} was
able to marginally resolve the two components. The SB2 character is not visible in the GOSSS data and we derive a classification of O8~Vz.

\paragraph{ALS~15\,360 = NGC~6611-222.}
\object[ALS 15360]{} 
This NGC~6611 star was classified as O7~III((f)) by \citet{Hilletal93b}. Using GOSSS data we obtain O7~V((f))z.

\paragraph{HD~168\,504 = ALS~4935.}
\object[HD 168504]{} 
This star was classified as O8 by \citet{Morgetal55} and as O8~III((f)) by \citet{Cont73}. We derive a classification of O7.5~V(n)z from GOSSS data.

\paragraph{ALS~4880.}
\object[ALS 4880]{} 
\citet{VijaDril93} classify this star as O5~V. We obtain a classification of O6~V((f)) with GOSSS data and we select it as a new standard star for this type.

\paragraph{HD~168\,461 = ALS~4931.}
\object[HD 168461]{} 
\citet{HiltJohn56} give a classification of O8 for this object. Our result using GOSSS data is O7.5~V((f))~Nstr.

\paragraph{BD~-10~4682 = ALS~9584.}
\object[BD -10 4682]{} 
\citet{HiltJohn56} classify this star as O7 and \citet{GarmVacc91} as O7~V. We classify it as O7~Vn((f)) with GOSSS data. 

\paragraph{BD~-04~4503 = ALS~9772.}
\object[BD -04 4503]{} 
\citet{HiltJohn56} classify this star as O7. We obtain O7~V using GOSSS data. 

\paragraph{HD~175\,514 = V1182~Aql = BD~+09~3928 = ALS~10\,048.}
\object[HD 175514]{} 
\citet{HiltJohn56} originally classified this star as O8~Vnn. The star was later discovered to be an eclipsing SB2 and \citet{Belletal87} obtained a spectral classification of
O9~Vnn~+~B3~V. We also see it as an SB2 with GOSSS but the new spectral classification, O7~V(n)((f))z~+~B implies a significantly earlier primary. The system was caught with a
$\Delta v$ of $\sim425$~km/s.

\paragraph{ALS~18\,929 = LSE~107.}
\object[ALS 18929]{} 
\object[Tyc 1036-00534-1]{}
This star appears as Osp\ldots\ in Simbad but there are no references listed for that classification. Indeed, the only references for this star that appear in the ADS are 
\citet{Reed03,Reed05} but there is no spectral type there either. One possibility is that the Simbad classification is a photometric one, not a real spectral 
classification\footnote{We discuss below other cases where the lack of references for Simbad classifications do not allow for their verification.}. Another possiblity is that 
the Simbad classification comes from the OB- classification in \citet{DrilBerg95}. Since 
this object is an O9.7 in the GOSSS data, this is the first time it receives an O-type spectral classification. The star has sharp lines, indicating a low $v\sin i$ but we are unable 
to give a luminosity class because it is a good example of the problem discussed in Appendix A.2 of \citet{Walbetal14}. For late-O stars there are two luminosity criteria, the ratios of 
\HeII{4686} to \HeII{4713} and of \SiIV{4089} to \HeI{4026}, respectively (Table~6 in paper I). For the majority of O9.5-O9.7 stars the two criteria agree but in some cases they yield 
different answers, with the second criterion providing results that are more consistent with the distance to 30 Doradus for the O9.5-O9.7 stars analyzed there. In the case of ALS~18\,929, the 
first criterion yields a luminosity class of II while the second one yields a luminosity class of V. Given the discrepancy, we do not assign a luminosity class to this object. When
observing ALS~18\,929, we placed a nearby star, Tyc~1036-00534-1, on the slit and it turned out to have an F spectral type.

\paragraph{HDE~344\,777 = ALS~10\,425.}
\object[HDE 344777]{} 
\citet{Turn79} classified this star as O9.5~III:. We agree that it is an O star but the GOSSS spectral classification is rather different, O7.5~Vz.

\paragraph{HDE~344\,758 = BD~+24~3843 = ALS~10\,421.}
\object[HDE 344758]{} 
\citet{HiltJohn56} classified this star as O8~V. The GOSSS spectral classification is O8.5~V(n)((f)).

\paragraph{HDE~338\,931 = BD~+24~3881 = ALS~10\,512.}
\object[HDE 338931]{} 
\citet{HiltJohn56} classified this star as O6f. The GOSSS spectral classification is O6~III(f) and we select it as the new standard star for this type.

\paragraph{HDE~338\,916 = BD~+25~3952 = ALS~10\,493.}
\object[HDE 338916]{} 
\citet{HiltJohn56} classified this star as O8. The GOSSS spectral classification is O7.5~Vz.

\paragraph{HDE~227\,465 = BD~+33~3717 = ALS~10\,789.}
\object[HDE 227465]{} 
\citet{HiltJohn56} classified this star as O7:. The GOSSS spectral classification is O7~V((f)).

\paragraph{HDE~227\,018 = BD~+34~3828 = ALS~10\,695.}
\object[HDE 227018]{} 
\citet{HiltJohn56} classified this star as O7. The GOSSS spectral classification is O6.5~V((f))z.

\paragraph{HDE~227\,245 = BD~+35~3924 = ALS~10~744.}
\object[HDE 227245]{} 
\citet{HiltJohn56} classified this star as O7. The GOSSS spectral classification is O7~V((f))z.

\paragraph{HDE~228\,779 = BD~+34~3961 = ALS~11\,098.}
\object[HDE 228779]{} 
\citet{HiltJohn56} classified this star as O9.5~Ib. The GOSSS spectral classification is O9~Iab.

\paragraph{HDE~228\,854 = V382~Cyg = BD~+35~4062 = ALS~11\,132.}
\object[HDE 228854]{} 
\citet{Pear52} assigned spectral types of O6.5 and O7.5 to the two components of this overcontact eclipsing binary with a period of just 1.8855 d.
The GOSSS spectral classification is O6~IVn~var~+~O5~Vn~var, with the O6 listed first because it is the brighter component. The system was caught at a $\Delta v$ of $\sim 575$~km/s.
The var suffix is used because we have another GOSSS observation near conjunction with a spectral type that is incompatible with the two observed near quadrature: the combined spectral 
subtype is O7 and there is a clear \NIII{4634-41-42} emission (that is absent in the near quadrature spectrum). 

\paragraph{ALS~18\,707 = S104~Anon~3.}
\object[ALS 18707]{} 
\object[ALS 18707 B]{}
\citet{Crametal78} classified this star as O6~V. The GOSSS spectral classification is O6.5~V((f))z. There is a companion $\sim$3 magnitudes fainter towards the SW at a distance
of 2\farcs1 that we were able to spatially resolve in our long-slit spectra: it is an early B star. The two objects appear to be the brightest stars in an obscured clusetr.

\paragraph{HD~193\,682 = ALS~11\,181.}
\object[HD 193682]{} 
\citet{HiltJohn56} classified this star as O5. Using GOSSS data we obtain O4.5~IV(f). It is one of the new O4-O5.5~IV stars and we use it as one of the new standards.

\paragraph{HD~193\,595 = ALS~11\,162.}
\object[HD 193595]{} 
\citet{Roma51} classified this star as O8 and \citet{Morgetal53b} as O7. The GOSSS spectral classification is O7~V((f)).

\paragraph{BD~+36~4145 = ALS~11\,453.}
\object[BD +36 4145]{} 
\citet{HiltJohn56} classified this star as O9~V and \citet{Neguetal04} as O8.5~V. The GOSSS spectral classification is O8.5~V(n).

\paragraph{HDE~229\,202 = BD~+39~4162 = ALS~11\,274.}
\object[HDE 229202]{} 
\citet{HiltJohn56} classified this star as O8:~V. The GOSSS spectral classification is O7.5~V(n)((f)).

\paragraph{ALS~11\,355 = LS~II~+39~53.}
\object[ALS 11355]{} 
\object[2MASS J20272914+3945054]{} 
\citet{VijaDril93} classified this star as O7~V:. The GOSSS spectral classification is O8~V(n)((f)). When observing this target, we placed a nearby star, 
2MASS~J20272914+3945054, on the slit and it turned out to have a G spectral type.

\paragraph{HD~194\,649~AB = ALS~11\,324.}
\object[HD 194649]{} 
\citet{HiltJohn56} classified this star as O6.5. The GOSSS spectral classification is O6.5~V((f)). The WDS lists a companion 0\farcs4 to the SW with a $\Delta m$ of 0.6
that we are unable to separate in the GOSSS spectra but that is easily seen in unpublished AstraLux lucky images \citep{Maiz10a}.

\paragraph{HDE~228\,759 = BD~+41~3689 = ALS~11\,083 = LS~III~+41~15.}
\object[HDE 228759]{} 
\citet{MayeMaca71} classified this star as O6. The GOSSS spectral classification is O6.5~V(n)((f))z. LS~III~+41~14 is located 30\arcsec\ away and was observed simultaneously.

\paragraph{LS~III~+41~14 = ALS~11\,081.}
\object[LS III +41 14]{} 
\citet{MayeMaca71} classified this star as O9~V. The GOSSS spectral classification is O9.5~V(n). HDE~228\,759 is located 30\arcsec\ away and was observed simultaneously.

\paragraph{BD~+40~4179 = ALS~11\,363.}
\object[BD +40 4179]{} 
\citet{HiltJohn56} classified this star as O8:~V. The GOSSS spectral classification is O8~Vz.

\paragraph{Cyg~OB2-B17 = V1827~Cyg = [CPR2002]~B17.}
\object[V1827 Cyg]{} 
\citet{Stroetal10} discovereed that this is an SB2 system composed of two O supergiants with spectral types O7~Iaf and O9~Iaf and a period of 4.0217~d. We obtained one GOSSS epoch in 
which the two components are clearly separated, as the system was caught with a $\Delta v$ of $\sim$450~km/s, and we obtained a spectral type of O6~Iaf~+~O9:~Iaf. The spectral type of 
the primary is earlier that the \citet{Stroetal10} result, possibly because of our better spectral resolution. The uncertainty in the secondary spectral type arises from inconsistencies
among some He\,{\sc ii} lines, possibly caused by the strong winds evidenced by the observed emission lines. Indeed, \HeII{4686} and \NIII{4634-41-42} for the primary appear as strong in
emission for the primary as for some O~Iafpe stars but we cannot give it that designation because \HeI{4471} does not show a P-Cygni profile. In any case, this system clearly deserves 
follow-up with a high-resolution spectrograph on a 10~m-class telescope.

\paragraph{Cyg~OB2-A24 = 2MASS~J20344410+4051584 = [CPR2002]~A24.}
\object[2MASS J20344410+4051584]{} 
\object[ALS 19626]{} 
\citet{Neguetal08a} classified this star as O6.5~III((f)). The GOSSS spectral classification is O6.5~III(f), see Table~2 in paper II. The star Cyg~OB2-A27 is located 13\farcs8 away and 
was placed on the slit. We obtain a spectral classification of B0~Ia, consistent with the \citet{Hans03} result.

\paragraph{2MASS~J20315961+4114504.}
\object[2MASS J20315961+4114504]{} 
\object[WR 144]{}
\citet{ComePasq12} classified this star as O7~V. The GOSSS spectral classification is O7.5~Vz. WR~144 is located 48\farcs7 away from this star and was placed on the slit.

\paragraph{Cyg~OB2-A11 = ALS~21\,079 = [MT91]~267 = [CPR2002]~A11.}
\object[ALS 21079]{} 
\object[Schulte 12]{}
\citet{Neguetal08a} classified this target as O7~Ib-II(f) and \citet{Kobuetal12} identified it as an SB1 system with a 15.511~d period. The GOSSS spectral classification is very similar,
O7~Ib(f). Cyg~OB2-A11 is 1\farcm8 away from Cyg~OB2-12, a well known, highly extinguished B hyper/supergiant \citep{Morgetal54,Walbetal15} which we placed on the slit and for which we found a 
spectral type of B5~Ia. Note, however, that this star is a photometric variable \citep{Salaetal15} and that for late B stars the \HeI{4471}/\MgII{4481} ratio is luminosity dependent, which leads 
to differing spectral subtypes in the literature \citep{Claretal12}.

\paragraph{ALS~15\,108~AB = [MT91]~516~AB.}
\object[ALS 15108]{} 
\citet{MassThom91} classified this star as O5.5~V((f)). The GOSSS spectral classification is O6~IV((f)) and was obtained with GTC data. \citet{Masoetal09} found a B component with $\Delta m$ of 0.4 magnitudes 0\farcs7 
away (see also \citealt{CabNetal14}). We were unable to spatially resolve the secondary with GOSSS but it is clearly seen in unpublished AstraLux lucky images \citep{Maiz10a}.

\paragraph{Cyg~OB2-5~B = V279~Cyg~B = BD~+40~4220~B = Schulte~5~B = ALS~11\,408~B.}
\object[BD +40 4220 B]{} 
Cyg~OB2-5~B is 3 magnitudes dimmer than Cyg~OB2-5~A and is located 0\farcs934 away \citep{Maiz10a}. In paper I we mentioned that it is a mid-O star but we were unable to give a precise
classification due to the low S/N of the spectrum (the low S/N was caused by the contamination from A). We observed the AB pair again with GOSSS and this time we were able to clearly 
separate the two stars, with the B component being an O6.5~Iabfp. The p suffix is used due to the anomalous \HeII{4686} profile.

\paragraph{ALS~15\,134 = [MT91]~534.}
\object[ALS 15134]{} 
\object[[MT91] 558]{}
\citet{MassThom91} classified this star as O7.5~V. The GOSSS spectral classification is O8~Vz. We also placed on the slit [MT91]~558, located 1\farcm0 away, and found it is an F star.

\paragraph{Cyg~OB2-22~D = Schulte~51 = [MT91]~425 = ALS~15\,149.}
\object[ALS 15149]{} 
Cyg~OB2-22~D is part of the Cyg~OB2-22 Trapezium-like system \citep{Maiz10a}, in which we classified three O stars in paper I: A, Ba, and C (=~V2185~Cyg).
\citet{MassThom91} classified the D component as B0~V and several subsequent works repeat this classification. In the GOSSS data, we classify it as O9.5~Vn, which is the first time it
has been recognized as an O star. 

\paragraph{ALS~15\,144 = [MT91]~378.}
\object[ALS 15144]{} 
\citet{MassThom91} classified this star as B0~V and according to \citet{Kobuetal14} it is an SB1. However, according to GOSSS data it is an O9.7~III(n) and, to our knowledge, it has never been
classified as an O star before.

\paragraph{ALS~15\,119 = [MT91]~588.}
\object[ALS 15119]{} 
\object[Schulte 18]{} 
\citet{MassThom91} classified this star as B0~V. However, according to GOSSS data it is an O9.5~IV(n) and, to our knowledge, it has never been
classified as an O star before. We also placed the nearby Cyg~OB2-18 on the slit: it is an early-B supergiant, in agreement with previous classifications (e.g. \citealt{Kimietal07}).

\paragraph{Cyg~OB2-17 = ALS~15\,105 = Schulte~17 = [MT91]~339.}
\object[ALS 15105]{} 
\citet{MassThom91} classified this star as O8.5~V and according to \citet{Kobuetal14} it is an SB1. The GOSSS spectral classification is O8~V. This star was observed placing ALS~15\,111 on the same slit.

\paragraph{Cyg~OB2-16 = LS~III~+41~33 = ALS~11\,415 = Schulte~16 = [MT91]~299. = Cyg~OB2-A43 = [CPR2002]~A43}
\object[ALS 11415]{} 
\citet{HiltJohn56} classified this star as O8~V. The GOSSS spectral classification is O7.5~IV(n). This star was observed placing Cyg~OB2-6 on the same slit.

\paragraph{Cyg~OB2-6 = LS~III~+41~35 = ALS~11\,418 = Schulte~6 = [MT91]~317 = BD~+40~4221.}
\object[ALS 11418]{} 
\citet{JohnMorg54} classified this star as O8~(V). The GOSSS spectral classification is O8.5~V(n). This star was observed placing Cyg~OB2-16 on the same slit.

\paragraph{ALS~15\,115 = [MT91]~485.}
\object[ALS 15115]{} 
\citet{MassThom91} classified this star as O8~V and we obtain the same spectral classification with GOSSS. \citet{Kobuetal14} find it is an eccentric SB1 with a period of 4066 d.

\paragraph{ALS~15\,111 = [MT91]~376.}
\object[ALS 15111]{} 
\citet{MassThom91} classified this star as O8~V and we obtain the same spectral classification with GOSSS. This star was observed placing Cyg~OB2-17 on the same slit.

\paragraph{Cyg~OB2-27~AB = ALS~15\,118~AB = Schulte~27~AB = [MT91]~696~AB.}
\object[ALS 15118]{} 
\object[2MASS J20335528+4118471]{} 
\citet{Kimietal15} classified this eclipsing SB2 as O9.5~V~+~B0.5~V and measured a distance of $1.32\pm0.07$ kpc. With GOSSS we caught this system with a $\Delta v$ of $\sim$450~km/s and obtain a classification of O9.7~V(n)~+~O9.7~V:(n). 
\citet{Salaetal15} measured the period of this eclipsing binary as 1.46917 d and \citet{Lauretal15} detected a small period change. \citet{CabNetal14} found a B companion with a $\Delta m$ = 0.94 and a separation of 0\farcs023. We also placed
the nearby [MT91]~674 on the slit and we found it is a K star.

\paragraph{Cyg~OB2-73 = Schulte~73.}
\object[Schulte 73]{} 
\citet{Kimietal09} classified this star as O8~III~+~O8~III and measured a period of 17.28~d, which were later revised to O8.5~III:~+~O9~III: and 34.88~d by \citet{Kobuetal14}. With GOSSS we caught the system at a $\Delta v$ of $\sim$175~km/s and
we obtain a classification of O8~Vz~+~O8~Vz, clearly excluding the prior giant classifications because the $z$ ratio is larger than 1.1 for both components.

\paragraph{Cyg~OB2-25~A = ALS~15\,104~A = Schulte~25~A = [MT91]~531~A.}
\object[ALS 15104 A]{} 
\object[ALS 15104 B]{}
\citet{MassThom91} classified this system as O8.5~V. The GOSSS spectral classification is O8~Vz and we are able to spatially separate the B component, which turns out to be an early-B star. \citet{CabNetal14} measure a separation between A and B
of 1\farcs45 and a visual $\Delta m$ of 0.5 magnitudes, which is consistent with the GOSSS data and with what we see in unpublished AstraLux lucky images \citep{Maiz10a}.

\paragraph{Cyg~OB2-10 = BD~+41~3804 = Schulte~10 = [MT91]~632 = ALS~11\,434.}
\object[BD +41 3804]{} 
\citet{HiltJohn56} classified this system as O9.5~Ia. The GOSSS spectral classification is O9.7~Iab. \citet{CabNetal14} detected a B component with a separation of 0\farcs22 and a $\Delta m$ of 2.0 which we were unable to spatially separate
in the GOSSS data but that can be seen in unpublished AstraLux lucky images \citep{Maiz10a}.

\paragraph{ALS~15\,125 = [MT91]~736.}
\object[ALS 15125]{} 
\citet{MassThom91} classified this star as O9~V. The GOSSS spectral classification is O9.5~IV:, with the uncertainty in the spectral type arising from the discrepant results for the
\HeII{4686} to \HeII{4713} and the \SiIV{4089} to \HeI{4026} ratios. This star was observed placing Cyg~OB2-29 on the same slit.

\paragraph{ALS~15\,114 = [MT91]~771.}
\object[ALS 15114]{} 
\object[[MT91] 775]{}
This system is an SB2 with O7~V~+~O9~V spectral types and a period of 2.82105~d according to \citet{Kobuetal14}. We only obtained one GOSSS epoch and it was at an unfavorable phase, so we can only derive a combined spectral type of O7.5~V(n)((f)).
We placed the neighbor [MT91]~775 and determined it is a K star.

\paragraph{Cyg~OB2-29 = ALS~15\,110 = Schulte~29 = [MT91]~745.}
\object[ALS 15110]{} 
\citet{MassThom91} classified this star as O7~V and \citet{Kobuetal14} determined it is an SB1 with a 151.2~d period. The GOSSS spectral classification is O7.5~V(n)((f))z. This star was observed placing ALS~15\,125 on the same slit.

\paragraph{BD~+43~3654 = ALS~11\,429.}
\object[BD +43 3654]{} 
\citet{ComePasq07} classified this object as O4~If and proposed it is a runaway ejected from Cyg~OB2. The GOSSS data yield the same spectral classification.

\paragraph{BD~+45~3216~A = ALS~11\,435~A.}
\object[BD +45 3216 A]{} 
\citet{HiltJohn56} classified this target as O8. GOSSS data yields a rather different spectral classification of O5~V and there is a good explanation for the difference. The WDS lists a B companion with a $\Delta m$ of 0.3 mag and a
separation of 0\farcs8 which is clearly seen in our unpublished AstraLux lucky images \citep{Maiz10a}. We were able to obtain GOSSS spectra under good seeing conditions and spatially separate both components. The O5~V((f))z classification
corresponds to the A component while the B component is an early B star, which explains the previous O8 classification (a reasonable classification for the combined spectrum).

\paragraph{Bajamar~Star = 2MASS~J20555125+4352246.}
\object[2MASS J20555125+4352246]{} 
\citet{ComePasq05} identified this star as the main ionizing source of the North America Nebula\footnote{We adopt the name ``Bajamar Star'' for the object due to its position relative to the North America Nebula, just to the east of
the ``Florida Peninsula''. ``Islas de Bajamar'', meaning ``low-tide islands'' in Spanish, was the original name of the Bahamas islands because many of them are only easily seen from a ship during low tide.} and classified it as O5~V. We 
observed this system several times and we noticed that the velocity of the He\,{\sc i} lines was not the same as that of the He\,{\sc ii} and that there were significant variations in the relative velocity ($\Delta v\sim 300$~km/s) 
between epochs. Therefore, the system is an SB2 composed of a star earlier than the average type obtained by \citet{ComePasq05} and a later star. Our best current spectrum is one where we caught the system with the GTC with a $\Delta v$ of
$\sim$100~km/s and from which we assign spectral types of O3.5~III((f*))~+~O8:. Note that prior to 2015 there were only two known Galactic O stars in the northern hemisphere earlier than O4, Cyg~OB2-7 and Cyg~OB2-22~A, see paper I. 
\citet{Maizetal15a} added two more (see below) and in this paper we identify the fifth and the sixth cases (the primary here and the primary of Sh~2-158~1).

\paragraph{LS~III~+46~12 = ALS~11\,449.}
\object[LS III +46 12]{} 
\citet{MayeMaca73} classified this star as O6. In \citet{Maizetal15a,Maizetal15c} we used GOSSS and other data to study it and classify it as O4.5~V((f)). Here, we change the luminosity classification to IV with the
introduction of that class for the spectral subtypes O4-O5.5. The change from IV to V partially (but not completely) alleviates the luminosity/distance discrepancy within Berkeley~90 discussed in \citet{Maizetal15a,Maizetal15c}. 
Therefore, we cannot yet discard that there is a hidden binary component in LS~III~+46~12.

\paragraph{LS~III~+46~11 = ALS~11\,448.}
\object[LS III +46 11]{} 
\citet{Motcetal97} classified this star as O3-5~III(f)e.  In \citet{Maizetal15a,Maizetal15c} we used GOSSS and other data to discover it is an eccentric massive SB2 binary with two similar components, each with spectral type O3.5~If*. 
As previously mentioned, the spectral classification of this 97.3~d binary raised the number of northern-hemisphere Galactic O stars earlier than O4 to four objects.

\paragraph{ALS~11\,761 = LS~III~+46~50.}
\object[ALS 11761]{} 
\citet{NeguMarc03} classified this star as O9.5~III. The GOSSS spectral classification is O9.2~II and we use it as one of the new standards.

\paragraph{ALS~12\,050 = LS~III~+57~18.}
\object[ALS 12050]{} 
\object[Tyc 3976-00299-1]{}
\citet{Russetal07} classified this star as O5~V. The GOSSS spectral classification is O5~V((f)). We also placed the nearby star Tyc~3976-00299-1 on the slit and we found it is an A star.

\paragraph{BD~+55~2722~A = ALS~12\,292~A = LS~III~+55~36~A.}
\object[BD +55 2722 A]{} 
BD~+55~2722 is a Trapezium-like system at the core of the open cluster Teutsch~127. The A component is the brightest one and was classified by \citet{MayeMaca73} as O9~V, though it is likely their result included also the nearby B 
component, which we spatially resolve (see below). The GOSSS spectral classification is O8~Vz.

\paragraph{BD~+55~2722~C = LS~III~+55~37 = ALS~12\,293.}
\object[BD +55 2722 C]{} 
BD~+55~2722~C is the wide component in the BD~+55~2722~ABC and is located 10\arcsec\ away from the center of the AB pair. \citet{Crametal78} classified it as O7~V. The GOSSS spectral classification is O7~V(n)z~+~B, which is the first time that this
system has been identified as an SB2 to our knowledge. The system was caught at a $\Delta v$ of $\sim$500~km/s.

\paragraph{BD~+55~2722~B = ALS~12\,292~B = LS~III~+55~36~B.}
\object[BD +55 2722 B]{} 
BD~+55~2722~B is the closer companion to BD~+55~2722~A and is located 1\farcs7 away with a $\Delta m$ of 0.3 mag according to the WDS and confirmed in our unpublished AstraLux lucky images \citep{Maiz10a}. Its existence is referrred to
in several of the papers on BD~+55~2722 but, to our knowledge, no spectral type has ever been published (see e.g. \citealt{Sauretal10}). We placed the slit along the AB position angle and we were able to spatially separate the two 
components. The GOSSS spectral type for BD~+55~2722~B is O9.5~V, making it the third O star in the Trapezium system.

\paragraph{ALS~12\,320 = LS~III~+55~45.}
\object[ALS 12320]{} 
\object[2MASS J22202471+5608080]{}
\citet{McCu55} classified this object as O8. The GOSSS spectral classification is O7~IV((f)) and we use it as one of the new standards. The nearby star 2MASS~J22202471+5608080 was also placed on the slit: it is a late-B star.

\paragraph{ALS~12\,370 = LS~III~+55~65.}
\object[ALS 12370]{} 
\citet{Hilt56} classified this object as O5. The GOSSS spectral classification is O6.5~Vnn((f)).

\paragraph{ALS~12\,619 = LS~III~+57~90.}
\object[ALS 12619]{} 
\citet{Massetal95a} classified this star as O8~V((f)). The GOSSS spectral classification is O7~V((f))z.

\paragraph{BD~+55~2840 = ALS~12\,685.}
\object[BD +55 2840]{} 
\citet{HiltJohn56} classified this star as O7.5p. The GOSSS spectral classification is O7.5~V(n).

\paragraph{ALS~12\,688 = LS~III~+56~109.}
\object[ALS 12688]{} 
This star appears as O\ldots\ in Simbad but we have not found any reference to any spectral classification, so we suspect the reason for the listing is a photometric classification. We have found it to be not only an O star but an SB2
as well. The GOSSS spectral type is O5.5~V(n)((fc))~+~B, with the companion having \HeI{4471} in absorption but no sign of \HeII{4542} (hence, we can only classify it as a B) and this system being caught on two different epochs with a
$\Delta v$ of 500~km/s. \citet{Lewaetal09} confirm the binary nature of the system by identifying it as an eclipsing binary with a 2.02170~d period.

\paragraph{BD~+62~2078 = ALS~12\,408.}
\object[BD +62 2078]{} 
\citet{HiltJohn56} classified this star as O7. The GOSSS spectral classification is O7~V((f))z.

\paragraph{HD~213\,023~A = BD~+62~2081 = ALS~12\,424.}
\object[HD 213023 A]{} 
\citet{Morgetal55} classified this star as O9~V:. The GOSSS spectral classification is O7.5~Vz, which is significantly different. We suspect that the reason is that the original spectral type included the B component, located at a
separation of 1\farcs7 with a $\Delta m$ of 0.4 according to the WDS. The GOSSS spectra spatially separated the two components and we obtained an early B spectral type for the companion.

\paragraph{ALS~12\,749 = LS~III~+56~119.}
\object[ALS 12749]{} 
\citet{Massetal95a} classified this star as O9.5~V((f)). The GOSSS spectral classification is O9~V.

\paragraph{Sh~2-158~2 = 2MASS~J23133024+6130103.}
\object[2MASS J23133024+6130103]{} 
\citet{Dehaetal79} correctly identified that this object and Sh~2-158~1 (see below) are the two ionizing stars of the H\,{\sc ii} region Sh~2-158~2 (or NGC 7538) while the bright nearby 2MASS~J23133680+6130395 is a K star 
(which we confirmed by placing one of our slits on it). However, they only provided rough spectral types of O7 for both sources. We have observed Sh~2-158~2 with GOSSS and found that it is an SB2 caught with a $\Delta v$ of
$\sim 275$~km/s and spectral types O9.5:~V~+~B0.5:~V. Those are significantly later than O7, so we suspect that the previous classification was intended for Sh~2-158~1 (though that is also inaccurate, see below).

\paragraph{Sh~2-158~1 = Tyc~4279-01463-1.}
\object[Tyc 4279-01463-1]{} 
As for the previous object, there is little information about Sh~2-158~1 in the literature. That is why we were surprised to find out with GOSSS that it is a very early type SB2 caught with the GTC with a $\Delta v$ of $\sim$375~km/s and 
spectral types of O3.5~V((f*))~+~O9.5:~V. As previously mentioned, the primary of this system is the fifth Galactic O star earlier than O4 in the northern hemisphere (the sixth is the primary of the Bajamar Star, see above). Most 
of the \HeI{4471} absorption originates in the secondary while most of the \HeII{4542}
absorption originates in the primary. If the absorption is considered to originate in a single object, the combined spectral type would be O4.5 (which is still too early to explain the previous classification as O7).

\paragraph{BD~+60~2635 = ALS~13\,256.}
\object[BD +60 2635]{} 
\citet{Mart72} classified this star as O9~III. The GOSSS spectral classification is O6~V((f)). The original classification was done with objective-prism data, so a possible explanation for the discrepancy is the low quality of the
original data. However, \citet{NeguMarc03} classified this star as ON7~III(f), which is also discrepant (we see a N enhancement in the GOSSS data but not significant enough to classify it as ON) so it is also possible that the star is
variable.

\paragraph{BD~+66~1661 = ALS~13\,323.}
\object[BD +66 1661]{} 
\citet{HiltJohn56} classified this star as O9~V. The GOSSS spectral classification is O9.2~V.

\paragraph{V747~Cep = BD~+66~1673 = ALS~13\,375 = NGC~7822-3.}
\object[V747 Cep]{} 
\object[2MASS J00015191+6731474]{}
This object is an eclipsing binary with a period of 5.33146~d in the open cluster Berkeley~59 that was classified as O5~Vn((f)) by \citet{Majaetal08}. The GOSSS spectral type is O5.5~V(n)((f)). No double lines were seen in our data. We placed the nearby
2MASS~J00015191+6731474 on the slit and found it is an early B star.

\paragraph{BD~+66~1675 = ALS~13\,379.}
\object[BD +66 1675]{} 
This object is the brightest star in the open cluster Berkeley~59 and was classified as O7 by \citet{HiltJohn56}. The GOSSS spectral classification is O7.5~Vz.

\paragraph{BD~+66~1674 = ALS~13\,378.}
\object[BD +66 1674]{} 
This object is the second brightest star in the open cluster Berkeley~59 and was classified as O9.5~V by \citet{Bisietal82}. The GOSSS spectral classification is O9.7~IV:. The uncertainty in the luminosity class is caused by 
the discrepancy between the \HeII{4686} to \HeII{4713} criterion (which yields III) and the \SiIV{4089} to \HeI{4026} criterion (which yields V), another example of the effect previously mentioned.

\paragraph{Tyc~4026-00424-1 = NGC~7822-29 = ALS~17\,957.}
\object[Tyc 4026-00424-1]{} 
\object[ALS 13 380]{}
This object is also located in Berkeley~59 and was classified as O9 by \citet{Walk65}. The GOSSS spectral classification is O7~V((f))z. We also placed a nearby star, ALS~13\,380, on the slit and found is an early B star.

\paragraph{ALS~6351 = LS~I~+62~139.}
\object[ALS 6351]{} 
\citet{HiltJohn56} classified this star as O7. The GOSSS spectral classification is O7~Vz.

\paragraph{BD~+60~134 = ALS~6405.}
\object[BD +60 134]{} 
\citet{Popp50} classified this star as O7. The GOSSS spectral classification is O5.5~V(n)((f)).

\paragraph{HD~5689 = BD~+62~178 = ALS~6425.}
\object[HD 5689]{} 
\citet{HiltJohn56} classified this star as O6. The GOSSS spectral classification is O7~Vn((f)).

\paragraph{ALS~6967 = LS~I~+59~112.}
\object[ALS 6967]{} 
\citet{HiltJohn56} classified this object as O9~V. We observed it star with GOSSS and found it is an SB2 with spectral types O8~V~+~B0:~V caught with a $\Delta v$ of 200~km/s. To our knowledge this is the first identification of this
object as a spectroscopic binary.

\paragraph{BD~+61~411~A = ALS~7203.}
\object[BD +61 411 A]{} 
\object[BD +61 411 B]{}
\citet{HiltJohn56} classified this star as O6. The GOSSS spectral classification is O6.5~V((f))z. There is a B component detected as a separate source in 2MASS located 5\farcs672 toward the NE. We were able to spatially separate it from
A and we found it is an early B star.

\paragraph{ALS~7833 = LS~I~+57~138.}
\object[ALS 7833]{} 
This is one of the two bright stars in the open cluster Alicante~1. \citet{NeguMarc03} classified it as O7~V. The GOSSS spectral classification is O8~Vz.

\paragraph{MY~Cam = BD~+56~864 = ALS~7836.}
\object[MY Cam]{} 
This is the second of the two bright stars in the open cluster Alicante~1. \citet{HiltJohn56} classified it as O6nn and \citet{Loreetal14} discovered it is an overcontact binary composed of two O stars with a period of 
1.1754514~d. In the GOSSS data the two components are separated by a $\Delta v$ of $\sim$500~km/s and we assign them spectral types O5.5~V(n)~+~O6.5~V(n).

\paragraph{BD~+50~886 = ALS~7868.}
\object[BD +50 886]{} 
This object is the main ionizing source of the H\,{\sc ii} region Sh~2-206 (=~NGC~1491) and was classified by \citet{Moffetal79} as O5neb. The GOSSS spectral classification is O4~V((fc)).

\paragraph{BD~+52~805 = ALS~7928.}
\object[BD +52 805]{} 
This object is the brightest star in the open cluster Waterloo~1 and was classified by \citet{Moffetal79} as O9.5~V. The GOSSS spectral classification is O8~V(n).

\paragraph{ALS~8272 = LS~V~+38~12.}
\object[ALS 8272]{} 
\object[HDE 277990]{} 
This object was classified by \citet{Georetal73} as O9~V. We observed it with GOSSS and found it is an SB2 with spectral types O7~V((f))~+~B0~III-V caught with a $\Delta v$ of 325~km/s. To our knowledge this is the 
first identification of this object as a spectroscopic binary. We placed the neighbor HDE~277\,990 on the slit and found it is an F star.

\paragraph{ALS~8294 = LS~V~+33~15 = NGC~1893-149.}
\object[ALS 8294]{} 
\citet{WalkHodg68} classified this star as O7. The GOSSS spectral classification is O7~V(n)z.

\paragraph{ALS~19\,265 = Rubin~\&~Losee~128.}
\object[ALS 19265]{} 
This object was classified as O7~V by \citet{Chro79}, who was searching for O stars at large Galactocentric radii near the Galactic anticenter. We find the spectrum to be
quite remarkable and we classify it as O4.5~V((c))z, with the ((c)) suffix indicating \CIII{4650} emission without apparent \NIII{4634-41-42} emission, an effect that we have not
seen in any other O star. Another peculiarity is the difference between the He\,{\sc i} and He\,{\sc ii} profiles.
We should caution that ALS~19\,265 could be an evolved low-mass star. For example, we also observed another one of the stars in \citet{Chro79}, ALS~19\,270, 
and its GOSSS spectrum looks quite similar to that of ALS~19\,265 though it is now known to be a planetary nebula nucleus or PNN \citep{Alleetal15}. To investigate this further we
downloaded the SDSS data for the target. In the images it appears as a blue object but no nebulosity is seen in the $r$ band, as it would be expected of a PN (the star could still be
a naked sdO). We have also processed the SDSS photometry through CHORIZOS \citep{Maiz04c,Maiz13a,Maizetal14a} assuming that ALS~19\,265 is a ZAMS O star. Under those circumstances, it
would have to be located at a distance close to 30 kpc and its extinction would have a large value of \rv\ (between 5 and 6, an indication that some of the extinction takes place in
an environment depleted of small dust grains such as an H\,{\sc ii} region or a PN). Such a large distance would be extreme for a Galactic O star (with the likely low metallicity
providing a possible explanation for the peculiar spectrum) but the available information does not exclude the possibility of the object being a lower luminosity, lower mass, closer 
evolved star. Gaia should be able to provide us with a distance measurement and decide between the two options.

\paragraph{HDE~256\,725~A = BD~+19~1339~A = ALS~46.}
\object[HDE 256725 A]{} 
This star is the brightest object of the Trapezium system HDE~256\,725 and was classified as O6 by \citet{Moffetal79}. The GOSSS spectral type is O5~V((fc)).

\paragraph{HDE~256\,725~B = BD~+19~1339~B = ALS~47.}
\object[HDE 256725 B]{} 
This star is the second brightest object of the Trapezium system HDE~256\,725 and was classified as O8~III by \citet{Moffetal79}. The GOSSS spectral type is O9.5~V.

\paragraph{Tyc~0737-01170-1 = NGC~2264~+10~60.}
\object[Tyc 0737-01170-1]{} 
\object[ALS 9044]{} 
\citet{Voroetal85} classified this star as O5:. The GOSSS spectral classification is O7~Vz. We also placed the nearby ALS~9044 and we found it is a late-B supergiant.

\paragraph{ALS~85.}
\object[ALS 85]{} 
This star was classified as O9.5~IV by \citet{Moffetal79}. \citet{Aldoetal15} find two dim companions. The GOSSS spectral type is O7.5~V. We have applied a CHORIZOS analysis to this star similar to
the one above for ALS~19\,265. If this target is a ZAMS O star it would be located at a distance of $\sim$7.5~kpc. The inclusion of the mentioned two dim companions would place it slightly closer but if the 
star is slightly evolved it should be slightly farther away. Therefore, this is a good candidate for an O star at large Galactocentric radius, something that could be confirmed once the Gaia distance 
becomes available.


\paragraph{ALS~207.}
\object[ALS 207]{} 
This object is the main ionizing source of the H\,{\sc ii} region Sh~2-301 (=~RCW~6) and was classified as O7 by \citet{Moffetal79}. The GOSSS spectral classification is O6.5~V((f)). 

\paragraph{BD~-15~1909 = ALS~552.}
\object[BD -15 1909]{} 
\citet{FitzMoff80} classified this star as O8. The GOSSS spectral classification is O6.5~V((f))z.

\paragraph{ALS~458 = Sh~2-306~4.}
\object[ALS 458]{} 
\citet{Moffetal79} classified this star as O5. The GOSSS spectral classification is O6.5~V((f))z.

\paragraph{V441~Pup = 4U~0728-25 = ALS~437.}
\object[V441 Pup]{} 
This X-ray binary was classified as O8/9~Ve by \citet{Neguetal96}. The GOSSS spectral classification is O5:~Ve. The reason for the difference in the classifications is that the GOSSS spectral types 
do not take into consideration the infilling of \HeI{4471} (i.e. they are strict, not corrected, spectral types), which is common among Oe stars. This is the earliest spectral type we have assigned to an Oe
star in GOSSS (the previous record holder was HD~39\,680 = V1382~Ori, an O6).

\paragraph{CPD~-26~2704 = ALS~830.}
\object[CPD -26 2704]{} 
\object[ALS 832]{} 
This object in the Haffner~18 open cluster was classified as O7k by \citet{FitzMoff74}. The GOSSS spectral type is O7~V(n). We placed another cluster star, ALS~832, on the slit and found it is an
early-B star.

\paragraph{V467~Vel = CPD~-45~2920 = ALS~1135.}
\object[V467 Vel]{} 
\object[Tyc 8151-01072-1]{}
This system is an SB2 with O6.5~V~+~B1~V spectral types and a period of 2.753~d according to \citet{FerLNiem06a}. We only obtained one GOSSS epoch and it was at an unfavorable phase, so we can only derive 
a combined spectral type of O6.5~V(n)((f)). Given the large magnitude difference, the secondary does not seem to affect the combined spectral type. We placed the nearby Tyc~8151-01072-1 on the slit and
found is an early-B star.

\paragraph{CPD~-49~2322 = ALS~1267.}
\object[CPD -49 2322]{} 
This object in the Pismis~11 open cluster was classified as O8~V by \citet{MarcNegu09}. The GOSSS spectral type is O7.5~V((f)). 

\paragraph{HD~90\,273.}
\object[HD 90273]{} 
\citet{Hoff56} classified this object as O7. The GOSSS spectral type is ON7~V((f)). We placed the nearby star HDE~302\,748 on the slit and found it is a mid-B star.

\paragraph{THA~35-II-42 = WR~21a.}
\object[THA 35-II-42]{} 
\citet{Niemetal06b} discovered that this star is a binary, assigned it spectral classifications of WN6h~+~O, and noted that its spectrum resembles that of HD~93\,162 (=~WR~25). However, as pointed out in
paper II, \citet{CrowWalb11} moved HD~93\,162 into the early Of/WN (or ``early-slash'') category on the basis of its P-Cygni H$\beta$ profile. Here we do the same with THA~35-II-42 and assign it a spectral
type of O2~If*/WN5. We do not see \HeI{4471} from the secondary, which is consistent with its recent identification by \citet{Trametal16} as an O3 star, but it is quite possible that our classification 
corresponds to a composite spectrum (we only have one epoch with a good S/N). We placed the nearby star Tyc~8608-00069-1 on the slit and found it is a mid-B star.

\paragraph{HD~89\,625 = CPD~-59~2044 = ALS~1492.}
\object[HD 89625]{} 
\citet{CannMaya49} classified this star as B0 (chart 135) and \citet{HoukSwif99} as B3/4~V:. The GOSSS spectral type is ON9.2~IVn and is discussed in more detail in \citet{Walbetal16}. To our knowledge this object had never been 
classified as an O star.

\paragraph{2MASS~J10224377-5930182 = [VRV91]~85.}
\object[2MASS J10224377-5930182]{} 
\object[Hen 3-406]{}
This object is located in the little-studied open cluster [KPS2012]~MWSC~1797 \citep{Kharetal13} and was classified as B0:~V by \citet{vanGetal91}. We observed it with GOSSS and we found it is an
O8~V(n), which is the first time it has been identified as an O star. We placed the nearby star Hen 3-406 on the slit and found is an early-type Be star. Note that \citet{vanGetal91} incorrectly
claimed that Hen 3-406 was an O-type dwarf.

\paragraph{2MASS~J10224096-5930305 = [VRV91]~82.}
\object[2MASS J10224096-5930305]{} 
This object is also located in [KPS2012]~MWSC~1797 and it was classified as O~V: by \citet{vanGetal91}. The GOSSS spectral type is O7~V((f))z.

\paragraph{ALS~18\,551.}
\object[ALS 18551]{} 
This star in Collinder~228 appears as O5 in Simbad but this is just a photometric classification from \citet{Wram76}, not a spectroscopic classification. The GOSSS spectral type is O4.5~V(n)z~+~O4.5~V(n)z 
with the system caught with a $\Delta v$ of $\sim$450~km/s. This is the first time this system is identified as being an SB2 and having an O-type spectral classification.

\paragraph{2MASS~J10584671-6105512.}
\object[2MASS J10584671-6105512]{} 
This target in Collinder~228 has no current entry in Simbad and, to our knowledge, has never been described in any publication. We observed it because we placed it on the same slit as ALS~18\,553 (see below) 
and we discovered it is an O8~Iabf, another example of a new O star.

\paragraph{ALS~18\,553.}
\object[ALS 18553]{} 
This star in Collinder~228 appears as O5+ in Simbad but this just a photometric classification from \citet{Wram76}, not a spectroscopic classification. The GOSS spectral type is O6~IIf, which is the
first time this system has received an O-type spectral classification.

\paragraph{2MASS~J10583238-6110565.}
\object[2MASS J10583238-6110565]{} 
This target in Collinder~228 has no current entry in Simbad and, to our knowledge, has never been described in any publication. We observed because we placed it on the same slit as ALS~18\,553 (see above) 
and we discovered it is an O5~V((f))~+~O7~V((f)) caught with a $\Delta v$ of $\sim$200~km/s. This is the first time this system is identified as being an SB2 and having an O-type spectral classification.

\paragraph{THA~35-II-153.}
\object[THA 35-II-153]{} 
This target in Collinder~228 has one single entry in Simbad \citep{The66} and no spectral classification. We observed it because we placed it on the same slit as a repeat observation for ALS~2063 
(see paper~II) and we found it to be an O3.5~If*/WN7. Early Of/WN stars are quite rare (there are only two in paper II and two more, including this one, in this paper) but to find a new example next to an 
O Iafpe star such as ALS~2063 (another rare class with just six examples in paper II) is simply remarkable. What is less surprising is that this coincidence takes place in the Carina Nebula, the richest 
stellar nursery within 3 kpc of the Sun. 

\paragraph{HD~97\,966 = CPD~-58~3372 = ALS~2276.}
\object[HD 97966]{} 
\citet{Morgetal55} classified this star as O7.5. The GOSSS spectral classification is O7~V((f))z and we select it as the new standard star for this type.

\paragraph{HD~97\,319 = CPD~-60~2606 = ALS~2217.}
\object[HD 97319]{} 
\object[HD 97352]{} 
\citet{Feasetal61} classified this star as O9.5~Ib. The GOSSS spectral classification is O7.5~IV((f)) and we use it as one of the new standards. We placed the nearby HD~97\,352 on the slit and found it
is an early-B star.

\paragraph{EM~Car = HD~97\,484 = CPD~-60~2638 = ALS~2232.}
\object[EM Car]{} 
\object[HDE 306190]{} 
\citet{SoliNiem86b} observed this previously known eclipsing SB2 and derived spectral types of O7.5~+~O8.5. The GOSSS spectral types are O7.5~V((f))~+~O7.5~V((f)), with the system caught at a $\Delta v$
of $\sim$475~km/s. We placed the nearby object HDE~306\,190 on the slit and found it is an early-B star.

\paragraph{NGC~3603~HST-51.}
\object[NGC 3603 56]{} 
\citet{Moff83} classified this star in NGC~3603 as O4~V(f) and \citet{Meleetal08} suggested an earlier type with a later companion. The GOSSS spectral type is O5.5~V(n) but the line profiles are anomalous,
indicating that it may indeed be a composite spectrum.

\paragraph{NGC~3603~HST-48.}
\object[NGC 3603 18]{} 
\citet{Meleetal08} classified this star in NGC~3603 as O3.5~If. The GOSSS spectral type is the same with the addition of the prescriptive * suffix, i.e. O3.5~If*. We use this object as one of the new 
standards.

\paragraph{NGC~3603~HST-24.}
\object[NGC 3603 47]{} 
\citet{Meleetal08} classified this star in NGC~3603 as O4~V. The GOSSS spectral type is O4~IV(f) and this is one of the new O4-O5.5 stars with luminosity class IV.

\paragraph{NGC~3603~MTT~25.}
\object[NGC 3603 19]{} 
\citet{Meleetal08} classified this star in NGC~3603 as O3~V((f)). The GOSSS spectral type is O5~V(n), with \HeI{4471} clearly detected.

\paragraph{HD~99\,546 = CPD~-58~3620 = ALS~2342.}
\object[HD 99546]{} 
\citet{Morgetal55} classified this star as O8. The GOSSS spectral type is O7.5~V((f))~Nstr.

\paragraph{HD~110\,360 = CPD~-59~4396 = ALS~2732.}
\object[HD 110360]{} 
\citet{Morgetal55} classified this star as O7 and \citet{Math89} identified it as an ON stra. \citet{Walbetal11} used GOSSS data to classify it as ON7~Vz. Here, with the new definition of the OVz phenomenon by \citet{Ariaetal16}
we reclassify it as ON7~V.

\paragraph{CPD~-61~3973 = ALS~3153.}
\object[CPD -61 3973]{} 
\object[Tyc 9008-01250-1]{}
\citet{VijaDril93} classified this star as O7~III. The GOSSS spectral type is O7.5~V((f)). We placed the slit on the nearby Tyc~9008-01250-1 and found it is an F star.

\paragraph{HD~122\,313 = CPD~-61~4286 = ALS~3187.}
\object[HD 122313]{} 
\citet{Houketal76} classified this star as O5/7. The GOSSS spectral type is O8.5~V. 

\paragraph{ALS~17\,591 = [OM80]~35.}
\object[ALS 17591]{} 
\object[Tyc 8706-00582-1]{}
This object is listed as OB+ in \citet{OrsaMuzz80} but, to our knowledge, it has never been classifed as an O star. With GOSSS we identify it as such and assign it a spectral classification of
O5:~n(f)p, noting that \HeII{4686} has a centrally reversed (and possibly variable) emission with the red peak stronger than the blue one. We placed the slit on the nearby Tyc~8706-00582-1 and
foud out it is a late-B star.

\paragraph{ALS~3386.}
\object[ALS 3386]{} 
\object[Tyc 8696-00095-1]{} 
\citet{Bassetal82} classified this star as O6~If. The GOSSS spectral type is O6~Iaf. We placed the slit on the nearby Tyc~8696-00095-1 and found it is an F star.

\paragraph{ALS~18\,049.}
\object[ALS 18049]{} 
\object[2MASS J15441197-5356471]{} 
\citet{Georetal94} derived a photometric classification of O7~V for this star but, to our knowledge, there are no published spectral classifications. The GOSSS spectral type is O9~V, making this another
case of a first-time O-type spectral classification. Note that ALS~18\,049 appears to be the ionizing source of a well defined H\,{\sc ii} region seen in both H$\alpha$ and WISE MIR images.
We placed the slit on the nearby 2MASS~J15441197-5356471 and found it is an F star.

\paragraph{Muzzio~III-9.}
\object[Muzzio III-9]{} 
\object[Muzzio III-10]{} 
\citet{Muzz74} derived a photometric classification of O5+: for this star but, to our knowledge, there are no published spectral classifications. The GOSSS spectral type is O8~Ib, making this another
case of a first-time O-type spectral classification. We placed the slit on the nearby Muzzio~III-10 and found it is a B star but the star is weak so the S/N of the spectrum is poor.

\paragraph{HD~145\,217 = CPD~-49~8996 = ALS~3499.}
\object[HD 145217]{} 
\citet{Feasetal61} classified this star as O8. The GOSSS spectral type is O8~V. However, some weak Si\,{\sc iii} and O\,{\sc ii} lines are visible, making it likely that this is a composite spectrum
of a slightly earlier O star and an early-B star.

\paragraph{HD~144\,647.}
\object[HD 144647]{} 
\citet{Feasetal61} classified this star as O8. The GOSSS spectral type is O8.5~V(n).

\paragraph{HDE~328\,209~AB = CPD~-44~7916 = ALS~3624.}
\object[HDE 328209]{}   
\object[HDE 328209 B]{} 
\citet{Feasetal61} classified this star as O9.5~I(a). The GOSSS spectral type is ON9~Ib-Iap and is discussed in \citet{Walbetal16}. OWN data \citep{Barbetal10} reveal it to be a short-period SB2.
Unpublished AstraLux Sur images reveal a visual companion (B) about 3\arcsec\ to the East that is unresolved in the GOSSS data.

\paragraph{HDE~329\,100~A = ALS 3815.}
\object[HDE 329100 A]{} 
\object[HDE 329100 B]{}
There is some confusion surrounding this double system, starting with the current classification of G0 in Simbad, which is clearly wrong. According to the WDS there are two components separated by 
4\farcs0 with a position angle of 105 degrees and a $\Delta m$ of 2 magnitudes. We have unpublished AstraLux Sur $iz$ data that are consistent (within small amounts) with that information and the 
2MASS PSC also points in the same direction. Simbad, however, incorrectly refers to the western component as B when it is really A (the brightest). \citet{Cram71} classified the system as O8.5. We were
able to separate the A (western) and B (eastern) components with GOSSS. HDE~329\,100~A is an O star with spectral type O8~V(n) while the B component is an early-B star.

\paragraph{HDE~326\,775 = CPD~-41~7848 = ALS~3906.}
\object[HDE 326775]{} 
\object[Tyc 7877-01317-1]{} 
\citet{VijaDril93} classified this star as O7~V. The GOSSS spectral type is O6.5~V(n)((f))z. We placed the nearby Tyc~7877-01317-1 on the slit and found it is a K star.

\paragraph{ALS~18\,769 = C1715-387-16 = HM~1-16.}
\object[ALS 18769]{} 
To our knowledge, this star in the open cluster Havlen-Moffat~1 has never received a spectral classification. We observed it with GOSSS and determined its spectral type to be O6~II(f), adding another
O star to the known Galactic sample.

\paragraph{HDE~323\,110 = ALS~4103.}
\object[HDE 323110]{} 
\citet{VijaDril93} classified this star as B0~IIIne. The GOSSS spectral type is ON9~Ia and is discussed in more detail in \citet{Walbetal16}. To our knowledge this object had never been classified 
as an O star. 

\paragraph{Tyc~7370-00460-1 = 2MASS~J17181540-3400061.}
\object[Tyc 7370-00460-1]{} 
\citet{Gvaretal11b} classified this star as O6.5:. The GOSSS spectral type is O6~V((f))~+~O8~V and we caught the system with a $\Delta v$ of $\sim$500~km/s. To our knowledge, this system had not been
characterized as an SB2 before. 

\paragraph{ALS~19\,693 = [N78]~51.}
\object[ALS 19693]{} 
\citet{Lortetal84} classified this star as O7:~V. The GOSSS spectral type is O6~Vn((f)). 

\paragraph{Pismis~24-15 = ALS~17\,700 = [N78]~46.}
\object[ALS 17700]{} 
\citet{Massetal01} classified this star as O8~V. The GOSSS spectral type is O7.5~Vz.

\paragraph{ALS~19\,692 = [N78]~49.}
\object[ALS 19692]{} 
\citet{Lortetal84} classified this star as O7:~V. The GOSSS spectral type is O5.5~IV((f)).

\addtocounter{figure}{-1}

\begin{figure*}
\centerline{\includegraphics*[width=\linewidth]{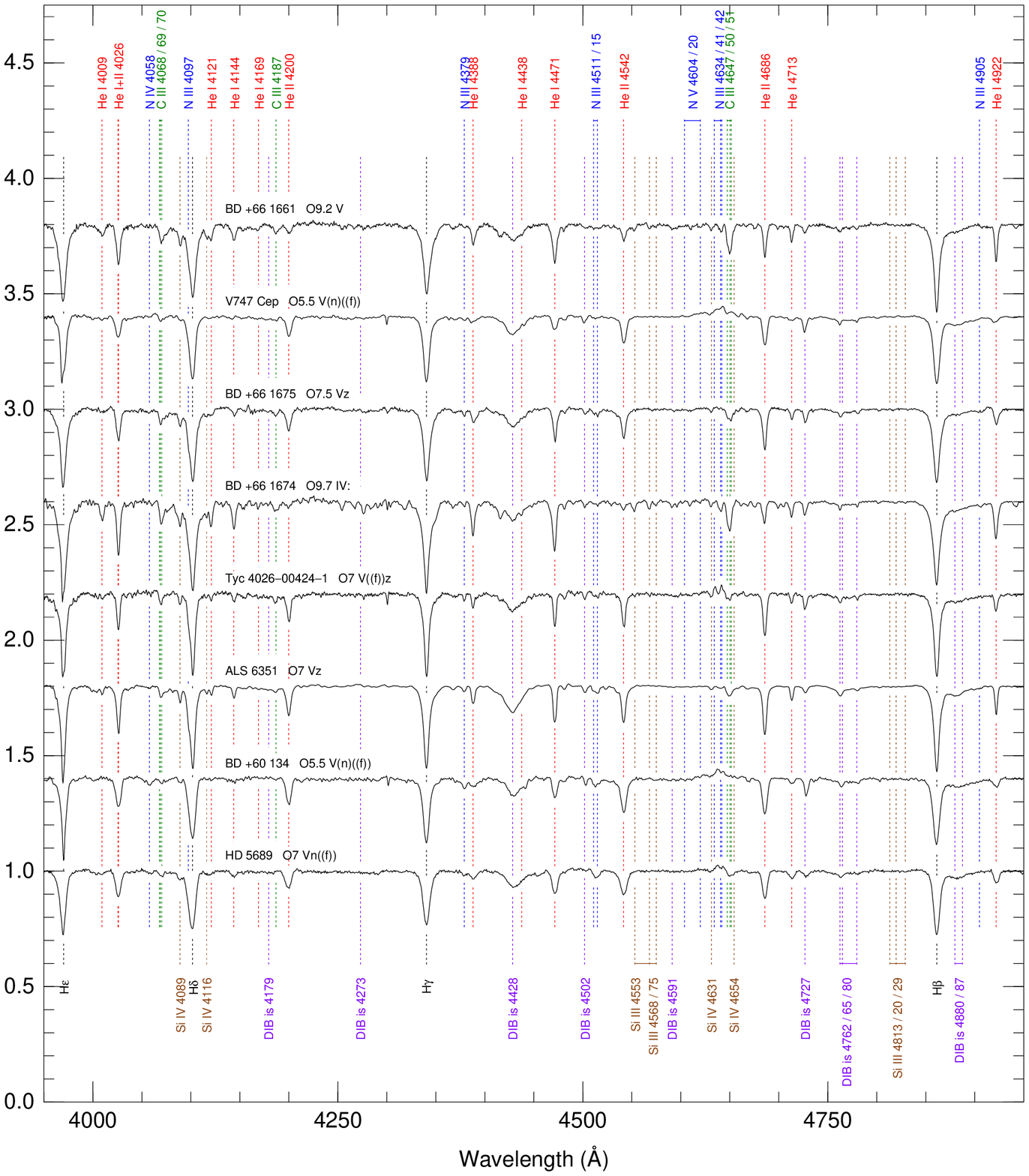}}
\caption{(continued).}
\end{figure*}	

\addtocounter{figure}{-1}

\begin{figure*}
\centerline{\includegraphics*[width=\linewidth]{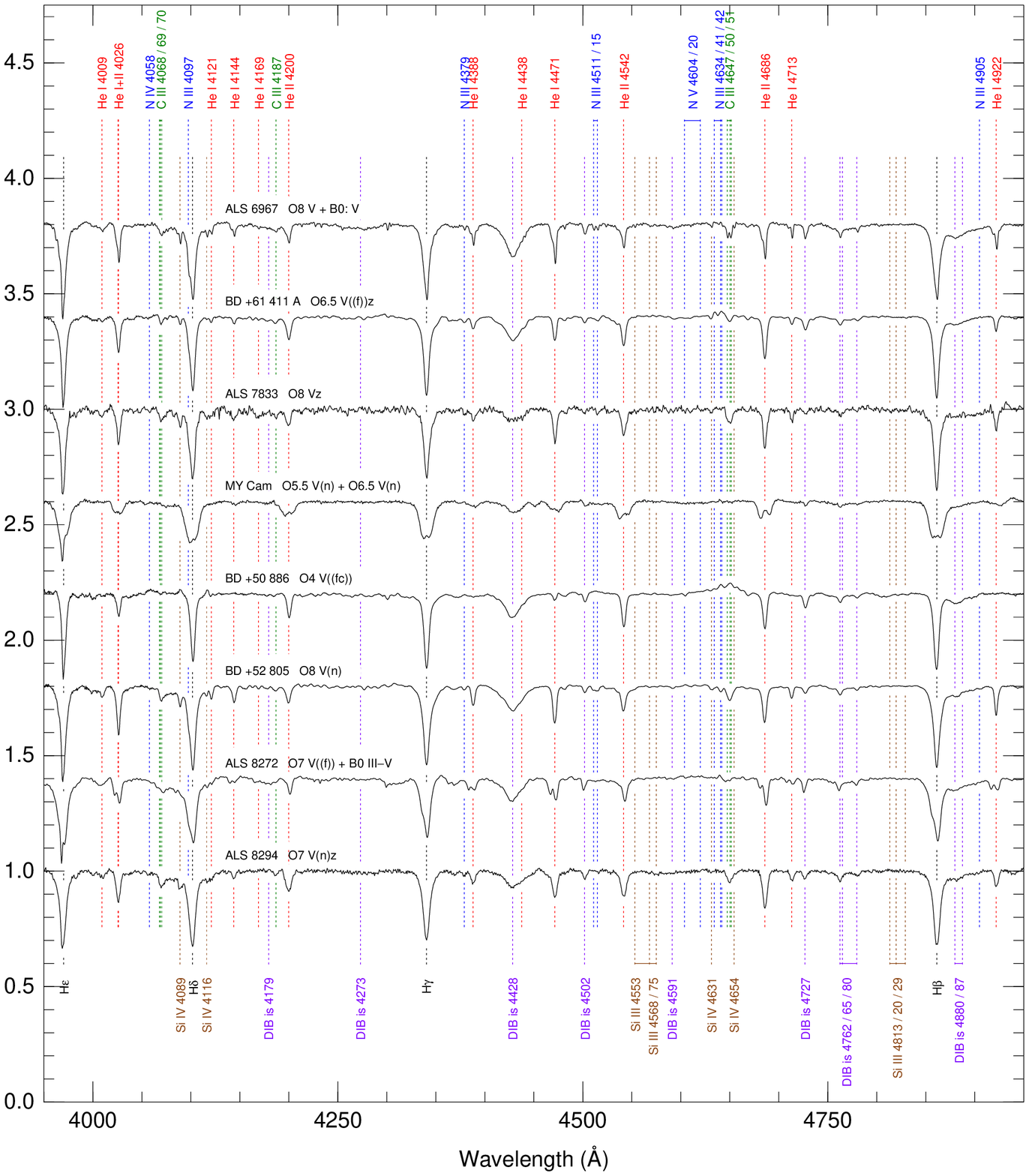}}
\caption{(continued).}
\end{figure*}	

\addtocounter{figure}{-1}

\begin{figure*}
\centerline{\includegraphics*[width=\linewidth]{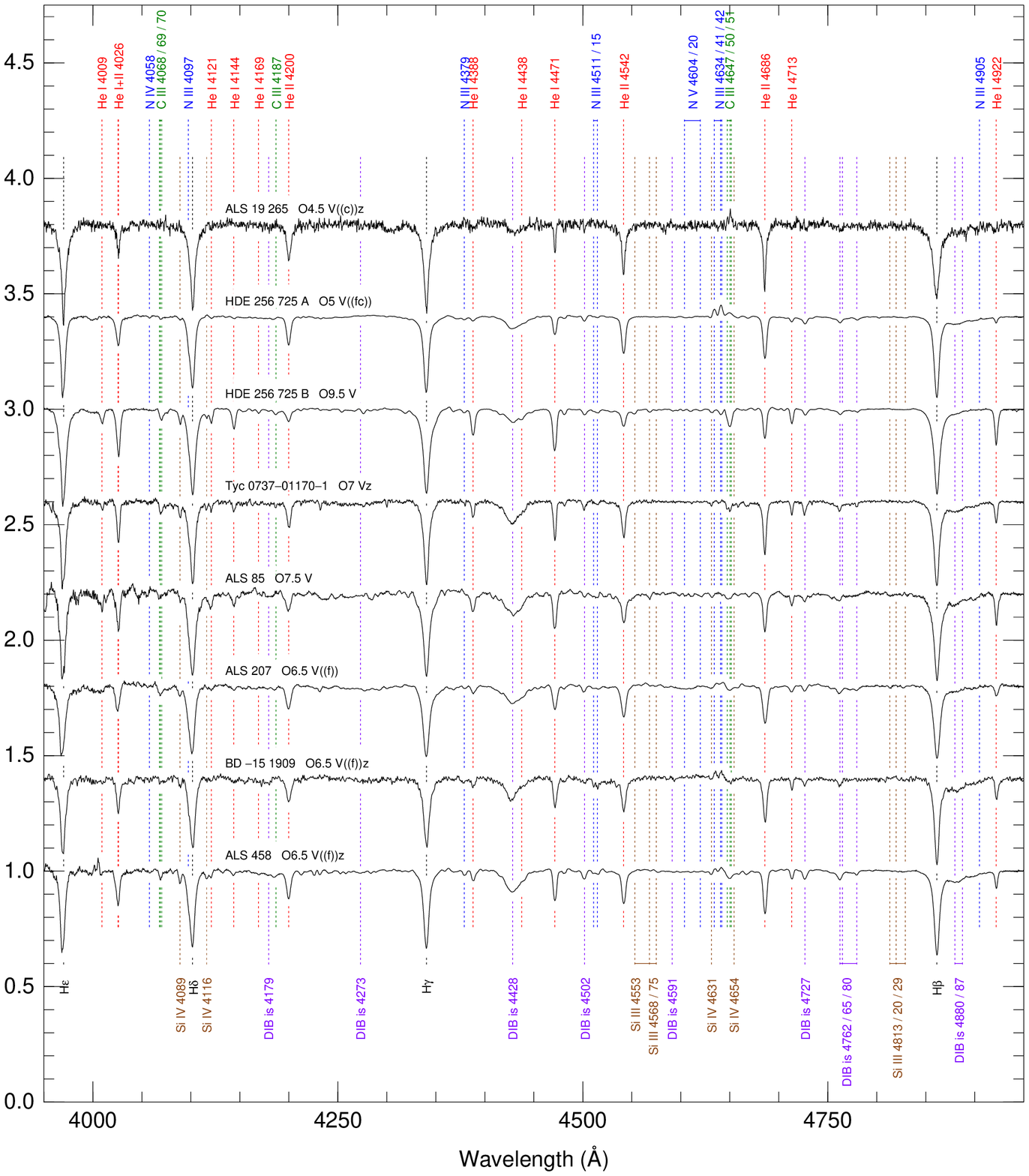}}
\caption{(continued).}
\end{figure*}	

\addtocounter{figure}{-1}

\begin{figure*}
\centerline{\includegraphics*[width=\linewidth]{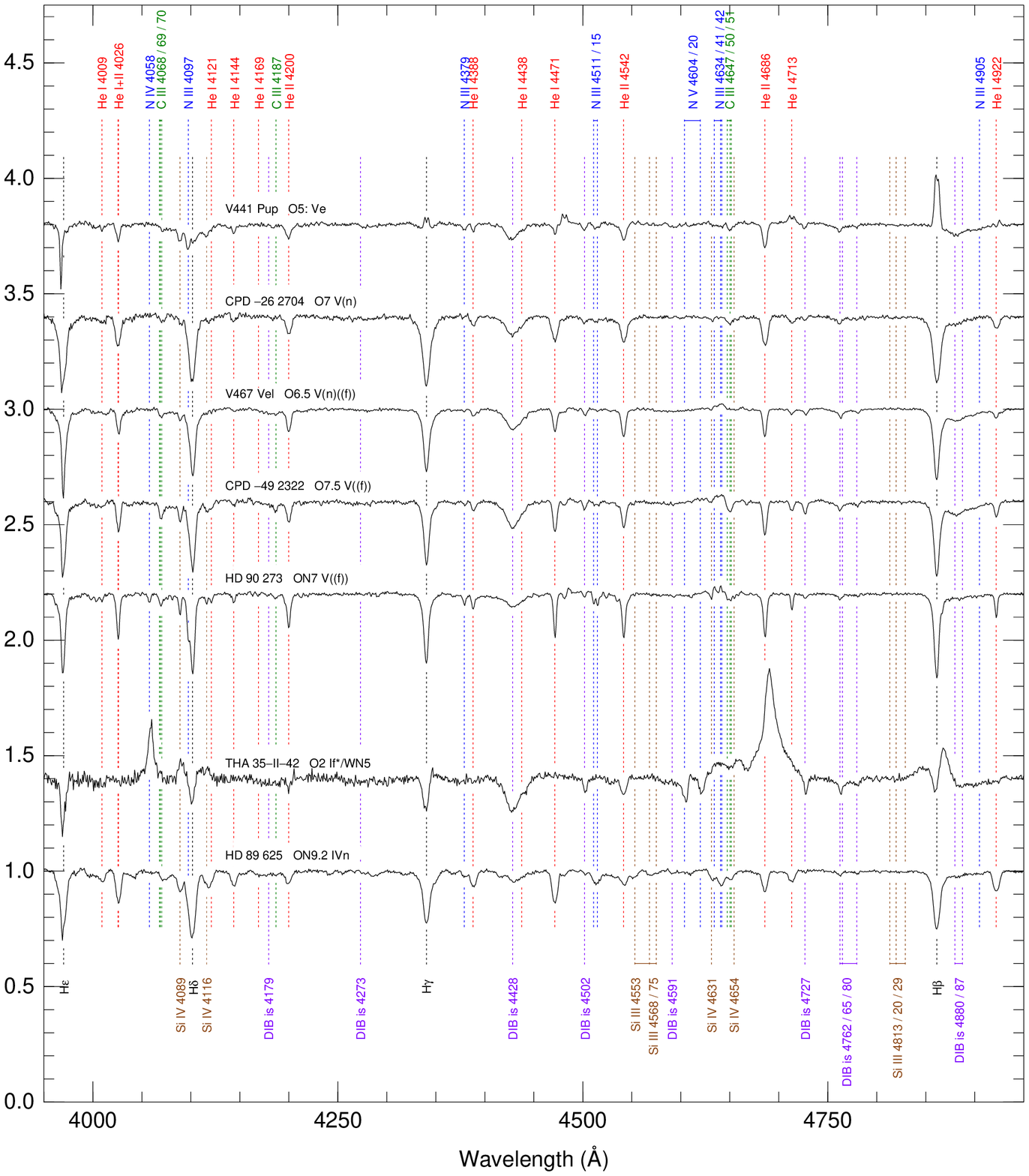}}
\caption{(continued).}
\end{figure*}	

\addtocounter{figure}{-1}

\begin{figure*}
\centerline{\includegraphics*[width=\linewidth]{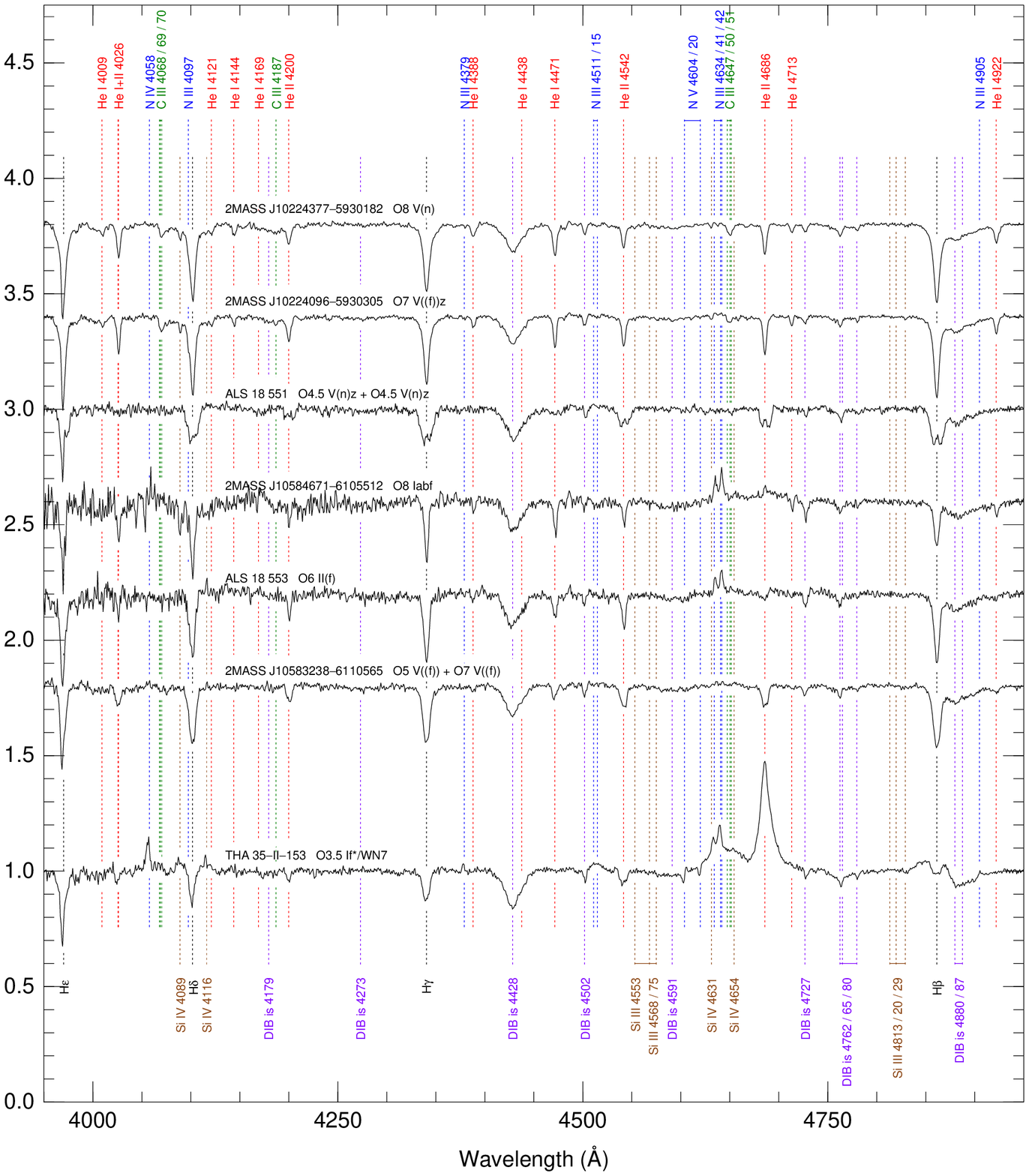}}
\caption{(continued).}
\end{figure*}	

\addtocounter{figure}{-1}

\begin{figure*}
\centerline{\includegraphics*[width=\linewidth]{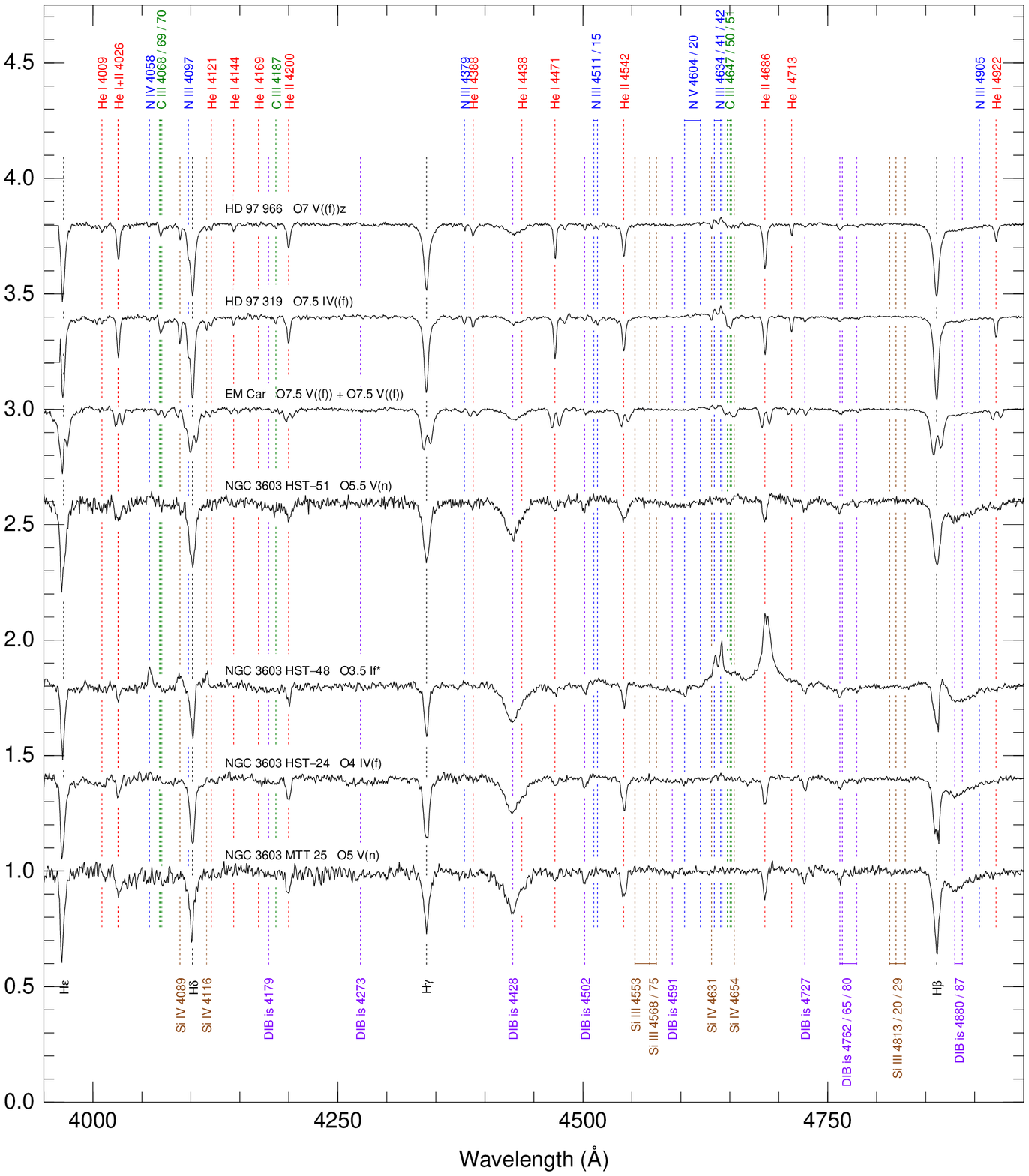}}
\caption{(continued).}
\end{figure*}	

\addtocounter{figure}{-1}

\begin{figure*}
\centerline{\includegraphics*[width=\linewidth]{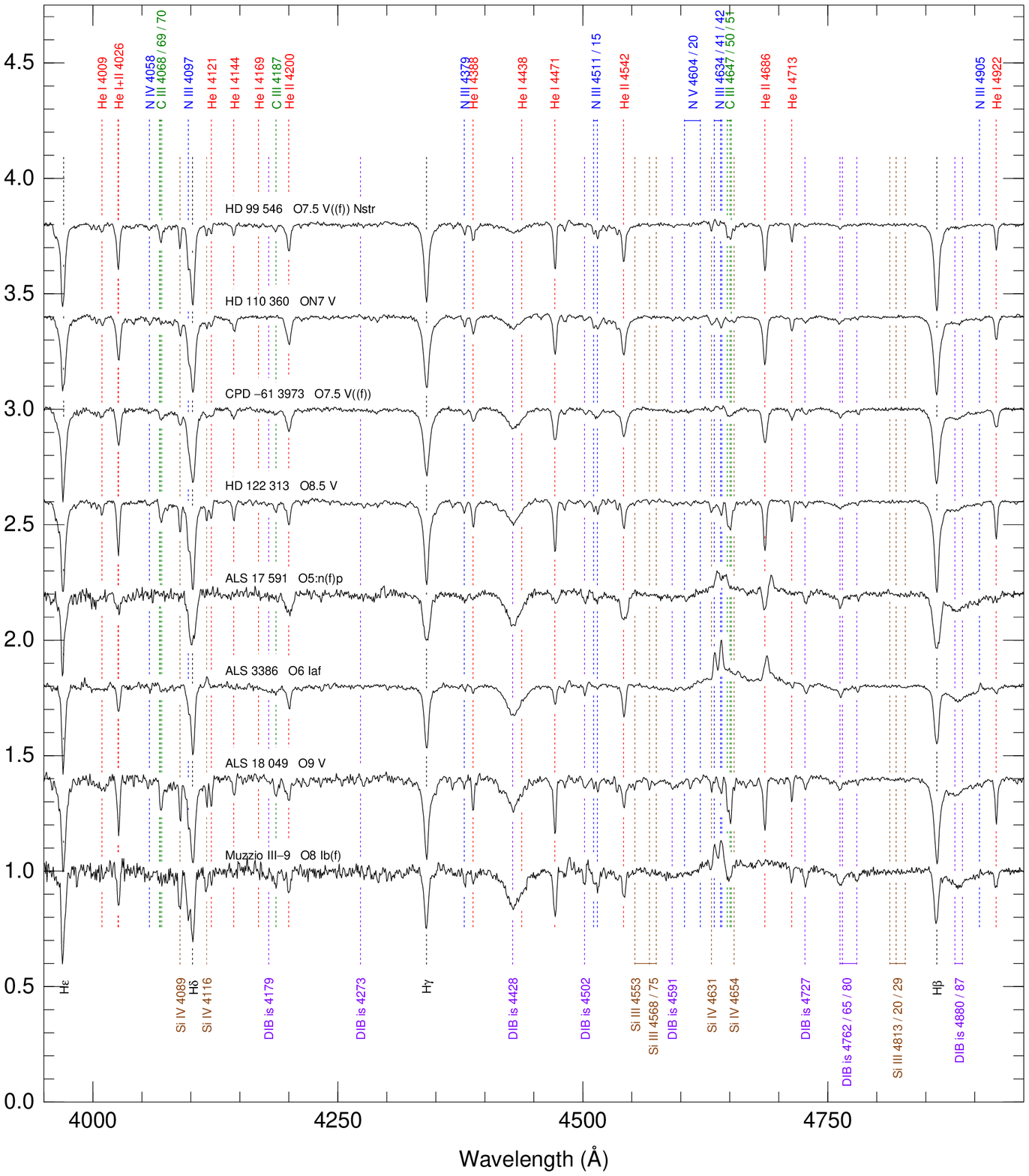}}
\caption{(continued).}
\end{figure*}	

\addtocounter{figure}{-1}

\begin{figure*}
\centerline{\includegraphics*[width=\linewidth]{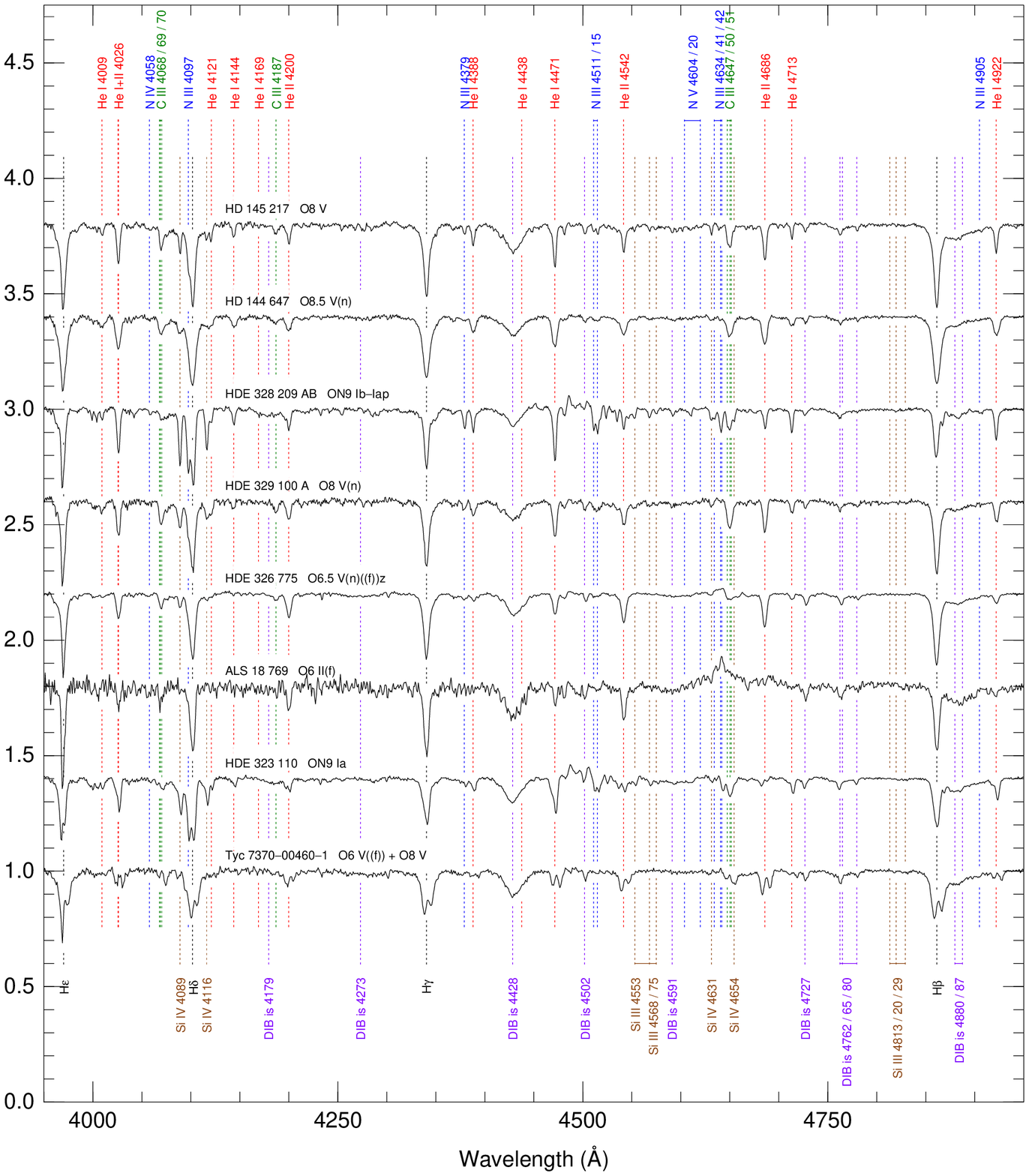}}
\caption{(continued).}
\end{figure*}	

\addtocounter{figure}{-1}

\begin{figure*}
\centerline{\includegraphics*[width=\linewidth]{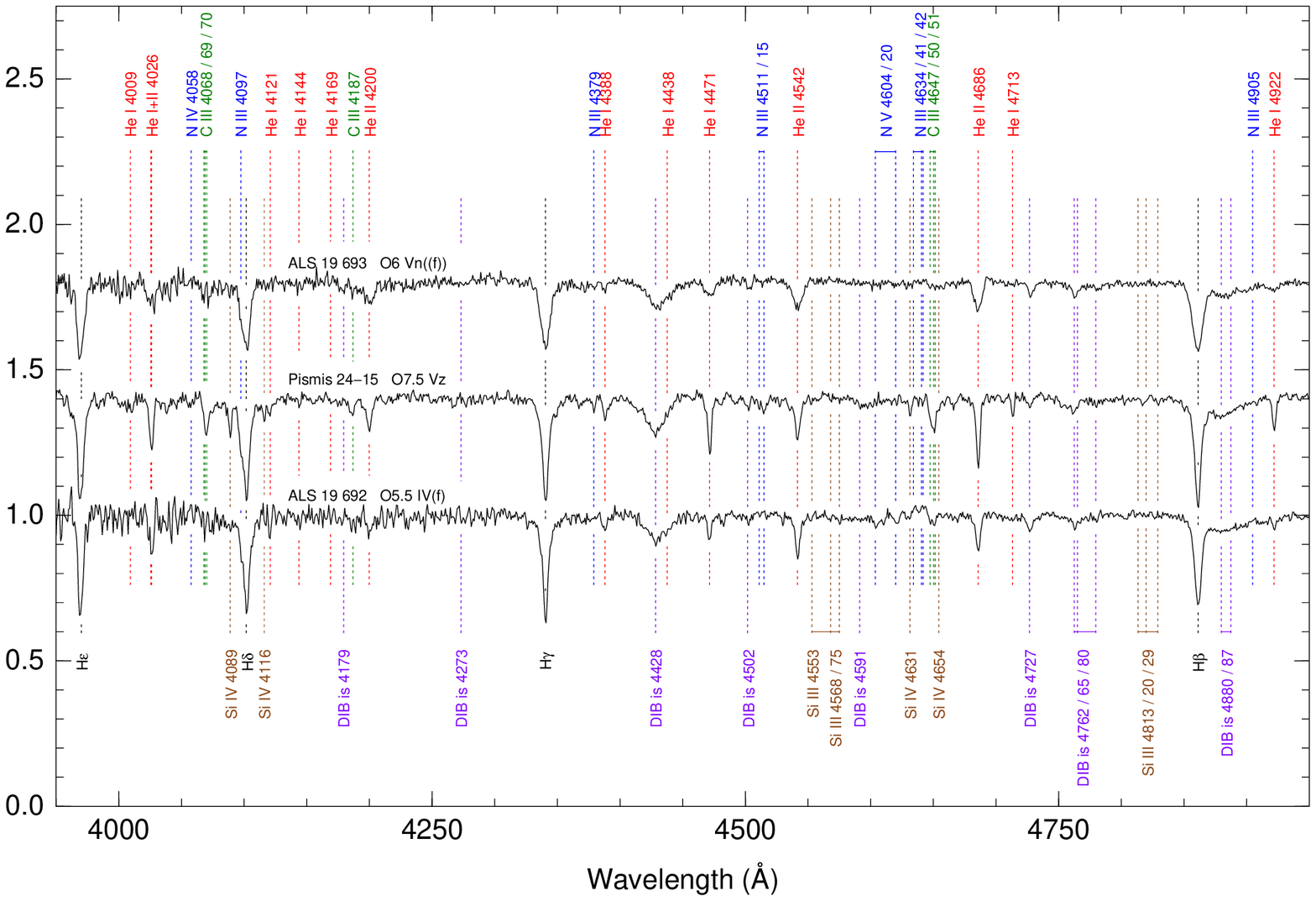}}
\caption{(continued).}
\end{figure*}	

\subsection{Spectral classification errors}
\label{sec:bad}

\begin{figure*}
\centerline{\includegraphics*[width=0.7\linewidth]{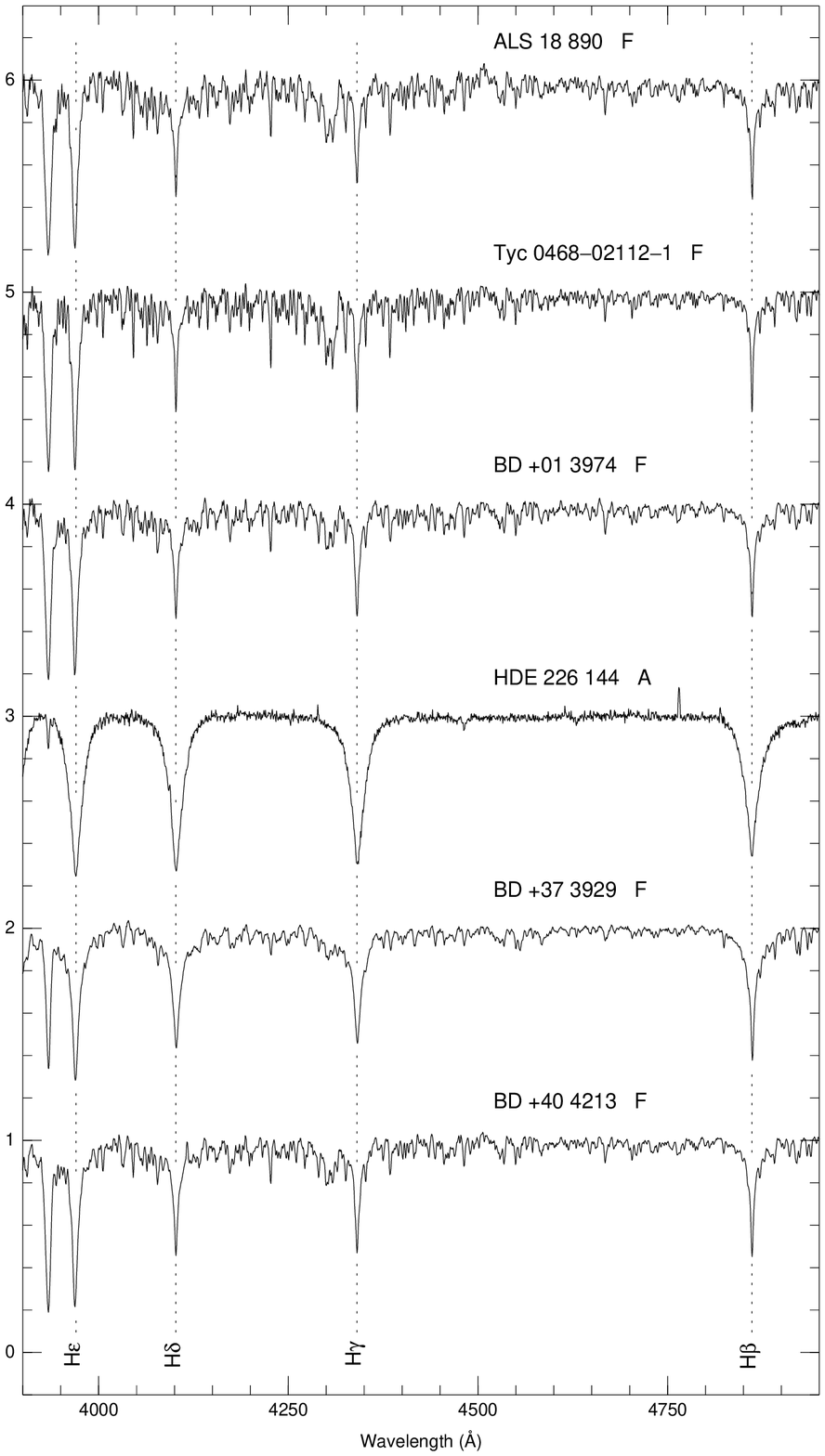}}
\caption{Spectrograms for late-type stars erroneously classifed as O stars. The targets are sorted by GOS ID. The spectral type is given after the name.
[See the electronic version of the journal for a color version of this figure.]}
\label{fig:bad}
\end{figure*}	

\addtocounter{figure}{-1}

\begin{figure*}
\centerline{\includegraphics*[width=0.7\linewidth]{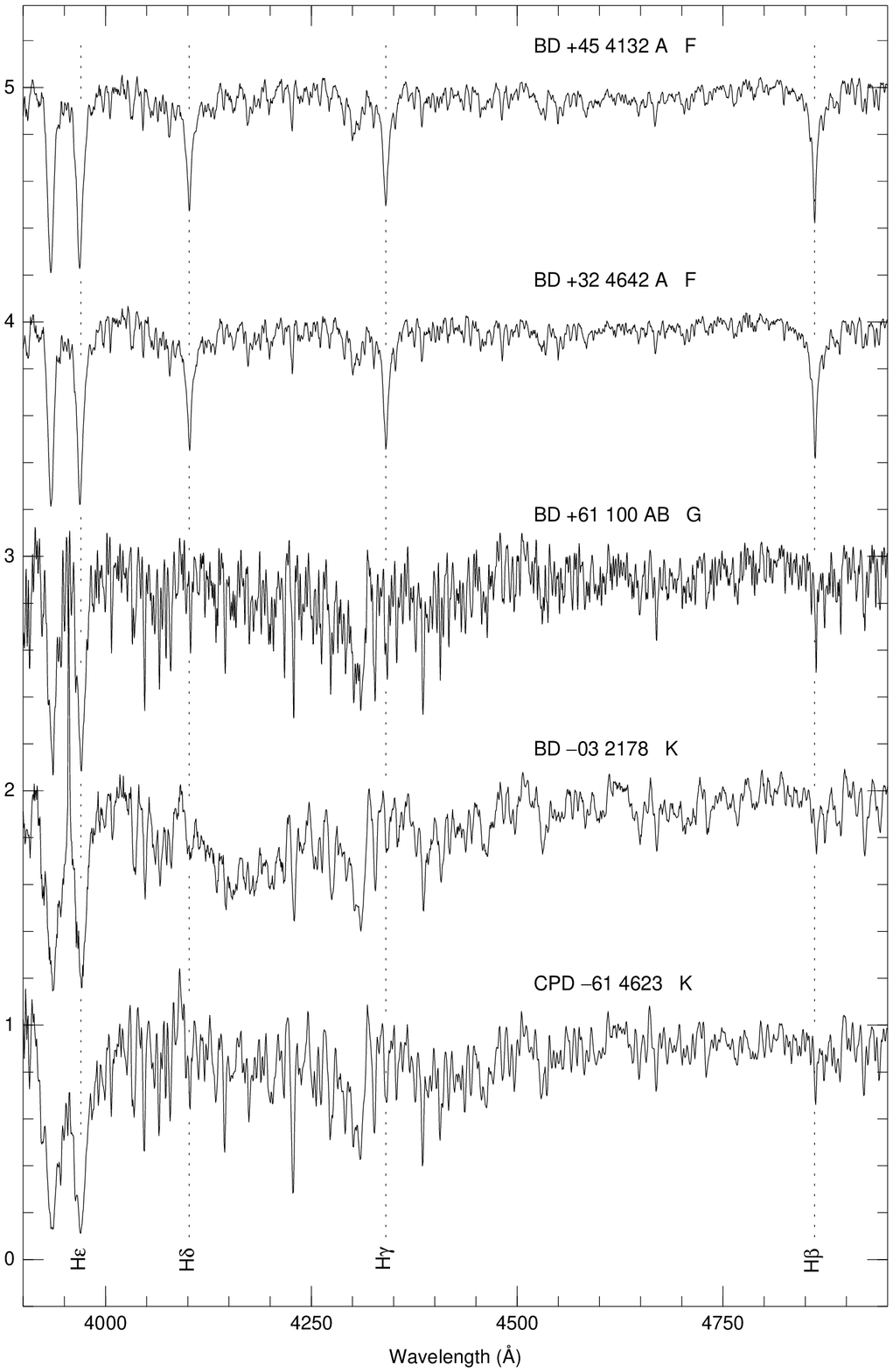}}
\caption{(continued).}
\end{figure*}	

One of the findings of GOSSS has been that the number of classification errors related to O stars in the literature is quite large. 
In \citet{Maizetal13}, we showed that the percentage of false positives (stellar systems erroneously classified as O type) in the GOSSS sample 
observed so far was 24.9\%. Since then, the number has risen above 30\% because, in general, dim stars have more uncertain classifications than bright 
ones and as GOSSS progresses we are moving (on average) into higher magnitude values (see below). 

The majority of the false positives turn out to be B stars, which is expected given the spectroscopic and photometric similitudes between O and B stars.
However, some of the false positives are egregious mistakes, since they turn out to be of A-K type. Those cases are sometimes caused by identification 
errors (the spectral type corresponds to a different star) or by photometric classifications reported as spectral types. 

We want to especially warn the reader about some of the results given by SIMBAD. The spectral classifications there are of highly variable quality.
Furthermore, some are ``legacy classifications'', which have been in SIMBAD before a reference was required and, as a result,
cannot be traced back to a source. SIMBAD classifications are corrected from time to time but some of the egregious false positives mentioned here are
still there at the time of this writing. On the other hand, we have seen some of them fixed in the last two years. 

We show in Fig.~\ref{fig:bad} the spectrograms of the eleven egregious false positives reported here. The corresponding spectral types and notes are listed in 
Table~\ref{spectralclasbad}. A preliminary version of this information was given by \citet{Maizetal15b}. In future papers we will present the spectrograms of 
additional false positives.

\object[ALS 18\,890]{}
\object[Tyc 0468-02112-1]{}
\object[BD +01 3974]{}
\object[HDE 226144]{}
\object[BD +37 3929]{}
\object[BD +40 4213]{}
\object[BD +45 4132 A]{}
\object[BD +32 4642 A]{}
\object[BD +61 100]{}
\object[BD -03 2178]{}
\object[CPD -61 4623]{}

\section{Analysis}   
\label{sec:Anal}

\begin{figure*}
\centerline{\includegraphics*[width=0.48\linewidth]{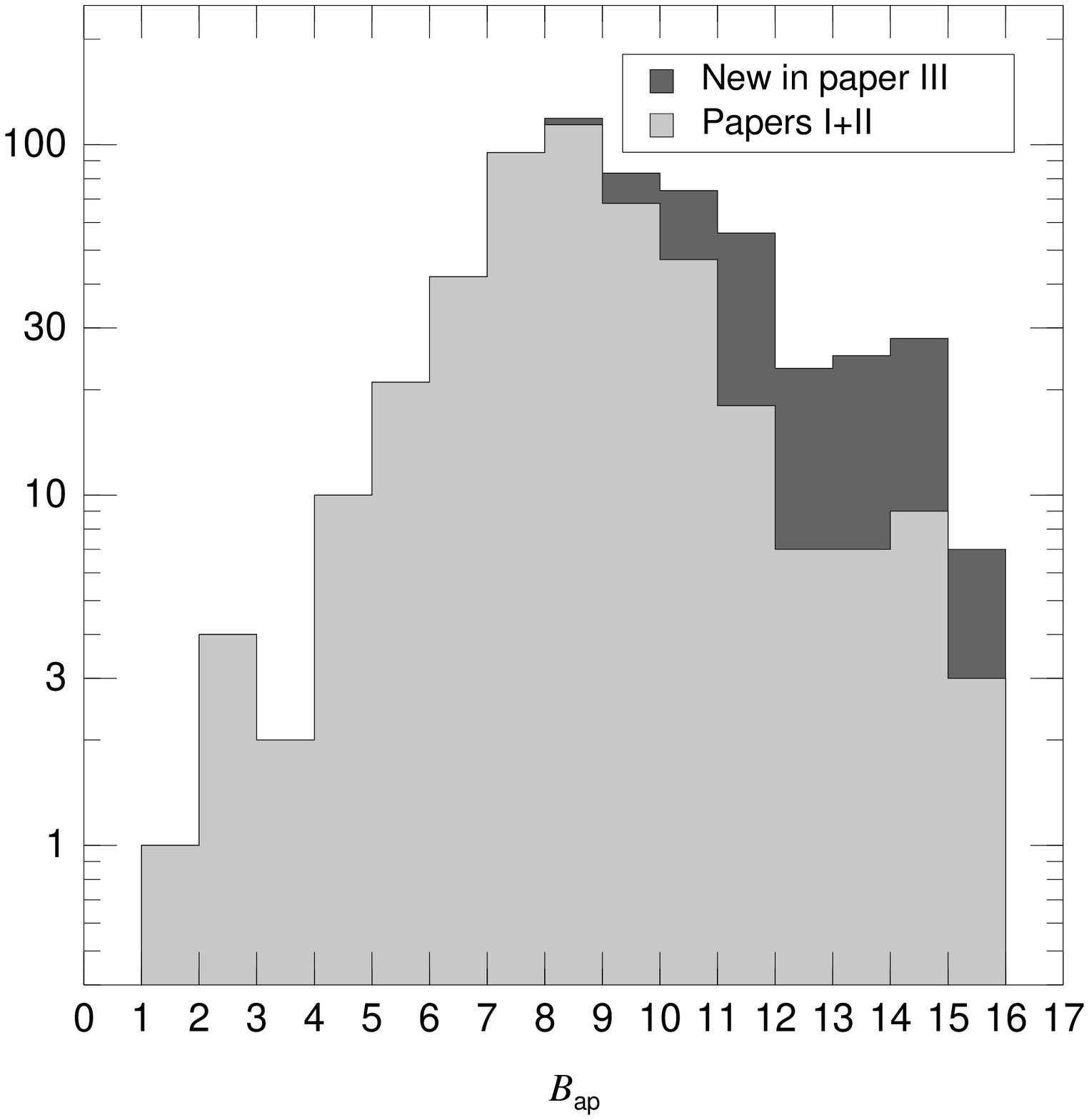} \
            \includegraphics*[width=0.48\linewidth]{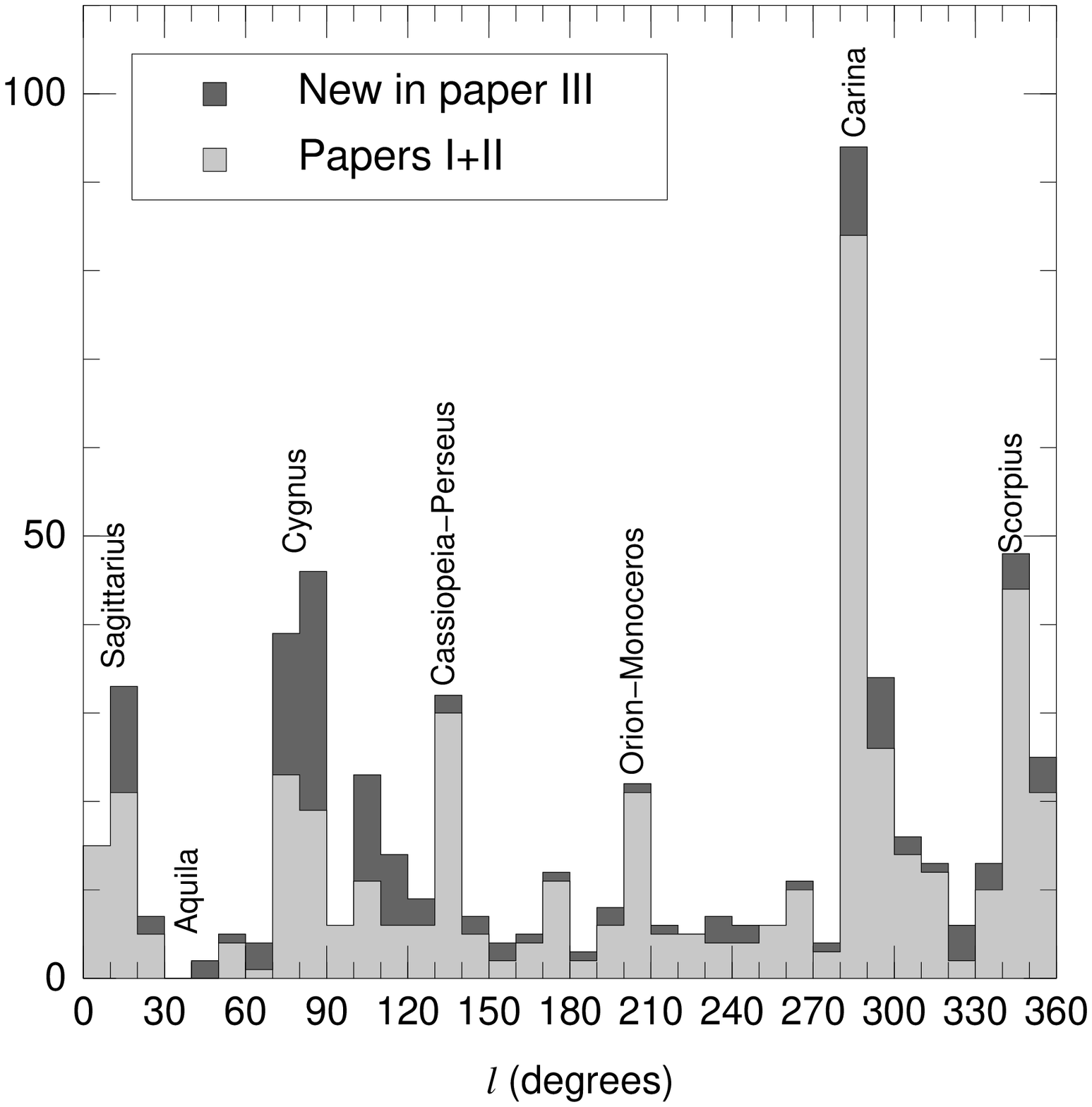}}
\caption{[left] \Bap\ and [right] $l$ histograms for the GOSSS new sample in this paper and in papers I and II.}
\label{fig:stats}
\end{figure*}	

In this paper we have added 142 O-type systems to the GOSSS sample, raising the total number to 590. Here we analyze some statistics and discuss the status and future of the project.

\begin{itemize}
 \item Twenty out of the 142 O-type systems (14\%) are first-time spectral classifications as O. Some of those had prior low-quality photometric classifications. As a comparison, a smaller number (18)
       of new O-type systems were found in papers I+II but the significantly larger sample there (448) yields a lower 4\%. The percentage difference between papers is expected, as papers I+II
       were dominated by the brightest and better studied O stars, leaving less room for new discoveries. Indeed, we expect that future GOSSS papers will include an even larger fraction of new O stars. 
 \item We have also discovered eleven new O-type SB2 systems\footnote{Twelve if you count the recently published system LS~III~+461~11, whose SB2 nature was also revealed by GOSSS \citep{Maizetal15a}.}, of 
       which six have O companions (O+O) and five have B companions (O+B), and a new SB3 system (O+O+B). We should point out that for most of the stars in this paper we have only obtained one epoch and that the GOSSS spectral 
       resolution only allows for the detection of SB2 systems with large velocity differences. Therefore, we strongly suspect that there are still many hidden spectroscopic binaries in this paper's 
       sample. The new SB2 and SB3 systems are being followed up (with GOSSS and in some cases with our high-spectral-resolution projects OWN, CAF\'E-BEANS, IACOB, and NoMaDS) to obtain their orbits.
 \item Figure~\ref{fig:stats} [left] shows the \Bap\ distribution of the previous and new samples. As expected, all new stars have $\Bap > 8$, so our previous claim that the 
       papers I+II sample was complete to \Bap = 8 is still valid. Our initial goal for this paper was to concentrate on the \Bap~=~8-11 magnitude range but we could not follow that path due to the
       technical problems of the Albireo spectrograph at the 1.5~m OSN telescope and the idiosyncrasies of some time allocation committees. Strangely enough, in the last years we have found easier to get 
       observing time for dim stars than for bright ones (even though we have tried both). As a result, we still have many stars to observe in the \Bap~=~8-11 magnitude range, as it is readily 
       apparent when comparing Fig.~\ref{fig:stats} [left] with Fig.~7 in \citet{Maizetal13}. We expect to reverse the situation in the next years, as one of the ultimte goals of GOSSS is to obtain
       the massive-star IMF in the solar neighborhood. 
 \item Figure~\ref{fig:stats} [right] shows the $l$ distribution of the previous and new samples. Some of the features of that plot are real, such as the well known prevalence of some 
       regions of the Galactic plane (Carina and Cygnus are the two areas that dominate the O-star population in the solar neighborhood) and the strong effect of foreground extinction in others (the Aquila rift being the most 
       clear one). Others are partially created by the different amounts of time we have had allocated in the two hemispheres (so far we have had more time in the north than in the south). It is interesting that
       the dichotomy between the first and fourth quadrants on one hand (toward the Galactic center) and the second and third on the other (away from it) is already visible, with more stars in the
       first case due to the larger stellar densities and disk depths. That is possible despite our easier telescope access towards the first three quadrants (those completely or partially visible 
       from the northern hemisphere), indicating the strength of the effect. 
 \item We have several additional hundreds of non-published stars with preliminary O-type classifications already observed with GOSSS and we are currently observing at a rate of $\sim$200 new ones per year. Our plan is to publish
       new blocks of O stars similar in quantity to the one in this paper every approximately 2 years. At this rate, we expect to run out of O stars with $\Bap < 14$ to observe in the second and third
       quadrants in a time scale of 2-3 years. Doing so in the first and fourth quadrants should take significantly longer. 
 \item Besides the developments derived from GOSSS associated with the stellar properties (e.g. \citealt{Ariaetal16}), we would like to point out that GOSSS is turning out to be a treasure trove to study the
       ISM \citep{Penaetal11,Penaetal13,Maiz13b,Maiz15,Maizetal14b,Maizetal15b}. We are currently working on an analysis of the extinction type and distribution based on the sample presented in the three GOSSS papers.
 \item GOSSS (including some of the new additions here) is being used as the basis for the sample selection for the OWN \citep{Barbetal10} and IACOB \citep{SimDetal11c} projects. The high-resolution spectra obtained
       there will be used to derive the properties of the likely single O stars through their quantitative analysis. 
\end{itemize}

\begin{acknowledgements}

We thank Brian Skiff for useful comments on a previous version of this manuscript.
J.M.A., A.S., and E.J.A. acknowledge support from [a] the Spanish Government Ministerio de Econom{\'\i}a y Competitividad (MINECO) through grants 
AYA2010-15\,081, AYA2010-17\,631, and AYA2013-40\,611-P and [b] the Consejer{\'\i}a de Educaci{\'o}n of the Junta de Andaluc{\'\i}a through 
grant P08-TIC-4075. 
J.I.A. and R.H.B. acknowledge support from FONDECYT Projects 1\,140\,076 and 11\,121\,550.
I.N. and A.M. acknowledge support from [a] the Spanish Government Ministerio de Econom\'{\i}a y Competitividad (MINECO) 
through grant AYA2012-39\,364-C02-01/02, [b] the European Union, and [c] the Generalitat Valenciana through grants ACOMP/2014/129 and BEST/2015/242.
A.H. and S.S.-D. acknowledge funding by [a] the Spanish Government Ministerio de Econom{\'\i}a y Competitividad (MINECO) through grants 
AYA2010-21\,697-C05-04, AYA2012-39\,364-C02-01, and Severo Ochoa SEV-2011-0187 and [b] the Canary Islands Government under grant PID2\,010\,119.
We would like to thank Nidia I. Morrell and Miguel Penad\'es Ordaz for their help in gathering the data for this paper.

\end{acknowledgements}

\bibliographystyle{apj}
\bibliography{general}

\eject
%
%
%
%
\addtolength{\textheight}{20mm}
\addtolength{\topmargin}{-5mm}
\addtolength{\textwidth}{10mm}
\addtolength{\oddsidemargin}{-5mm}
\addtolength{\evensidemargin}{-5mm}

\begin{deluxetable}{lllcl}
\tablewidth{0pt}
\tabletypesize{\scriptsize}
\tablecaption{Stars in papers I+II with recently discovered bright companions unresolved in the GOSSS spectra.}
\tablehead{\colhead{Name} & \colhead{GOSSS ID} & Ref. & \colhead{New} & Comments}
\startdata    
9~Sgr~AB             & 006.01$-$01.20\_01 & S14       & B  & \\
HD~168\,112~AB       & 018.44$+$01.62\_01 & S14       & B  & \\
HD~14\,633~AaAb      & 140.78$-$18.20\_01 & A15       & Ab & \\
HD~46\,966~AaAb      & 205.81$-$00.55\_01 & S14,A15   & Ab & \\
$\sigma$~Ori~AaAbB   & 206.82$-$17.34\_01 & H13,S13c  & Ab & B was already in paper I, S15 obtain a more accurate spectroscopic orbit. \\
HD~54\,662~AB        & 224.17$-$00.78\_01 & S14       & B  & \\
CPD~$-$35~2105~AaAbB & 253.64$-$00.45\_01 & A15       & Ab & B was already considered in paper II. \\
HD~75\,759~AB        & 262.80$+$01.25\_01 & S14       & B  & \\
CPD~$-$47~2963~AB    & 267.98$-$01.36\_01 & S14       & B  & \\
HD~93\,160~AB        & 287.44$-$00.59\_01 & S14       & B  & WDS 10441$-$5935 CaCb, HD~93\,161 AB is WDS 10441$-$5935 AB. \\
QZ~Car~AaAc          & 287.67$-$00.94\_01 & S14       & Ac & We use the WDS component nomenclature instead of that of the paper. \\
HD~93\,222~AB        & 287.74$-$01.02\_01 & S14       & B  & \\
HD~96\,670~AB        & 290.20$+$00.40\_01 & S14       & B  & \\
HD~97\,253~AB        & 290.79$+$00.09\_01 & S14       & B  & \\
HD~101\,190~AaAb     & 294.78$-$01.49\_01 & S14       & Ab & \\
HD~101\,131~AB       & 294.78$-$01.62\_01 & S14       & B  & \\
HD~101\,545~AaAb     & 294.88$-$00.81\_01 & S14       & Ab & A separate spectrum for the B component was obtained in paper II. \\
HD~101\,413~AB       & 295.03$-$01.71\_01 & S14       & B  & \\
HD~123\,590~AB       & 311.95$-$01.00\_01 & S14       & B  & \\
HD~124\,314~AaAb     & 312.67$-$00.42\_01 & S14       & Ab & A separate spectrum for the BaBb component was obtained in paper II. \\
HD~125\,206~AB       & 313.45$-$00.03\_01 & S14       & B  & \\
$\delta$~Cir~AaAbAc  & 319.69$-$02.91\_01 & S14       & Ac & AaAb is spatially unresolved but is catalogued that way in the WDS. \\
HD 148\,937~AaAb     & 336.37$-$00.22\_01 & S14       & Ab & \\
HD~150\,135~AaAb     & 336.71$-$01.57\_01 & S14       & Ab & WDS 6413$-$4846 CaCb, HD~150\,136 AaAb is WDS 6413$-$4846 AaAb. \\
HD~150\,136~AaAb     & 336.71$-$01.57\_02 & S13a,S13b & Ab & The inner pair is still unresolved. \\
HD~151\,003~AB       & 342.72$+$02.41\_01 & S14       & B  & \\
HD~152\,233~AaAb     & 343.48$+$01.22\_01 & S14       & Ab & WDS 16540$-$4148 FaFb, HD~152\,234 A is WDS 16540$-$4148 A. \\
HD~152\,314~AaAb     & 343.52$+$01.14\_01 & S14       & Ab & \\
HD~152\,247~AaAb     & 343.61$+$01.30\_01 & S14       & Ab & \\
HD~152\,246~AaAb     & 344.03$+$01.67\_01 & S14       & Ab & \\
HD~152\,623~AaAbB    & 344.62$+$01.61\_01 & S14       & Ab & B was already considered in paper II. \\
HD~156\,738~AB       & 351.18$+$00.48\_01 & S14       & B  & \\
HD~155\,806~AB       & 352.59$+$02.87\_01 & S14       & B  & \\
\enddata
\tablerefs{A15: \citet{Aldoetal15}, H13: \citet{Hummetal13b}, S13a: \citet{Sanaetal13a}, S13b: \citet{SanBetal13}, S13c: \citet{Scha13}, S14: \citet{Sanaetal14}, S15: \citet{SimDetal15a}}
\label{literature}
\end{deluxetable}

\begin{deluxetable}{llllllll}
\tablecaption{Spectral reclassifications for stars already present in papers I and II.}
\tablewidth{0pt}
\tabletypesize{\small}
\tablehead{\colhead{Name} & \colhead{GOSSS ID} & \colhead{RA (J2000)} & \colhead{dec (J2000)} & \colhead{ST} & \colhead{LC} & \colhead{Qual.} & \colhead{Second.}}
\startdata
HD 164\,536                 & GOS 005.96$-$00.91\_01 & 18:02:38.619 & $-$24:15:19.39 & O7.5    & V       & (n)         & \nodata        \\
9 Sgr AB                    & GOS 006.01$-$01.20\_01 & 18:03:52.446 & $-$24:21:38.64 & O4      & V       & ((f))       & \nodata        \\
HD 165\,052                 & GOS 006.12$-$01.48\_01 & 18:05:10.551 & $-$24:23:54.85 & O6      & V       & z           & O8 Vz          \\
HD 168\,075                 & GOS 016.94$+$00.84\_01 & 18:18:36.043 & $-$13:47:36.46 & O6.5    & V       & ((f))       & \nodata        \\
HD 168\,076 AB              & GOS 016.94$+$00.84\_02 & 18:18:36.421 & $-$13:48:02.38 & O4      & IV      & (f)         & \nodata        \\
HD 168\,112 AB              & GOS 018.44$+$01.62\_01 & 18:18:40.868 & $-$12:06:23.39 & O5      & IV      & (f)         & \nodata        \\
HD 173\,010                 & GOS 023.73$-$02.49\_01 & 18:43:29.710 & $-$09:19:12.61 & O9.7    & Ia+     & var         & \nodata        \\
HDE 344\,784 A              & GOS 059.40$-$00.15\_01 & 19:43:10.970 & $+$23:17:45.38 & O6.5    & V       & ((f))z      & \nodata        \\
HD 192\,281                 & GOS 077.12$+$03.40\_01 & 20:12:33.121 & $+$40:16:05.45 & O4.5    & IV      & (n)(f)      & \nodata        \\
HDE 229\,232 AB             & GOS 077.40$+$00.93\_01 & 20:23:59.183 & $+$39:06:15.27 & O4      & V:      & n((f))      & \nodata        \\
HD 191\,978                 & GOS 077.87$+$04.25\_01 & 20:10:58.281 & $+$41:21:09.91 & O8      & V       & \nodata     & \nodata        \\
HD 207\,198                 & GOS 103.14$+$06.99\_01 & 21:44:53.278 & $+$62:27:38.04 & O8.5    & II      & ((f))       & \nodata        \\
BD +60 513                  & GOS 134.90$+$00.92\_01 & 02:34:02.530 & $+$61:23:10.87 & O7      & V       & n           & \nodata        \\
HD 14\,434                  & GOS 135.08$-$03.82\_01 & 02:21:52.413 & $+$56:54:18.03 & O5.5    & IV      & nn(f)p      & \nodata        \\
HD 17\,520 A                & GOS 137.22$+$00.88\_01 & 02:51:14.434 & $+$60:23:09.97 & O8      & V       & \nodata     & \nodata        \\
HD 35\,619                  & GOS 173.04$-$00.09\_01 & 05:27:36.147 & $+$34:45:18.96 & O7.5    & V       & ((f))       & \nodata        \\
HD 36\,879                  & GOS 185.22$-$05.89\_01 & 05:35:40.527 & $+$21:24:11.72 & O7      & V       & (n)((f))    & \nodata        \\
HD 48\,099                  & GOS 206.21$+$00.80\_01 & 06:41:59.231 & $+$06:20:43.54 & O5.5    & V       & ((f))z      & O9 V           \\
HD 46\,485                  & GOS 206.90$-$01.84\_01 & 06:33:50.957 & $+$04:31:31.62 & O7      & V       & ((f))n var? & \nodata        \\
HD 48\,279 A                & GOS 210.41$-$01.17\_01 & 06:42:40.548 & $+$01:42:58.23 & O8.5    & V       & Nstr var?   & \nodata        \\
HD 64\,315 AB               & GOS 243.16$+$00.36\_01 & 07:52:20.284 & $-$26:25:46.69 & O5.5    & V       & \nodata     & O7 V           \\
CPD -47 2962                & GOS 268.00$-$01.38\_01 & 08:57:51.661 & $-$47:45:43.94 & O7      & V       & ((f))       & \nodata        \\
HD 92\,206 B                & GOS 286.22$-$00.17\_02 & 10:37:22.957 & $-$58:37:23.04 & O6      & V       & ((f))       & \nodata        \\
HD 92\,206 C                & GOS 286.22$-$00.18\_01 & 10:37:18.627 & $-$58:37:41.73 & O8      & V       & (n)z        & B0: V          \\
ALS 15\,204                 & GOS 287.40$-$00.63\_02 & 10:43:41.237 & $-$59:35:48.18 & O7.5    & V       & z           & O9: V          \\
HD 93\,129 B                & GOS 287.41$-$00.57\_02 & 10:43:57.638 & $-$59:32:53.50 & O3.5    & V       & ((f))z      & \nodata        \\
HD 93\,250 AB               & GOS 287.51$-$00.54\_01 & 10:44:45.027 & $-$59:33:54.67 & O4      & IV      & (fc)        & \nodata        \\
V572 Car                    & GOS 287.59$-$00.69\_01 & 10:44:47.307 & $-$59:43:53.23 & O7.5    & V       & (n)         & B0 V(n)        \\
CPD -59 2591                & GOS 287.60$-$00.75\_01 & 10:44:36.688 & $-$59:47:29.63 & O8      & V       & z           & B0.5: V:       \\
CPD -59 2626 AB             & GOS 287.63$-$00.69\_01 & 10:45:05.794 & $-$59:45:19.60 & O7.5    & V       & (n)         & \nodata        \\
HD 93\,343                  & GOS 287.64$-$00.68\_01 & 10:45:12.217 & $-$59:45:00.42 & O8      & V       & \nodata     & \nodata        \\
HD 93\,146 A                & GOS 287.67$-$01.05\_01 & 10:44:00.158 & $-$60:05:09.86 & O7      & V       & ((f))       & \nodata        \\
V662 Car                    & GOS 287.71$-$00.71\_01 & 10:45:36.318 & $-$59:48:23.37 & O5      & V       & (n)z        & B0: V          \\
HD 93\,222 AB               & GOS 287.74$-$01.02\_01 & 10:44:36.250 & $-$60:05:28.88 & O7      & V       & ((f))       & \nodata        \\
HDE 305\,525                & GOS 287.79$-$00.71\_01 & 10:46:05.704 & $-$59:50:49.45 & O5.5    & V       & ((f))z      & O7.5 V + B     \\
TU Mus                      & GOS 294.81$-$04.14\_01 & 11:31:10.927 & $-$65:44:32.10 & O8      & V       & (n)         & B0 V(n)        \\
$\delta$ Cir AaAbAc         & GOS 319.69$-$02.91\_01 & 15:16:56.894 & $-$60:57:26.12 & O7      & IV      & ((f))       & B              \\
CPD -41 7721 A              & GOS 343.44$+$01.17\_01 & 16:54:06.709 & $-$41:51:07.21 & O9.7    & V:      & (n)         & \nodata        \\
V1034 Sco                   & GOS 343.48$+$01.15\_01 & 16:54:19.845 & $-$41:50:09.36 & O9.2    & IV      & \nodata     & B1: V          \\
HD 152\,623 AaAbB           & GOS 344.62$+$01.61\_01 & 16:56:15.026 & $-$40:39:35.76 & O7      & V       & (n)((f))    & \nodata        \\
HD 156\,292                 & GOS 345.35$-$03.08\_01 & 17:18:45.814 & $-$42:53:29.92 & O9.7    & III     & \nodata     & B              \\
ALS 18\,770                 & GOS 348.71$-$00.79\_01 & 17:19:00.800 & $-$38:49:23.13 & O7      & V       & ((f))       & \nodata        \\
HDE 319\,703 A              & GOS 351.03$+$00.65\_01 & 17:19:46.156 & $-$36:05:52.34 & O7      & V       & ((f))       & O9.5 V         \\
HD 155\,806 AB              & GOS 352.59$+$02.87\_01 & 17:15:19.247 & $-$33:32:54.30 & O7.5    & V       & ((f))(e)    & \nodata        \\
HD 158\,186                 & GOS 355.91$+$01.60\_01 & 17:29:12.925 & $-$31:32:03.44 & O9.2    & V       & \nodata     & B1: V          \\
\enddata
\tablecomments{{\it GOSSS ID} is the identification for each star with ``GOS'' standing for ``Galactic O Star''.}
\label{spectralclasold}
\end{deluxetable}

\begin{deluxetable}{llllllll}
\tablecaption{Spectral classifications for new GOSSS stars.}
\tablewidth{0pt}
\tabletypesize{\small}
\tablehead{\colhead{Name} & \colhead{GOSSS ID} & \colhead{RA (J2000)} & \colhead{dec (J2000)} & \colhead{ST} & \colhead{LC} & \colhead{Qual.} & \colhead{Second.}}
\startdata
ALS 19\,618                 & GOS 015.07$-$00.70\_01 & 18:20:34.493 & $-$16:10:11.85 & O4      & V       & (n)((fc))   & \nodata        \\
BD -16 4826                 & GOS 015.26$-$00.73\_01 & 18:21:02.231 & $-$16:01:00.94 & O5.5    & V       & ((f))z      & \nodata        \\
ALS 4923                    & GOS 015.70$-$00.06\_01 & 18:19:28.435 & $-$15:18:46.27 & O8.5    & V       & \nodata     & O8.5 V         \\
ALS 4626                    & GOS 015.88$+$04.22\_01 & 18:04:17.885 & $-$13:06:13.76 & ON6     & V       & ((f))       & \nodata        \\
BD -14 5014                 & GOS 016.65$-$00.35\_01 & 18:22:22.310 & $-$14:37:08.46 & O7.5    & V       & (n)((f))    & \nodata        \\
V479 Sct                    & GOS 016.88$-$01.29\_01 & 18:26:15.045 & $-$14:50:54.33 & ON6     & V       & ((f))z      & \nodata        \\
BD -14 5040                 & GOS 016.90$-$01.12\_01 & 18:25:38.896 & $-$14:45:05.70 & O5.5    & V       & (n)((f))    & \nodata        \\
HD 168\,137 AaAb            & GOS 016.97$+$00.76\_01 & 18:18:56.189 & $-$13:48:31.08 & O8      & V       & z           & \nodata        \\
ALS 15\,360                 & GOS 017.00$+$00.87\_01 & 18:18:37.494 & $-$13:43:39.39 & O7      & V       & ((f))z      & \nodata        \\
HD 168\,504                 & GOS 017.03$+$00.35\_01 & 18:20:34.096 & $-$13:57:15.75 & O7.5    & V       & (n)z        & \nodata        \\
ALS 4880                    & GOS 018.32$+$01.87\_01 & 18:17:33.672 & $-$12:05:42.80 & O6      & V       & ((f))       & \nodata        \\
HD 168\,461                 & GOS 018.57$+$01.25\_01 & 18:20:17.179 & $-$12:10:19.19 & O7.5    & V       & ((f)) Nstr  & \nodata        \\
BD -10 4682                 & GOS 020.24$+$01.01\_01 & 18:24:20.651 & $-$10:48:34.29 & O7      & V       & n((f))      & \nodata        \\
BD -04 4503                 & GOS 026.85$+$01.34\_01 & 18:35:32.534 & $-$04:47:55.39 & O7      & V       & \nodata     & \nodata        \\
HD 175\,514                 & GOS 041.71$+$03.38\_01 & 18:55:23.124 & $+$09:20:48.07 & O7      & V       & (n)((f))z   & B              \\
ALS 18\,929                 & GOS 042.79$+$10.57\_01 & 18:31:01.379 & $+$13:30:12.85 & O9.7    & \nodata & \nodata     & \nodata        \\
HDE 344\,777                & GOS 059.41$+$00.11\_01 & 19:42:11.470 & $+$23:26:00.52 & O7.5    & V       & z           & \nodata        \\
HDE 344\,758                & GOS 060.17$+$00.63\_01 & 19:41:52.721 & $+$24:20:51.07 & O8.5    & V       & (n)((f))    & \nodata        \\
HDE 338\,931                & GOS 061.19$-$00.14\_01 & 19:47:02.739 & $+$24:50:55.57 & O6      & III     & (f)         & \nodata        \\
HDE 338\,916                & GOS 061.47$+$00.38\_01 & 19:45:42.114 & $+$25:21:16.45 & O7.5    & V       & z           & \nodata        \\
HDE 227\,465                & GOS 070.73$+$01.21\_01 & 20:04:27.225 & $+$33:42:18.40 & O7      & V       & ((f))       & \nodata        \\
HDE 227\,018                & GOS 071.58$+$02.87\_01 & 19:59:49.103 & $+$35:18:33.53 & O6.5    & V       & ((f))z      & \nodata        \\
HDE 227\,245                & GOS 072.17$+$02.62\_01 & 20:02:21.713 & $+$35:40:29.84 & O7      & V       & ((f))z      & \nodata        \\
HDE 228\,779                & GOS 073.18$-$00.51\_01 & 20:17:54.189 & $+$34:49:02.03 & O9      & Iab     & \nodata     & \nodata        \\
HDE 228\,854                & GOS 074.54$+$00.20\_01 & 20:18:47.219 & $+$36:20:26.08 & O6      & IV      & n var       & O5 Vn var      \\
ALS 18\,707                 & GOS 074.76$+$00.62\_01 & 20:17:41.846 & $+$36:45:26.42 & O6.5    & V       & ((f))z      & \nodata        \\
HD 193\,682                 & GOS 075.92$+$00.82\_01 & 20:20:08.937 & $+$37:49:51.30 & O4.5    & IV      & (f)         & \nodata        \\
HD 193\,595                 & GOS 076.86$+$01.62\_01 & 20:19:31.327 & $+$39:03:26.21 & O7      & V       & ((f))       & \nodata        \\
BD +36 4145                 & GOS 077.45$-$02.02\_01 & 20:36:18.208 & $+$37:25:02.79 & O8.5    & V       & (n)         & \nodata        \\
HDE 229\,202                & GOS 078.19$+$01.63\_01 & 20:23:22.842 & $+$40:09:22.52 & O7.5    & V       & (n)((f))    & \nodata        \\
ALS 11\,355                 & GOS 078.29$+$00.78\_01 & 20:27:17.572 & $+$39:44:32.55 & O8      & V       & (n)((f))    & \nodata        \\
HD 194\,649 AB              & GOS 078.46$+$01.35\_01 & 20:25:22.124 & $+$40:13:01.07 & O6.5    & V       & ((f))       & \nodata        \\
HDE 228\,759                & GOS 079.01$+$03.62\_01 & 20:17:07.539 & $+$41:57:26.51 & O6.5    & V       & (n)((f))z   & \nodata        \\
LS III\,+41 14              & GOS 079.01$+$03.63\_01 & 20:17:05.515 & $+$41:57:46.89 & O9.5    & V       & (n)         & \nodata        \\
BD +40 4179                 & GOS 079.03$+$01.21\_01 & 20:27:43.617 & $+$40:35:43.53 & O8      & V       & z           & \nodata        \\
Cyg OB2-B17                 & GOS 079.84$+$01.16\_01 & 20:30:27.302 & $+$41:13:25.31 & O6      & Ia      & f           & O9: Ia:        \\
Cyg OB2-A24                 & GOS 080.03$+$00.30\_01 & 20:34:44.106 & $+$40:51:58.50 & O6.5    & III     & (f)         & \nodata        \\
2MASS J20315961+4114504     & GOS 080.03$+$00.94\_01 & 20:31:59.613 & $+$41:14:50.50 & O7.5    & V       & z           & \nodata        \\
Cyg OB2-A11                 & GOS 080.08$+$00.85\_01 & 20:32:31.543 & $+$41:14:08.21 & O7      & Ib      & (f)         & \nodata        \\
ALS 15\,108 AB              & GOS 080.11$+$00.67\_01 & 20:33:23.460 & $+$41:09:13.02 & O6      & IV      & ((f))       & \nodata        \\
Cyg OB2-5 B                 & GOS 080.12$+$00.91\_02 & 20:32:22.489 & $+$41:18:19.45 & O6.5    & Iab     & fp          & \nodata        \\
ALS 15\,134                 & GOS 080.14$+$00.68\_01 & 20:33:26.749 & $+$41:10:59.51 & O8      & V       & z           & \nodata        \\
Cyg OB2-22 D                & GOS 080.14$+$00.74\_02 & 20:33:10.115 & $+$41:13:10.10 & O9.5    & V       & n           & \nodata        \\
ALS 15\,144                 & GOS 080.15$+$00.79\_01 & 20:32:59.643 & $+$41:15:14.67 & O9.7    & III     & (n)         & \nodata        \\
ALS 15\,119                 & GOS 080.23$+$00.71\_01 & 20:33:37.001 & $+$41:16:11.30 & O9.5    & IV      & (n)         & \nodata        \\
Cyg OB2-17                  & GOS 080.24$+$00.90\_01 & 20:32:50.011 & $+$41:23:44.71 & O8      & V       & \nodata     & \nodata        \\
Cyg OB2-16                  & GOS 080.24$+$00.94\_01 & 20:32:38.575 & $+$41:25:13.76 & O7.5    & IV      & (n)         & \nodata        \\
Cyg OB2-6                   & GOS 080.26$+$00.93\_01 & 20:32:45.447 & $+$41:25:37.50 & O8.5    & V       & (n)         & \nodata        \\
ALS 15\,115                 & GOS 080.27$+$00.81\_01 & 20:33:18.046 & $+$41:21:36.90 & O8      & V       & \nodata     & \nodata        \\
ALS 15\,111                 & GOS 080.27$+$00.88\_01 & 20:32:59.190 & $+$41:24:25.47 & O8      & V       & \nodata     & \nodata        \\
Cyg OB2-27 AB               & GOS 080.29$+$00.66\_01 & 20:33:59.528 & $+$41:17:35.48 & O9.7    & V       & (n)         & O9.7 V:(n)     \\
Cyg OB2-73                  & GOS 080.32$+$00.60\_01 & 20:34:21.930 & $+$41:17:01.60 & O8      & V       & z           & O8 Vz          \\
Cyg OB2-25 A                & GOS 080.44$+$00.91\_01 & 20:33:25.539 & $+$41:33:26.74 & O8      & V       & z           & \nodata        \\
Cyg OB2-10                  & GOS 080.47$+$00.85\_01 & 20:33:46.111 & $+$41:33:01.05 & O9.7    & Iab     & \nodata     & \nodata        \\
ALS 15\,125                 & GOS 080.53$+$00.80\_01 & 20:34:09.519 & $+$41:34:13.69 & O9.5    & IV:     & \nodata     & \nodata        \\
ALS 15\,114                 & GOS 080.54$+$00.73\_01 & 20:34:29.601 & $+$41:31:45.42 & O7.5    & V       & (n)((f))    & \nodata        \\
Cyg OB2-29                  & GOS 080.55$+$00.80\_01 & 20:34:13.505 & $+$41:35:03.01 & O7.5    & V       & (n)((f))z   & \nodata        \\
BD +43 3654                 & GOS 082.41$+$02.33\_01 & 20:33:36.080 & $+$43:59:07.41 & O4      & I       & f           & \nodata        \\
BD +45 3216 A               & GOS 083.78$+$03.29\_01 & 20:33:50.366 & $+$45:39:40.95 & O5      & V       & ((f))z      & \nodata        \\
Bajamar Star                & GOS 084.81$-$00.88\_01 & 20:55:51.255 & $+$43:52:24.67 & O3.5    & III     & (f*)        & O8:            \\
LS III\,+46 12              & GOS 084.88$+$03.78\_01 & 20:35:18.566 & $+$46:50:02.90 & O4.5    & IV      & (f)         & \nodata        \\
LS III\,+46 11              & GOS 084.88$+$03.81\_01 & 20:35:12.642 & $+$46:51:12.12 & O3.5    & I       & f*          & O3.5 If*       \\
ALS 11\,761                 & GOS 088.81$-$01.12\_01 & 21:12:00.455 & $+$46:41:51.31 & O9.2    & II      & \nodata     & \nodata        \\
ALS 12\,050                 & GOS 101.08$+$02.47\_01 & 21:55:15.291 & $+$57:39:45.66 & O5      & V       & ((f))       & \nodata        \\
BD +55 2722 A               & GOS 102.81$-$00.67\_01 & 22:18:58.629 & $+$56:07:23.47 & O8      & V       & z           & \nodata        \\
BD +55 2722 C               & GOS 102.81$-$00.67\_02 & 22:18:59.876 & $+$56:07:18.92 & O7      & V       & (n)z        & B              \\
BD +55 2722 B               & GOS 102.81$-$00.67\_03 & 22:18:58.832 & $+$56:07:23.47 & O9.5    & V       & \nodata     & \nodata        \\
ALS 12\,320                 & GOS 102.98$-$00.76\_01 & 22:20:21.783 & $+$56:08:52.21 & O7      & IV      & ((f))       & \nodata        \\
ALS 12\,370                 & GOS 103.05$-$01.41\_01 & 22:23:17.417 & $+$55:38:02.31 & O6.5    & V       & nn((f))     & \nodata        \\
ALS 12\,619                 & GOS 107.18$-$00.95\_01 & 22:47:50.595 & $+$58:05:12.39 & O7      & V       & ((f))z      & \nodata        \\
BD +55 2840                 & GOS 107.30$-$02.92\_01 & 22:55:08.492 & $+$56:22:58.88 & O7.5    & V       & (n)         & \nodata        \\
ALS 12\,688                 & GOS 107.42$-$02.87\_01 & 22:55:44.944 & $+$56:28:36.70 & O5.5    & V       & (n)((fc))   & B              \\
BD +62 2078                 & GOS 107.45$+$05.02\_01 & 22:25:33.579 & $+$63:25:02.62 & O7      & V       & ((f))z      & \nodata        \\
HD 213\,023 A               & GOS 107.73$+$05.20\_01 & 22:26:52.362 & $+$63:43:04.87 & O7.5    & V       & z           & \nodata        \\
ALS 12\,749                 & GOS 108.54$-$02.74\_01 & 23:02:44.556 & $+$57:03:50.21 & O9      & V       & \nodata     & \nodata        \\
Sh 2-158 2                  & GOS 111.53$+$00.82\_01 & 23:13:30.243 & $+$61:30:10.34 & O9.5:   & V       & \nodata     & B0.5: V        \\
Sh 2-158 1                  & GOS 111.53$+$00.82\_02 & 23:13:34.435 & $+$61:30:14.73 & O3.5    & V       & ((f*))      & O9.5: V        \\
BD +60 2635                 & GOS 115.90$-$01.16\_01 & 23:53:05.205 & $+$60:54:44.62 & O6      & V       & ((f))       & \nodata        \\
BD +66 1661                 & GOS 117.81$+$05.22\_01 & 23:57:32.603 & $+$67:33:15.28 & O9.2    & V       & \nodata     & \nodata        \\
V747 Cep                    & GOS 118.20$+$05.09\_01 & 00:01:46.870 & $+$67:30:25.13 & O5.5    & V       & (n)((f))    & \nodata        \\
BD +66 1675                 & GOS 118.21$+$04.99\_01 & 00:02:10.287 & $+$67:24:32.22 & O7.5    & V       & z           & \nodata        \\
BD +66 1674                 & GOS 118.22$+$05.01\_01 & 00:02:10.236 & $+$67:25:45.21 & O9.7    & IV:     & \nodata     & \nodata        \\
Tyc 4026-00424-1            & GOS 118.23$+$05.01\_01 & 00:02:19.027 & $+$67:25:38.55 & O7      & V       & ((f))z      & \nodata        \\
ALS 6351                    & GOS 122.57$+$00.12\_01 & 00:48:12.548 & $+$62:59:24.84 & O7      & V       & z           & \nodata        \\
BD +60 134                  & GOS 123.50$-$01.11\_01 & 00:56:14.216 & $+$61:45:36.91 & O5.5    & V       & (n)((f))    & \nodata        \\
HD 5689                     & GOS 123.86$+$00.75\_01 & 00:59:47.589 & $+$63:36:28.27 & O7      & V       & n((f))      & \nodata        \\
ALS 6967                    & GOS 132.94$-$01.39\_01 & 02:12:29.974 & $+$59:54:04.12 & O8      & V       & \nodata     & B0: V          \\
BD +61 411 A                & GOS 133.84$+$01.17\_01 & 02:26:34.379 & $+$62:00:42.32 & O6.5    & V       & ((f))z      & \nodata        \\
ALS 7833                    & GOS 146.25$+$03.12\_01 & 03:59:07.494 & $+$57:14:11.69 & O8      & V       & z           & \nodata        \\
MY Cam                      & GOS 146.27$+$03.14\_01 & 03:59:18.290 & $+$57:14:13.72 & O5.5    & V       & (n)         & O6.5 V(n)      \\
BD +50 886                  & GOS 150.60$-$00.94\_01 & 04:03:20.736 & $+$51:18:52.46 & O4      & V       & ((fc))      & \nodata        \\
BD +52 805                  & GOS 151.26$+$01.79\_01 & 04:18:35.640 & $+$52:51:54.21 & O8      & V       & (n)         & \nodata        \\
ALS 8272                    & GOS 168.75$+$01.00\_01 & 05:20:00.634 & $+$38:54:43.54 & O7      & V       & ((f))       & B0 III-V       \\
ALS 8294                    & GOS 173.61$-$01.72\_01 & 05:22:39.690 & $+$33:22:18.23 & O7      & V       & (n)z        & \nodata        \\
ALS 19\,265                 & GOS 186.10$+$06.56\_01 & 06:24:59.866 & $+$26:49:19.41 & O4.5    & V       & ((c))z      & \nodata        \\
HDE 256\,725 A              & GOS 192.32$+$03.36\_01 & 06:25:01.300 & $+$19:50:56.07 & O5      & V       & ((fc))      & \nodata        \\
HDE 256\,725 B              & GOS 192.32$+$03.36\_02 & 06:25:01.900 & $+$19:50:54.52 & O9.5    & V       & \nodata     & \nodata        \\
Tyc 0737-01170-1            & GOS 201.61$+$01.64\_01 & 06:36:29.003 & $+$10:49:20.73 & O7      & V       & z           & \nodata        \\
ALS 85                      & GOS 218.82$-$04.57\_01 & 06:45:48.800 & $-$07:18:46.63 & O7.5    & V       & \nodata     & \nodata        \\
ALS 207                     & GOS 231.49$-$04.40\_01 & 07:09:55.206 & $-$18:30:07.88 & O6.5    & V       & ((f))       & \nodata        \\
BD -15 1909                 & GOS 232.31$+$01.94\_01 & 07:34:58.463 & $-$16:14:23.22 & O6.5    & V       & ((f))z      & \nodata        \\
ALS 458                     & GOS 234.28$-$00.50\_01 & 07:30:01.272 & $-$19:08:34.98 & O6.5    & V       & ((f))z      & \nodata        \\
V441 Pup                    & GOS 240.28$-$04.05\_01 & 07:28:53.586 & $-$26:06:28.89 & O5:     & V       & e           & \nodata        \\
CPD -26 2704                & GOS 243.14$+$00.44\_01 & 07:52:36.593 & $-$26:22:41.99 & O7      & V       & (n)         & \nodata        \\
V467 Vel                    & GOS 265.20$-$02.18\_01 & 08:43:49.809 & $-$46:07:08.78 & O6.5    & V       & (n)((f))    & \nodata        \\
CPD -49 2322                & GOS 271.65$-$00.70\_01 & 09:15:52.787 & $-$50:00:43.82 & O7.5    & V       & ((f))       & \nodata        \\
HD 90\,273                  & GOS 284.18$-$00.25\_01 & 10:23:44.454 & $-$57:38:31.55 & ON7     & V       & ((f))       & \nodata        \\
THA 35-II-42                & GOS 284.52$-$00.24\_01 & 10:25:56.505 & $-$57:48:43.50 & O2      & I       & f*/WN5      & \nodata        \\
HD 89\,625                  & GOS 284.81$-$02.37\_01 & 10:18:58.251 & $-$59:46:04.30 & ON9.2   & IV      & n           & \nodata        \\
2MASS J10224377-5930182     & GOS 285.06$-$01.89\_01 & 10:22:43.774 & $-$59:30:18.22 & O8      & V       & (n)         & \nodata        \\
2MASS J10224096-5930305     & GOS 285.06$-$01.90\_01 & 10:22:40.961 & $-$59:30:30.58 & O7      & V       & ((f))z      & \nodata        \\
ALS 18\,551                 & GOS 289.73$-$01.26\_01 & 10:58:17.678 & $-$61:12:03.48 & O4.5    & V       & (n)z        & O4.5 V(n)z     \\
2MASS J10584671-6105512     & GOS 289.74$-$01.14\_01 & 10:58:46.716 & $-$61:05:51.22 & O8      & Iab     & f           & \nodata        \\
ALS 18\,553                 & GOS 289.74$-$01.18\_01 & 10:58:37.773 & $-$61:08:00.35 & O6      & II      & (f)         & \nodata        \\
2MASS J10583238-6110565     & GOS 289.75$-$01.23\_01 & 10:58:32.389 & $-$61:10:56.50 & O5      & V       & ((f))       & O7 V((f))      \\
THA 35-II-153               & GOS 289.79$-$01.18\_01 & 10:59:00.805 & $-$61:08:50.24 & O3.5    & I       & f*/WN7      & \nodata        \\
HD 97\,966                  & GOS 290.96$+$01.20\_01 & 11:15:11.779 & $-$59:24:58.28 & O7      & V       & ((f))z      & \nodata        \\
HD 97\,319                  & GOS 291.12$-$00.57\_01 & 11:11:06.156 & $-$61:07:04.56 & O7.5    & IV      & ((f))       & \nodata        \\
EM Car                      & GOS 291.22$-$00.50\_01 & 11:12:04.503 & $-$61:05:42.94 & O7.5    & V       & ((f))       & O7.5 V((f))    \\
NGC 3603 HST-51             & GOS 291.62$-$00.52\_15 & 11:15:07.498 & $-$61:15:46.35 & O5.5    & V       & (n)         & \nodata        \\
NGC 3603 HST-48             & GOS 291.62$-$00.53\_02 & 11:15:08.712 & $-$61:15:59.95 & O3.5    & I       & f*          & \nodata        \\
NGC 3603 HST-24             & GOS 291.62$-$00.53\_03 & 11:15:09.353 & $-$61:16:02.07 & O4      & IV      & (f)         & \nodata        \\
NGC 3603 MTT 25             & GOS 291.63$-$00.52\_01 & 11:15:11.317 & $-$61:15:55.63 & O5      & V       & (n)         & \nodata        \\
HD 99\,546                  & GOS 292.33$+$01.68\_01 & 11:26:36.905 & $-$59:26:13.61 & O7.5    & V       & ((f)) Nstr  & \nodata        \\
HD 110\,360                 & GOS 301.80$+$02.20\_01 & 12:42:12.700 & $-$60:39:08.71 & ON7     & V       & \nodata     & \nodata        \\
CPD -61 3973                & GOS 309.13$-$00.20\_01 & 13:45:21.103 & $-$62:25:35.37 & O7.5    & V       & ((f))       & \nodata        \\
HD 122\,313                 & GOS 311.18$-$00.54\_01 & 14:03:12.987 & $-$62:15:38.60 & O8.5    & V       & \nodata     & \nodata        \\
ALS 17\,591                 & GOS 320.32$-$01.16\_01 & 15:13:55.206 & $-$59:07:51.61 & O5:     & \nodata & n(f)p       & \nodata        \\
ALS 3386                    & GOS 326.31$+$00.74\_01 & 15:42:12.037 & $-$54:11:21.27 & O6      & Ia      & f           & \nodata        \\
ALS 18\,049                 & GOS 326.73$+$00.77\_01 & 15:44:20.206 & $-$53:54:41.31 & O9      & V       & \nodata     & \nodata        \\
Muzzio III-9                & GOS 327.39$-$00.62\_01 & 15:53:48.597 & $-$54:35:10.64 & O8      & Ib      & (f)         & \nodata        \\
HD 145\,217                 & GOS 332.29$+$00.77\_01 & 16:12:00.298 & $-$50:18:20.48 & O8      & V       & \nodata     & \nodata        \\
HD 144\,647                 & GOS 332.45$+$01.58\_01 & 16:09:16.197 & $-$49:36:21.75 & O8.5    & V       & (n)         & \nodata        \\
HDE 328\,209 AB             & GOS 338.49$+$02.85\_01 & 16:29:19.165 & $-$44:28:14.27 & ON9     & Ib-Ia   & p           & \nodata        \\
HDE 329\,100 A              & GOS 340.86$-$01.05\_01 & 16:54:42.304 & $-$45:15:14.80 & O8      & V       & (n)         & \nodata        \\
HDE 326\,775                & GOS 345.01$-$00.30\_01 & 17:05:31.316 & $-$41:31:20.12 & O6.5    & V       & (n)((f))z   & \nodata        \\
ALS 18\,769                 & GOS 348.67$-$00.79\_01 & 17:18:53.372 & $-$38:51:13.23 & O6      & II      & (f)         & \nodata        \\
HDE 323\,110                & GOS 349.65$-$00.67\_01 & 17:21:15.794 & $-$37:59:09.58 & ON9     & Ia      & \nodata     & \nodata        \\
Tyc 7370-00460-1            & GOS 352.57$+$02.11\_01 & 17:18:15.396 & $-$34:00:05.94 & O6      & V       & ((f))       & O8 V           \\
ALS 19\,693                 & GOS 353.07$+$00.64\_01 & 17:25:29.167 & $-$34:25:15.74 & O6      & V       & n((f))      & \nodata        \\
Pismis 24-15                & GOS 353.10$+$00.91\_01 & 17:24:28.952 & $-$34:14:50.68 & O7.5    & V       & z           & \nodata        \\
ALS 19\,692                 & GOS 353.11$+$00.65\_01 & 17:25:34.213 & $-$34:23:11.68 & O5.5    & IV      & (f)         & \nodata        \\
\enddata
\tablecomments{{\it GOSSS ID} is the identification for each star with ``GOS'' standing for ``Galactic O Star''.}
\label{spectralclasnew}
\end{deluxetable}

\eject

\addtolength{\topmargin}{35mm}
\addtolength{\textheight}{-25mm}

\begin{deluxetable}{llllcccl}
\rotate
\tablecaption{Spectral classifications for late-type stars erroneously classifed as O stars.}
\tablewidth{0pt}
\tablehead{\colhead{Name} & \colhead{RA (J2000)} & \colhead{dec (J2000)} & \colhead{Simbad} & \colhead{GOSSS} & \colhead{Ref.} & \colhead{Fixed?} & Comment}
\startdata
ALS~18\,890      & 19:35:23.587 & $-$16:19:46.78 & O+\ldots & F & None & No  &                                                                      \\ 
Tyc~0468-02112-1 & 19:16:44.466 & $+$02:28:41.60 & O\ldots  & F & None & Yes & Currently K86 given as reference for an F? spectral type.            \\ 
BD~+01~3974      & 19:22:01.484 & $+$02:12:01.90 & O        & F & K86  & No  &                                                                      \\ 
HDE~226\,144     & 19:50:59.376 & $+$36:00:03.24 & O9 V     & A & M80  & No  &                                                                      \\ 
BD~+37~3929      & 20:25:20.022 & $+$37:42:23.25 & O8f      & F & H56  & No  & Confusion with BD~+37~3927.                                          \\ 
BD~+40~4213      & 20:31:46.006 & $+$41:17:27.07 & O9.5 I   & F & M91  & No  & Not in the original reference, likely transcription error in SIMBAD. \\ 
BD~+45~4132 A    & 23:04:14.816 & $+$46:36:33.63 & O        & F & None & Yes &                                                                      \\ 
BD~+32~4642 A    & 23:25:38.697 & $+$33:26:16.94 & O        & F & None & Yes &                                                                      \\ 
BD~+61~100 AB    & 00:30:32.445 & $+$62:34:00.93 & O/B2     & G & R89  & No  & A different reference is given now but the link is broken.           \\ 
BD~-03~2178      & 08:02:10.340 & $-$04:01:36.39 & O5       & K & M76  & Yes & Confusion with BD~-03~2179, a sdO.                                   \\ 
CPD~$-$61~4623   & 14:35:36.520 & $-$61:34:12.77 & O        & K & None & Yes &                                                                      \\ 
\enddata
\tablerefs{B60:\citet{Brod60}, H56:\citet{HiltJohn56}, K86:\citet{KellKilk86}, M76: \citet{MacCBide76}, M80:\citet{MikoMiko80}, M91:\citet{MassThom91}, R89:\cite{Rado89}}
\label{spectralclasbad}
\end{deluxetable}

\end{document}